\documentclass[12pt]{article}
\usepackage[margin=2 cm]{geometry}
\usepackage{comment}
\usepackage{graphicx}
\usepackage[font={small,it}]{caption}
\usepackage{mwe}
\usepackage[nottoc]{tocbibind}
\usepackage{amsmath,amssymb,extarrows,mathtools,graphicx,subfigure,setspace}
\usepackage{cite}
\usepackage{braket}
\usepackage{bbm}
\usepackage{mathrsfs}
\usepackage{epsfig}
\usepackage[section]{placeins}
\usepackage{slashed}
\usepackage{color}
\usepackage{caption}
\usepackage{amsmath}
\usepackage{hyperref}
\makeatother

\newcommand{\be}{\begin{equation}}
\newcommand{\bea}{\begin{eqnarray}}
\newcommand{\eea}{\end{eqnarray}}
\newcommand{\ba}{\begin{array}}
\newcommand{\ea}{\end{array}}
\newcommand{\ee}{\end{equation}}
\newcommand{\bes}{\begin{equation*}}
\newcommand{\beas}{\begin{eqnarray*}}
\newcommand{\eeas}{\end{eqnarray*}}
\newcommand{\bas}{\begin{array*}}
\newcommand{\eas}{\end{array*}}
\newcommand{\ees}{\end{equation*}}

\numberwithin{equation}{section}

\begin{document}

\onehalfspacing
\vfill
\begin{titlepage}
\vspace{10mm}

\begin{center}

\vspace*{10mm}
\vspace*{1mm}
{\Large  \textbf{Entanglement wedge reconstruction and correlation measures in mixed states: modular flows versus quantum recovery channels}} 
 \vspace*{1cm}
 
{$\text{Mahdis Ghodrati}^{a}$}

\vspace*{8mm}
{ \textsl{
$^a $Department of Physics, Sharif University of Technology, Tehran 1458889694, Iran}

\textsl{
$^b $Shing-Tung Yau Center of Southeast University, Nanjing 210096, China}

\textsl{
$^c $Center for Gravitation and Cosmology,Yangzhou University, Yangzhou 225009, China}

} 
 \vspace*{1cm}

\textsl{E-mails: {\href{mahdisg@yzu.edu.cn}{mahdisg@yzu.edu.cn}}}
 \vspace*{2mm}

\vspace*{1.7cm}

\end{center}

\begin{abstract}

In this work we study the nature of correlations among mixed states in the setup of two symmetric strips. We use various tools to determine how the bulk geometry could be reconstructed from the boundary mixed information. These tools would be the modular Hamiltonian and modular flow, OPE blocks, quantum recovery channels such as Petz map, Uhlmann holonomy and Wilson lines. We comment on the similarities and connections between these approaches in our symmetric setup of a mixed system. Specially, we use parameters such as dissipation which is being modeled by the mass of graviton, and also the same sign charge of the two strips to find connections between these different approaches.  Then, using Uhlmann fidelity as the correlation measure, we look into the various types of correlations in mixed systems such as discord. Next, we use simple results of modular Hamiltonian for fermions to get insights about the relations between modular flow and entanglement and complexity of purification (EoP/CoP), and also behavior of modular flows in confining geometries. Finally, we study the dynamics of correlations using various information speeds and also model of void formation in CFTs and again we comment on their relationships with the behavior of EoP and CoP.

 \end{abstract}

\end{titlepage}

\tableofcontents


\section{Introduction}

In the setup of holography, out of information and entanglement in the boundary field theory side, the one dimension higher bulk geometry could be reconstructed. Various models of bulk reconstruction has been discussed in the literature, for instance HKLL \cite{Hamilton:2006az}, DHW \cite{Dong:2016eik}, ADH \cite{Almheiri:2014lwa}, and FL \cite{Faulkner:2017vdd} and also reference \cite{Bao:2019bib} which uses extremal area variations. In other works such as \cite{Roy:2018ehv}, a recipe for extracting the specific bulk \textit{metric} from the boundary state has been proposed. Various models of tensor networks and also quantum error corrections have been employed for studying different features of bulk reconstruction. Most of these methods specifically use the subregion/subregion duality and they were studied in a fixed background.

Recent works on the connections between geometry in the bulk and information on the boundary CFT, in the setup of modular Berry connection and Berry curvature of modular Hamiltonians \cite{Czech:2019vih}, see also \cite{Czech:2018kvg, Czech:2017zfq}, have also been presented. Our motivation here is that by using the results of our previous work \cite{Ghodrati:2019hnn}, to show the connections between measures of purifications and the two methods of bulk reconstructions, modular flow and quantum error correction. Specifically  we would like  to demonstrate how in the mixed quantum systems, modular flow or quantum recovery channels would change by changing the parameters of the system. To attain this goal, we implement these two approaches in several backgrounds, charged case and massive gravity backgrounds where the graviton has a finite mass that simulates dissipation in the model. By changing the mass of graviton $m$ or the charge of the strips $q$, in these two bulk reconstruction approaches, we track their effects which would lead to the connections between modular flows and quantum error correction. 

Here, similar to \cite{Ghodrati:2019hnn} we consider a symmetric setup where two strips have the same width $l$, with a distance $D$ between them. We concentrate on how the modular flow, modular zero modes, soft modes or edge modes and modular Hamiltonian would change when the two subregions move from far distances (when $I = 0$ or $\text{EoP} = 0$ to a closer distance to each other ($I \ne 0$ and $\text{EoP} \ne 0$), where the system undergo a phase transition from two pure sub-spaces to one mixed system. This would be done by tracking the extremal surfaces at each stage of the evolution. In general, the quantum information are being processed in quantum channels where by passing time, more errors would be accumulated. On the other hand, under modular evolution, the modular Hamiltonian become more and more non-local and get larger commutators with all other operators of the system.  Another hint for their connections came from the value of conical $2\pi$ at the $x$-axis in the connected entanglement wedge part  which is related to the $2\pi$ constant of CoP in $2d$ that was found in \cite{Ghodrati:2019hnn}. 

In addition, in \cite{Cotler:2017erl}, some specific connections between universal recovery channels and modular Hamiltonian have been mentioned as they could show that by perturbing a bulk state in a direction of a bulk operator which is within a boundary subregion causal wedge,  the modular Hamiltonian of the boundary would correspondingly respond. This would be related to the non-commutative version of Bayes rule which then could be used to reconstruct the lost information similar to a quantum error correction system. Note also that, the connections between thermal quantum chaos and quantum recovery channels have already been discussed in quantum information literature, so we expect such connections also work for modular chaos which then could give us further information about the mechanisms of bulk emergence. Therefore, the connections between various universal recovery channels such as \textit{twirled Petz map} or normal Petz map and modular Berry flow could be discussed.

Most of the discussions in bulk reconstructions have been done for pure states but here we are interested in a mixed setup. The connections between modular Berry connection and entanglement wedge for pure systems has been discussed in \cite{Czech:2019vih, Czech:2018kvg,Czech:2016xec}. As for the case of mixed states, in \cite{Kirklin:2019ror}, it has been proposed that the Berry phase along the ``Uhlmann parallel paths'' would be the integral of a connection which its curvature would be the symplectic form of the entanglement wedge ($E_W$). So Berry phase and the symplectic form of $E_W$, has been linked in that work. In \cite{Chen:2019gbt}, the reconstruction of entanglement wedge using Petz map has been discussed. Using the results of these works, one then could catch the connections between Petz map and Berry and Uhlmann phase and also the symplectic form of $E_W$. Most of the discussions of holographic bulk reconstructions using error corrections have been done for the two dimensional JT gravity  background, while our setup of two intervals in the $\text{AdS}_3$ background, would be richer than those discussed in \cite{Penington:2019kki,Chen:2019gbt}. Also, the connections between quantum error correcting properties of holography and modular flow, modular connection, and symplectic form of entanglement wedges could then be perceived.

One other compelling phenomenon would be the phase transitions between zero mutual information (MI) between the strips when they are far apart and a jump in MI when they get closer than a critical distance $D_c$.  As  Berry curvature of modular Hamiltonians could sews together the orthonormal coordinate systems along the HRRT surfaces and lead to the bulk reconstruction, \cite{Czech:2019vih}, we would like to examine how this proposal would work for the case of two disconnected versus connected subsystems and then investigate the properties of the modular Hamiltonian and modular flows during the phase transitions. 

\section{The Setup}

\begin{figure}[ht!]
 \centering
  \includegraphics[width=5cm] {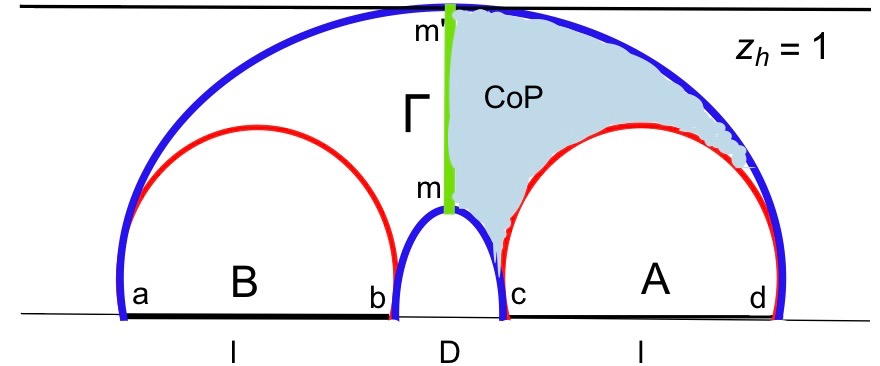}
  \caption{Two strips of $A$ and $B$ with length $l$ and with the distance $D$ between them is shown. The two turning points would correspond to region $ad$ and $bc$ which are $m$ and $m^\prime$, and $\Gamma$ is the minimal cross section of the ``connected'' entanglement wedge.}
 \label{fig:strips}
\end{figure}

Our setup consists of two subregions $A$ and $B$ which are infinite strips with the width of $l$ separated by the distance $D$ on the same side of the boundary as
\begin{gather}
A:= \{l+D/2 >x_1>D/2, -\infty <x_i<\infty, i=2,3,...,d-1\}, \nonumber\\
B:= \{ -l-D/2 < x_1 < -D/2, -\infty < x_i < \infty, i=2,3,...,d-1\},
\end{gather}
which is shown in figure \ref{fig:strips}.
\vspace*{6px}
The critical distance $D_c$ for each dimension could be found by setting in each case  the mutual information zero, $I(D,l)=0$,  as we did in our previous work \cite{Ghodrati:2019hnn}.  The background was chosen to be the Schwarzchild AdS black brane in the form
\begin{gather}
ds^2=\frac{1}{z^2} \Big [ -f(z) dt^2+\frac{dz^2}{f(z)}+d \vec{x}^2_{d-1} \Big], \ \ \ \ \ f(z):=1-z^d/z_h^d.
\end{gather}  
The entanglement of purification (EoP) between the two states could be computed using the area of the surface $\Gamma$ shown in green and the complexity of purification (CoP) would be the volume of blue region as shown in \cite{Ghodrati:2019hnn}. 

The effects of charge $q$ and mass of graviton $m$ on EoP and CoP could then be studied by considering the background of charged-massive BTZ black hole in the following form \cite{Hendi:2016pvx}
\begin{equation}
ds^2=\frac{1}{z^2}[-f(z)dt^2+\frac{dz^2}{f(z)}+dx^ { 2 }]~~~\mathrm{with} ~~~f(z)=-\Lambda-m_0 z^2-2q^2 z^2 \ln (\frac{1}{z \ell})+m^2 c c_1 z,
\label{Metric}
\end{equation}%
which is a solution to Einstein equations for the three dimensional Einstein-massive gravity with the action \cite{Hendi:2016pvx}
\begin{equation}
\mathcal{I}=-\frac{1}{16\pi }\int d^{3}x\sqrt{-g}\left[ \mathcal{R}-2\Lambda+L(\mathcal{F})+m^{2}\sum_{i}^{4}c_{i}\mathcal{U}_{i}(g,h)\right],
\label{Action}
\end{equation}%
where $\mathcal{R}$ is the scalar curvature, $L(\mathcal{F})$ is an arbitrary Lagrangian of electrodynamics and $\Lambda$ is the cosmological constant.

In relation \ref{Metric}, $m_0$ is an integration constant which is related to the total mass of black hole and we could set it as $m_0=1$. In addition, one could set $c=c_1=1$, without any loss of generality. In the action, $m$ is the mass of graviton in the theory. In the holographic framework, the massive terms in the gravitational action break the diffeomorphism symmetry in the bulk, which then would correspond to momentum dissipations in the dual boundary field theory as shown in \cite{Blake:2013bqa}. So the solution would be a  ``massive-charged BTZ black hole'' in a ``massive gravity theory''. Here we are interested in studying the effects of the parameters $m$  and $q$ on various bulk reconstruction formulations such as quantum error corrections, modular flow, CC flow, OPE block, and the dynamics.

We first bring here some of the results we found in \cite{Ghodrati:2019hnn} to summarize the effects of $m$, $q$ and $d$ on EoP and CoP.  The figures in this section are brought from our previous work \cite{Ghodrati:2019hnn} to review the results that we are going to use here.
  \begin{figure}[ht!]
 \centering
  \includegraphics[width=5cm] {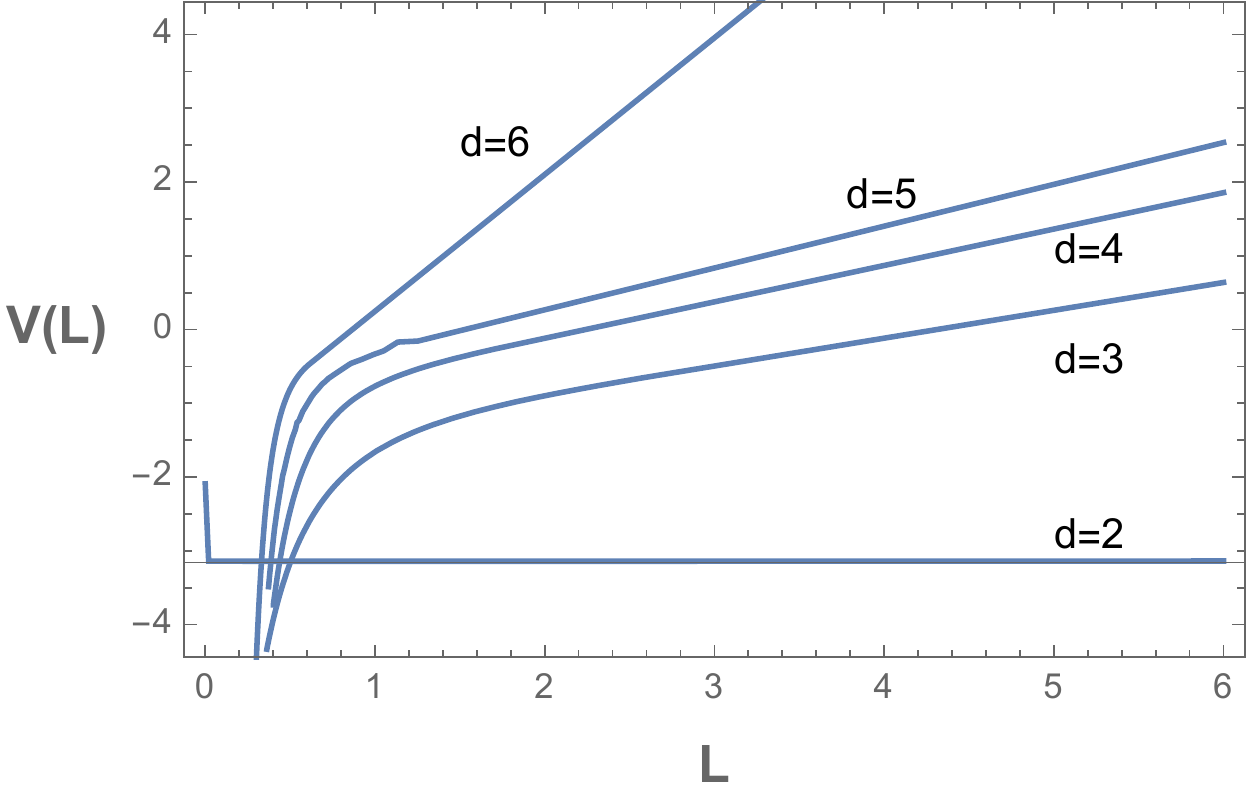}
    \includegraphics[width=5cm] {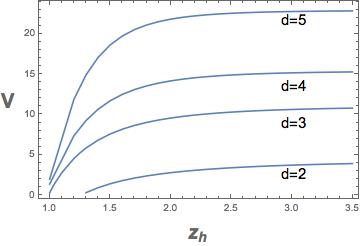}
  \caption{The volume $V(L)$ corresponding to each length of strip with the width $L$, for various dimensions $d$. }
 \label{fig:CoPvolume}
\end{figure}
 \begin{figure}[ht!]
 \centering
    \includegraphics[width=5cm] {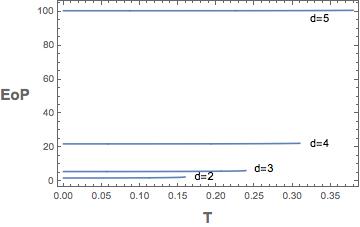}
  \caption{ The EoP curves for different dimensions are shown here. For both cases we took $l=20$ and $D=0.3$.}
 \label{fig:Eopp1122}
\end{figure}
\begin{figure}[ht!]
\centering
\includegraphics[width=0.25\textwidth]{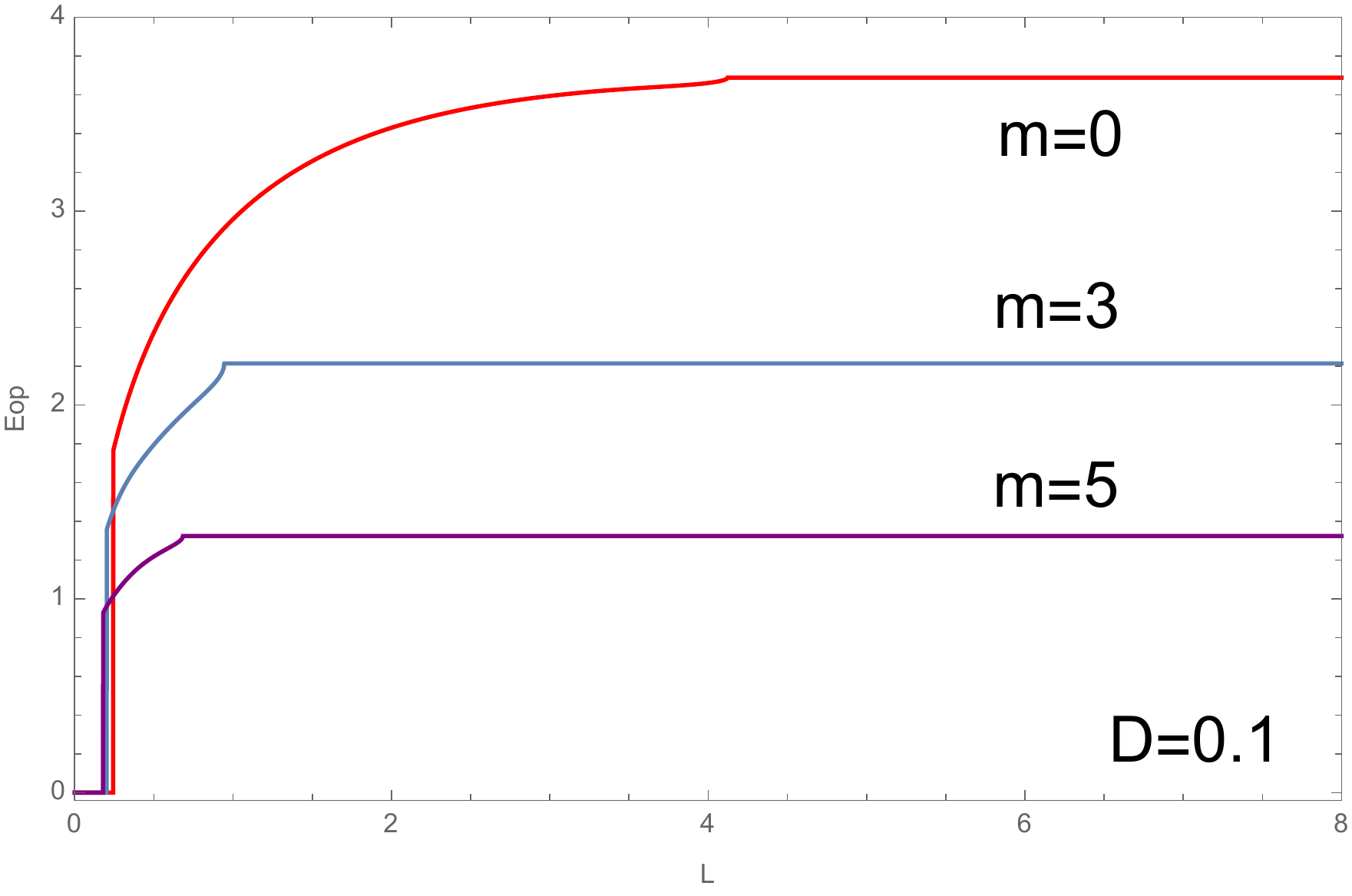}
\caption{ EoP as a function of $l$ with $D=0.1$.}\label{fig-D-EOP}
\end{figure}
 \begin{figure}[ht!]
\centering
\includegraphics[width=0.25\textwidth]{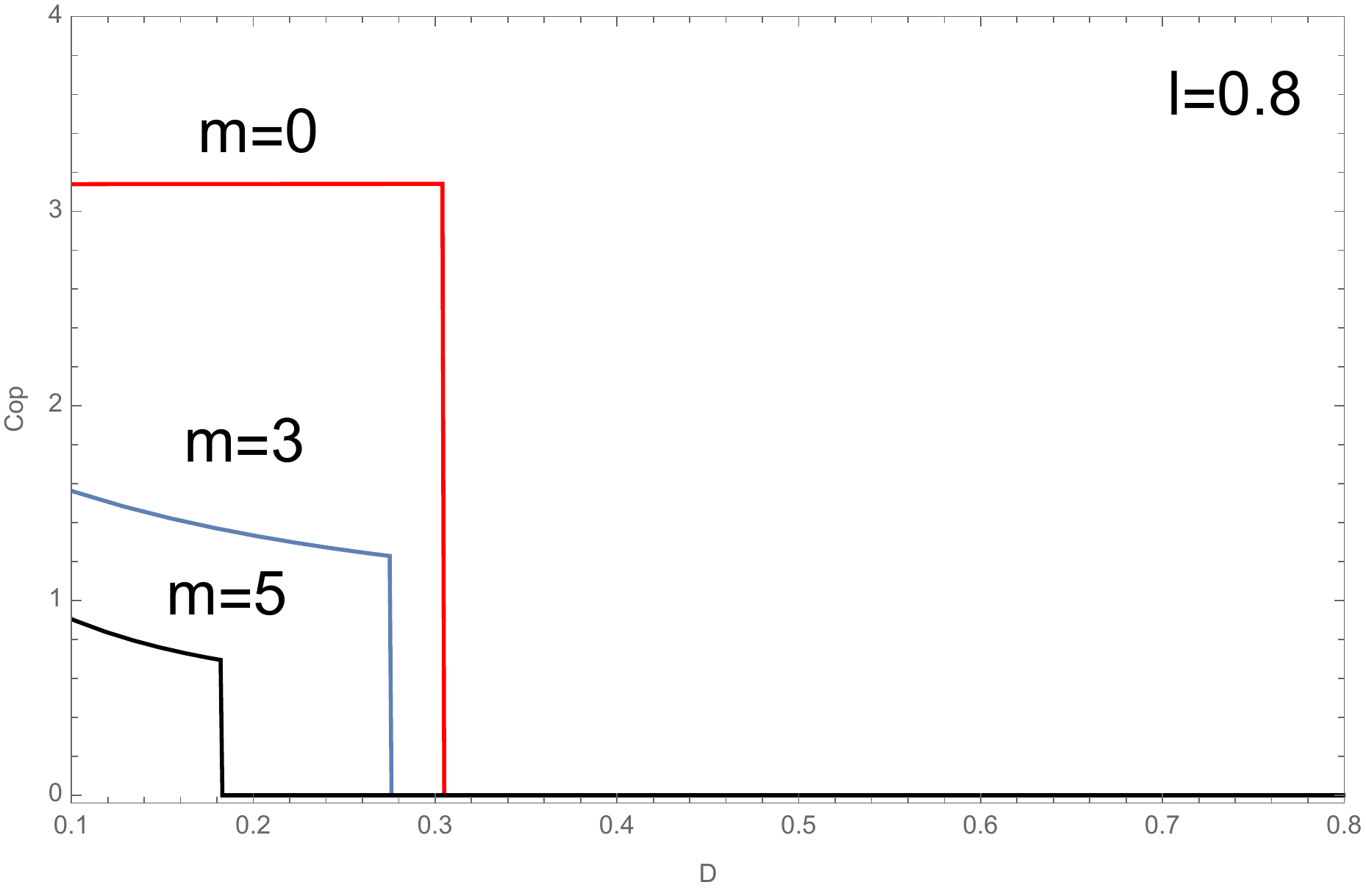}\hspace{0.5cm}
\includegraphics[width=0.25\textwidth]{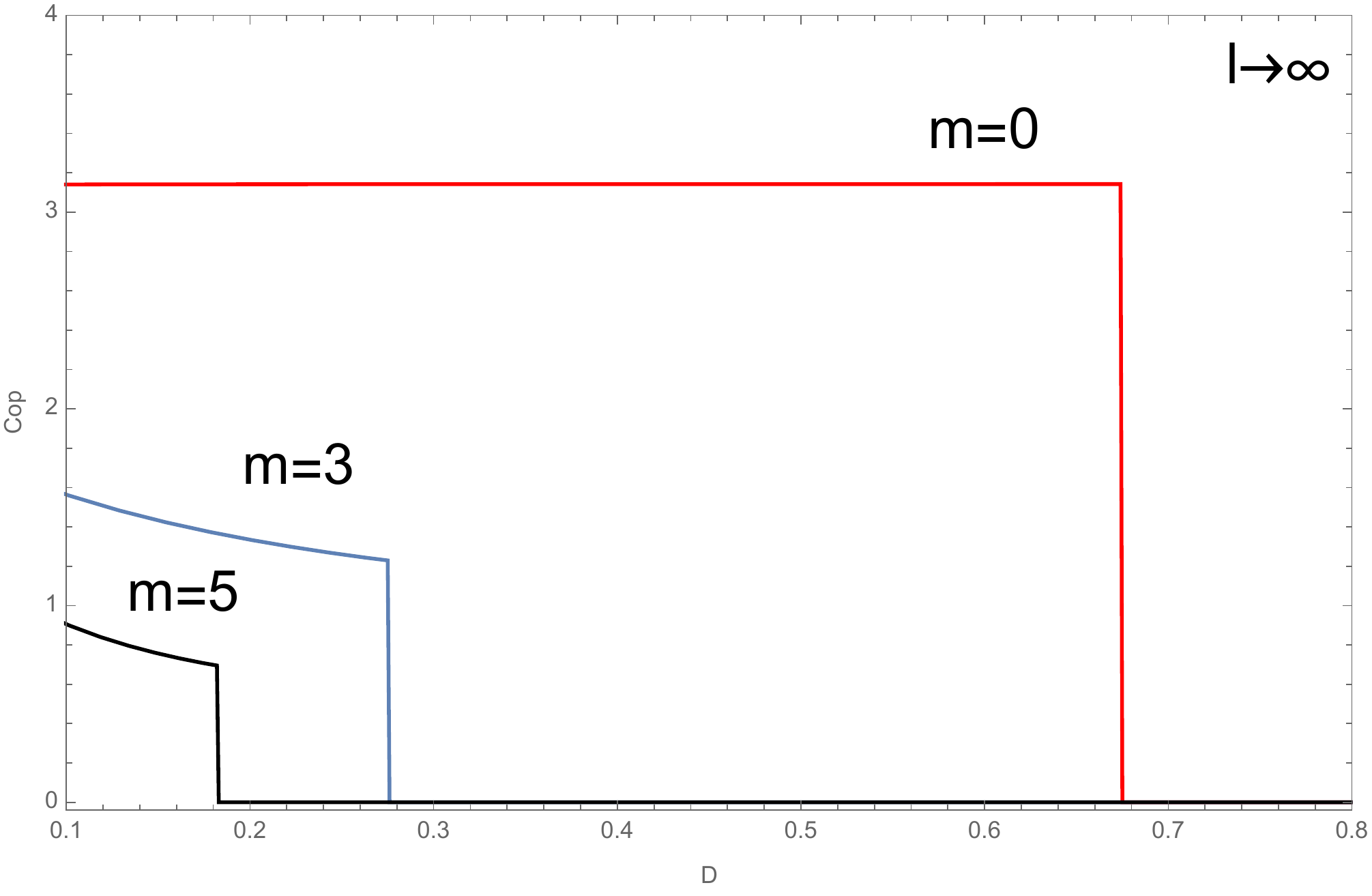}\hspace{0.5cm}
\includegraphics[width=0.25\textwidth]{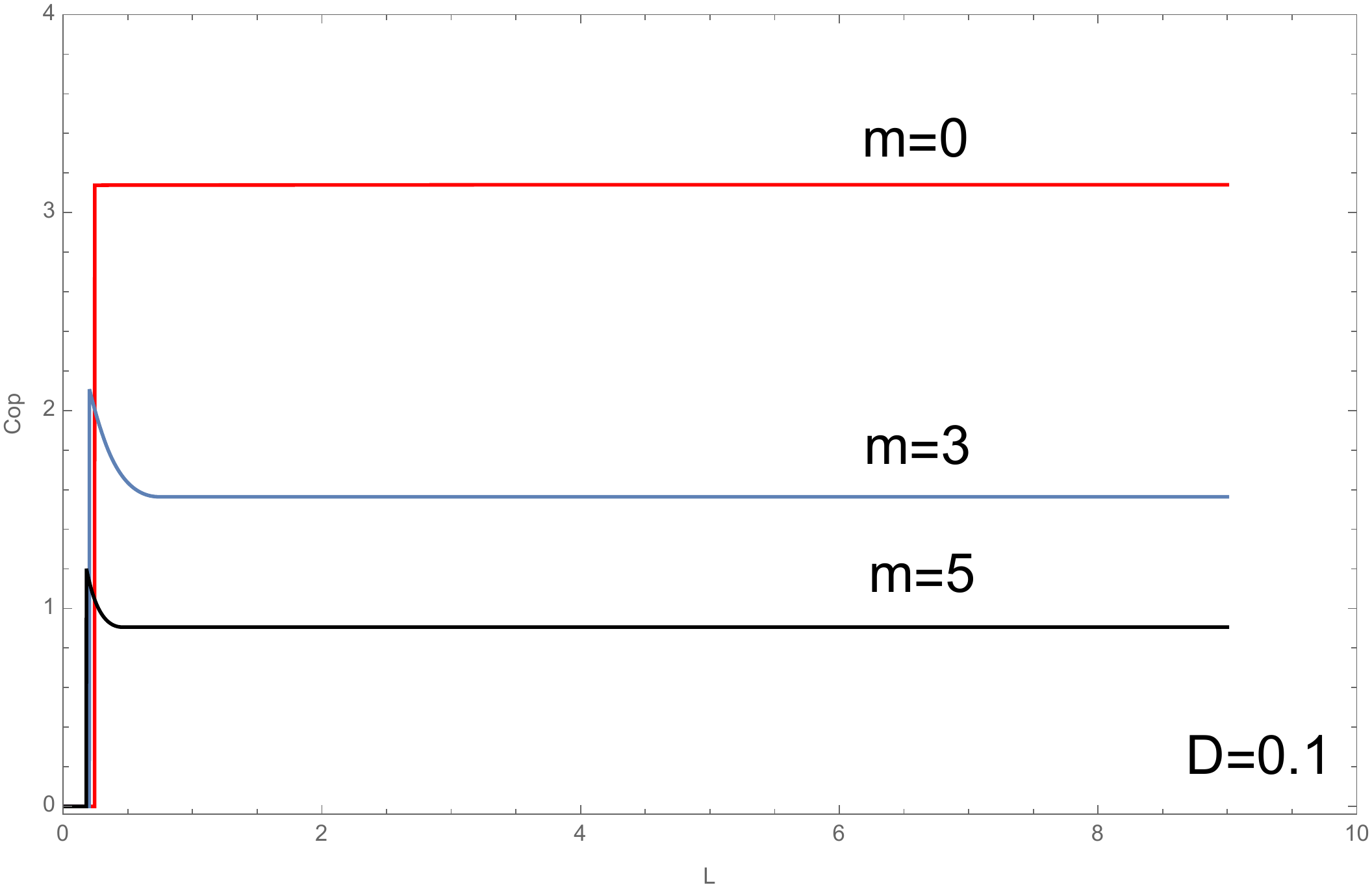}
\caption{The CoP as a function of  $D$ with $l=0.8$ (left)  and $l=\infty$ (middle), and CoP as a function of $l$ with $D=0.1$ (right) is shown here.}\label{fig-D-COP}
\end{figure}
\begin{figure}[ht!]
\centering
\includegraphics[width=0.25\textwidth]{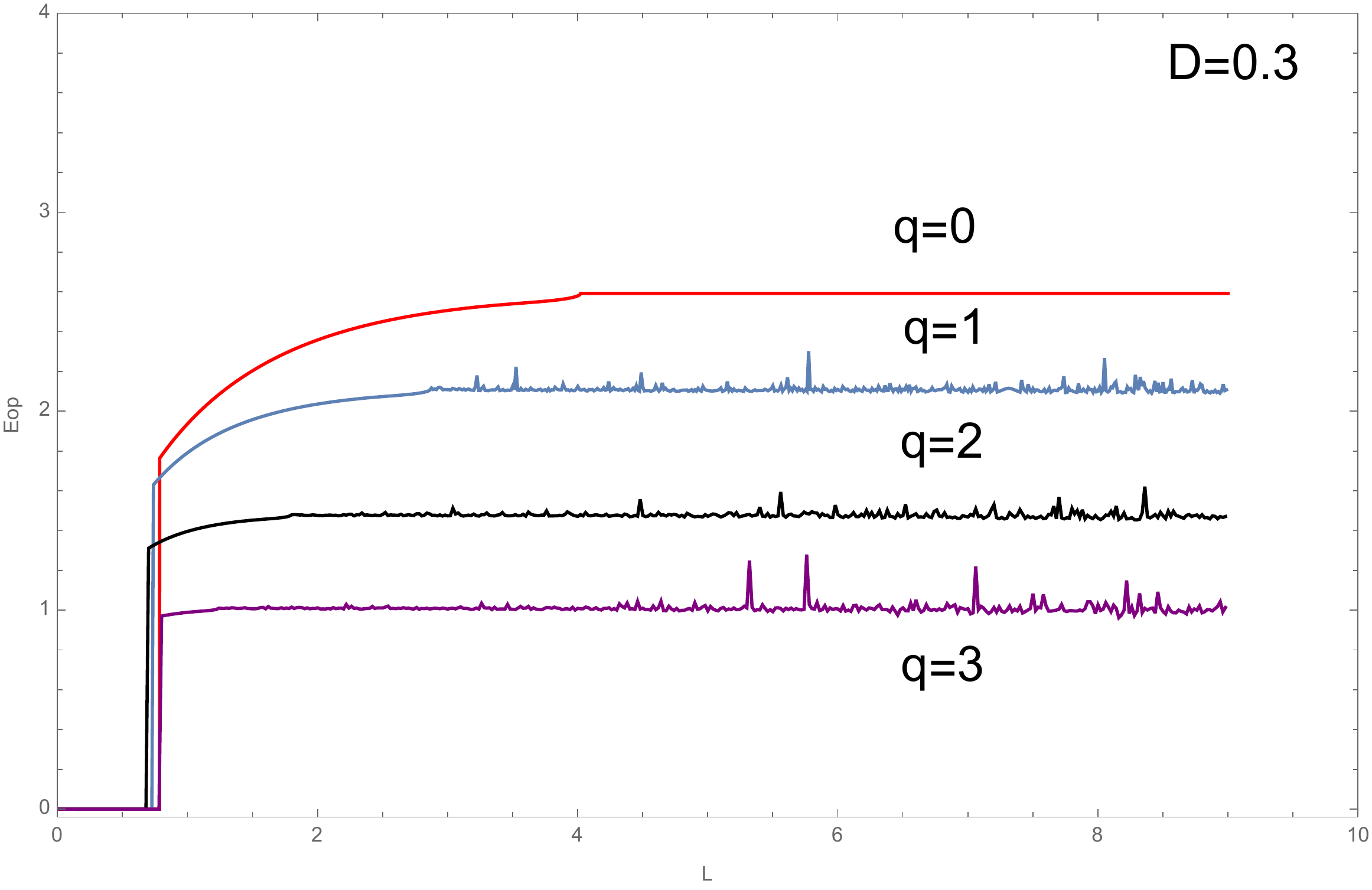}
\caption{The relationship between EoP and $l$ with $D=0.3$.}\label{fig-D-EOP2}
\end{figure}
\begin{figure}[ht!]
\centering
\includegraphics[width=0.25\textwidth]{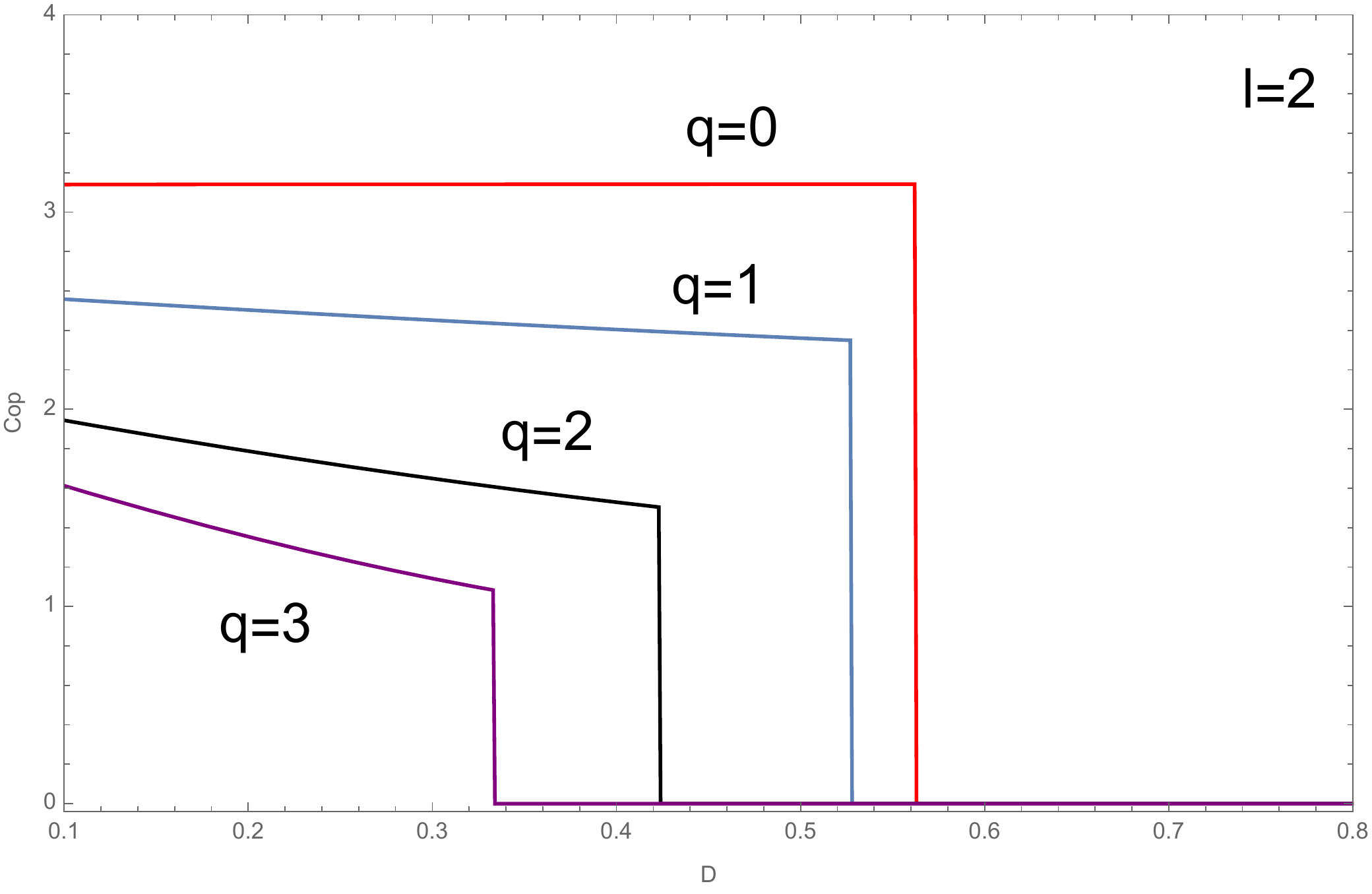}\hspace{1cm}
\includegraphics[width=0.25\textwidth]{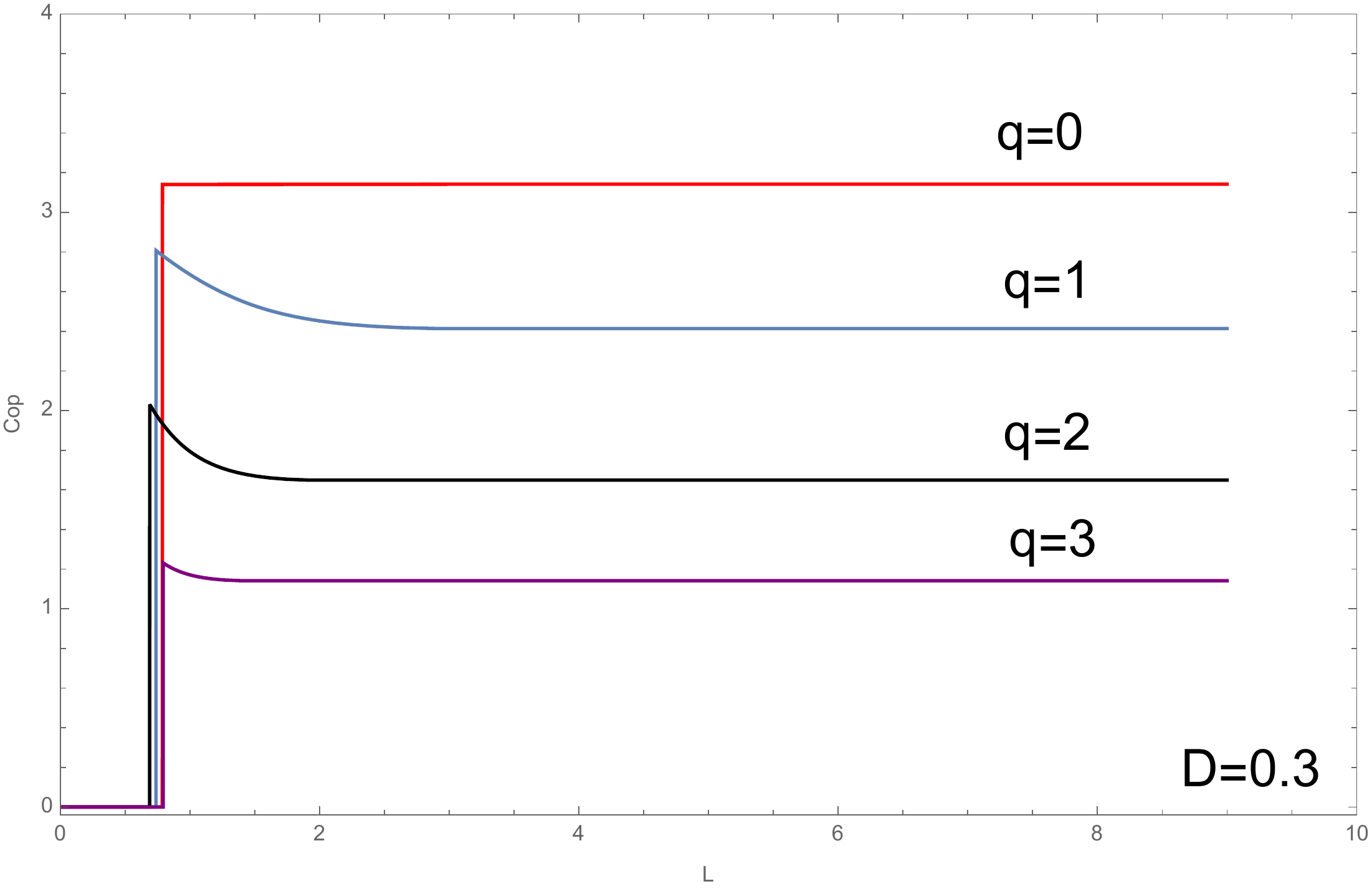}
\caption{The relationship between CoP and $l$ with $D=0.3$.}\label{fig-D-COP2}
\end{figure}

 As you could see from the figures, we found in \cite{Ghodrati:2019hnn} that in the case of massive gravity where the graviton gain a small mass, EoP/CoP would be lower than the massless case. We expect to see this decreasing effects in specific stages of various bulk reconstruction methods, e.g, tensor networks, modular Berry connection and quantum error corrections.

 \begin{figure}[ht!]
 \centering
  \includegraphics[width=5 cm] {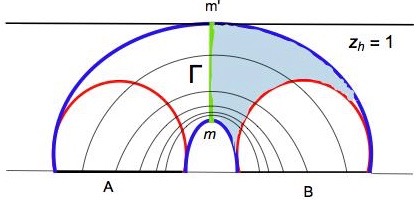} \hspace{1cm}
    \includegraphics[width=5cm] {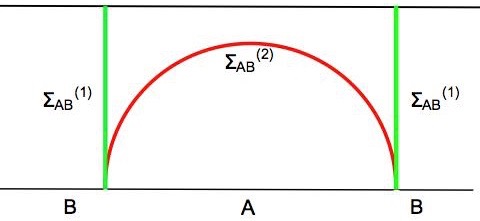}
  \caption{The modular flow and edge modes along the minimal wedge cross section could be studied to depict the relation between the flow of modular zero-modes, the holonomy which they create and their effects on EoP and CoP of mixed states.}
 \label{fig:Bitthread2}
\end{figure}

In \cite{Ghodrati:2019hnn}, we have proposed that the correlation strengths and density of bit threads would have a decreasing behavior along $\Gamma$ from the turning point $m$ toward the turning point $m^\prime$. We would like to check this observation using other bulk reconstruction methods as well. We could show that modular zero modes would have a decaying behavior along this path from m to $m^\prime$.

We give a short overview of modular Hamiltonian and modular Berry phase, quantum recovery channels and the connections between them and then we study these procedures in our setup. We also study the effects of the parameters we mentioned on the formulations of bulk reconstruction.

 \section{Entanglement Wedge Cross Section and Modular Flow}
One of the main question of holography is that how using the subregion duality and the operator algebra of the boundary CFT, the physics of bulk entanglement wedge would emerge. One of the guiding principle would be a duality found in \cite{Takayanagi:2017knl}, which relates td and form a mixed state entanglement of purification between mixed states to the area of the minimal entanglement wedge cross section, i.e, the $E_W=E_P$ conjecture. This duality would have connections with the behavior of modular zero-modes and modular Hamiltonian. We would like to connect $E_W=E_P$ conjecture to the setup of \cite{Czech:2019vih} using the duality between the modular Hamiltonians of the bulk and boundary within the code subspace, i.e, the JLMS formula \cite{Jafferis:2015del} $H_{\text{mod}}^{\text{CFT}}=\frac{A}{4 G_N}+H_{\text{mod}}^{\text{bulk}}+...+O(G_N)$. 

Modular Hamiltonian is defined as $H_{\text{mod}}=-\log \rho$, and $A$ is the area operator of HRRT surface. The gravity dual of modular Hamiltonian operators has been discussed in \cite{Jafferis:2014lza}.  For a fixed spatial region $R$, modular Hamiltonian could be written as $H=-\log \rho_R \otimes I$,  and the modular evolution of a density matrix $\rho$ could be written as $ \rho_\alpha \equiv e^{-i \alpha H} \rho \ e^{i \alpha H}$. Here $H$ is a state-dependent operator and also due to a kink at the boundary of $R$, it would be a non-smooth operator. However, the full modular Hamiltonian for the region $V$  written as $ \hat{H}_V=H_V -H_{\bar{V}}$, would be smooth, \cite{Jafferis:2014lza}, since it has support on all regions of space. Also, this operator always annihilates the vacuum, $\hat{H}_V \ket{0}=0$. 

 In \cite{Casini:2017roe}, it has been shown that the full modular Hamiltonian could be written in terms of energy momentum tensor in the form as
   \begin{gather}
  \hat{H}_\gamma= 2\pi \int d^{d-2} x^{\perp} \int_{-\infty}^{\infty} d\lambda (\lambda-\gamma(x^\perp)) T_{\lambda \lambda} (\lambda, x^\perp).
  \end{gather}
So the general form of modular Hamiltonian for a region $A$ would be as $H_A= \int_A d \sigma^\mu T_{\mu \nu} \eta^\nu$,
where $\eta^\nu$ is a time-like vector that generates the modular flow for the region $A$.

Based on the perspective of \cite{Jafferis:2015del}, the bulk and boundary modular flows and their corresponding relative entropies are dual to each other.  The goal would be using this duality, and also the intuitions from other formalisms such as tensor network and bit thread picture, to check how zero-modes and edge modes would behave along the surface $\Gamma$. Also, we would like to study how as these modes are correlated and form a mixed state, they could glue the entanglement wedges of each boundary CFT, and then how they could reconstruct the curvature of the metric in the bulk. From the intuitions of bit thread, tensor network and $E_W=E_P$ conjecture we claim several statements here.

First, note that the recent studies in \cite{Czech:2019vih}  were focused on edge modes along the Hubeny-Rangamani-Ryu-Takayanagi (HRRT) surface. Here, as we study the correlations for mixed states and the connections between modular flow, CoP and EoP, we should focus our studies on not only the zero-modes, but the flow of the whole ``normal modes", and not just along HRRT surface but along the ``\textit{minimal wedge cross section}" \ between the two mixed CFTs, $\Gamma$.

\textit{Claim: Through the entanglement wedge cross section, the maximum amount of flow of \textbf{edge modes} \cite{Wong:2017pdm} would pass.}

Note that the flow of these modes could be formulated using amplitude of the wavefunction between states on $A$ and $B$ as $ \bra{\phi_A} e^{-\pi K_A} \mathcal{J} \ket{\phi_B} $ where $K_A$ is related to modular Hamiltonian and would implement the CPT transformation as $\mathcal {J}: H_B \to H_A$. The operator $K_A$ in a special case where the angular coordinate $\theta$ around the entangling surface is zero, could be written as $K_A= \int_A d^{D-1} x f( x) T_{00}$, where $T_{00}$ is the energy density and $f(x)$ is a weight function. So along $\Gamma$, this weight function would make the amplitude of transformation maximum.  Also note that the pattern of \textbf{entanglement of purification} between mixed boundary CFTs would create the bulk curvature.  As MI and EoP have decreasing gradients along the surface $\Gamma$, the bulk would be negatively curved.

The next claim involves the structure of modular Hamiltonian and modular zero modes using the structure of \cite{Czech:2019vih}.  \textit{Claim: In our setup, moving from the HRRT surface with turning point $m$ to the HRRT surface with turning point $m'$, the change in the local gauge and the modular Hamiltonian,  would be proportional to the area of minimal wedge cross section $\Gamma$. }So the structure of entanglement of purification and the connections between the algebra of the two Hilbert space of \textit{mixed states} would dictate the structure of bulk curvatures. For instance, similar to  \cite{Czech:2019vih}, one could imagine that for each pair of qubit on the two subregions $A$, $B$, a map between two Hilbert space would transform under the action of a local $SU(2)$ symmetry as $\ket{i}_A \to \ket{\tilde{i} }_B=    i = \sum_j= W_{ij} \ket{j}_B,\   \text{where} \ 
W_{ij} \to U^{\dagger}_{A,ik} W_{kl} U_{B,lj}$.
The matrix $W_{ij}$ could be considered as the Wilson loop between the two regions connecting qubits on each side. Based on the idea of ER$=$EPR, each of these would be a wormhole connecting the qubits. The pattern of \textit{``Entanglement of Purification"} would then dictate the local density of threads and their gradient along the surface $\Gamma$, moving from turning point $m$ to $m'$. 

Then we have the following picture: \textit{Claim: As the correlations between pairs in the two subregions are stronger where they are closer to each other, the density of threads would be higher around those points, (also the gradients of change of correlations would be higher), the densities of bit threads, modular flow and edge modes are bigger around $m$, the holonomy is bigger around the $m$ compared to $m'$, the density of the symmetry operators or symmetry generators is higher, the Bures metric has a bigger absolute value and also the curvature in the bulk is bigger around the HRRT surface associated to turning point $m$ compared to the one passing through $m'$.}

These claims indicate that the gradient of modular flow would be similar to the behavior of bit threads, or even a gradient of an electric field between two charged strips, as the flow between two equal and symmetric strips as shown in figure \ref{fig:Bitthread2} would be symmetric and decreasing along $\Gamma$, moving further away from the strips. Also, we expect that similar to bit threads formalism, the modular flow has a bound of $J \le \frac{1}{4G_N}$ which has been observed in \cite{deBoer:2019uem} too. So, the gradient of bit threads could be found by considering the behavior of modular scrambling modes around various RT surface along the minimal entanglement wedge cross section. In a dynamical setup, the change in modular Hamiltonian of the boundary region (for instance by closing the two strips to each other in our case), would act as a vector flow close to RT surface which would depend on the boost vector and gradient of fields in the bulk.  The effects of varying the physical parameters of the boundary system, on the bulk reconstruction, could also be studied through the structure of modular Hamiltonian.

We show that mass of graviton $m$ or charge of the system, $q$, within the ``code subspace'', would decrease the rate of change of the matrix elements of modular flow $\delta H_{\text{mod}} (s) / \delta s $, and bring the bound lower than $2\pi$. As the mass of graviton corresponds to the viscosity parameter in the field theory side, one would expect that increasing $m$ would damp the ``modular scrambling modes''.
 
 The change in the modular Hamiltonian of the system due to the viscosity and dissipations could be tracked by considering the change in the energy momentum tensor which in terms of the viscosity coefficients could be written as \cite{Pimentel:2016jlm}
 \begin{gather}
 T^{\alpha \beta}= \rho u^\alpha u^\beta + q^\alpha u^\beta+q^\beta u^\alpha +(p-\zeta \Theta) h^{\alpha \beta}-2 \eta \sigma^{\alpha \beta},
 \end{gather}
 where $\rho$ is the energy density, $u^\alpha$ is the velocity of the  ``comoving'' observer and $q^\alpha$ is the spacelike   ``heat'' flux vector that satisfy $q^\alpha u_\alpha=0$. Also, note that $\zeta >0$ and $\eta >0$, which are the bulk and shear viscosity respectively. The parameter $\Theta= u^\alpha_{; \alpha}$ is the expansion and  $\sigma^{\alpha \beta}$ is the shear tensor which has the relation
  \begin{gather}
 \sigma^{\alpha \beta} =\frac{1}{2} \Big( u^\alpha_{;\mu} h^{\mu \beta} + u^\beta_{; \mu} h^{\mu \alpha} \Big ) -\frac{1}{3} \Theta h^{\alpha \beta}.
 \end{gather}
 
The stress tensor could be written as the sum of three terms in the following form \cite{Pimentel:2016jlm}
 \begin{gather}
 T^{\alpha \beta}= T^{\alpha \beta}_{pf}+T^{\alpha \beta}_{heat}+T^{\alpha \beta}_{visc},\nonumber\\
 \text{where\ \ }T^{\alpha \beta}_{pf}= \rho u^\alpha u^\beta+ p h^{\alpha \beta}, \ \ \ T^{\alpha \beta}_{heat}= q^\alpha u^\beta + q^\beta u^\alpha, \ \ \ T^{\alpha \beta}_{visc}= - \zeta \Theta h^{\alpha \beta}-2 \eta \sigma^{\alpha \beta}. 
 \end{gather}
From these relations, one could see that the mass parameter $m$ can increase the coefficients $\zeta$ and $\eta$ and therefore decreases the matrix elements of energy momentum tensor and their derivatives which then leads to the damping of the modular scrambling modes.  Also, the same-sign charge of the system, due to the repulsions between the internal degrees of freedom, would lead to the suppression of the modular scrambling modes.

 One may directly calculate the modular Hamiltonian and the growth rate of modular flows, in the presence of these additional parameters such as charge and mass and so prove this conjecture. The modular chaos bound could specifically be calculated for dissipative systems using the field theory model introduced in \cite{Baggioli:2020whu}. Their structure could model dissipative systems with gapped momentum states using the two-field Lagrangian. Using this model, then one can show how the modular chaos bound would decrease from the maximum value of $2\pi$, by a factor proportional to the dissipation parameter $\tau=\frac{\eta}{G}$, where here $\eta$ is the viscosity and $G$ is the shear modulus. Note that due to the dissipations, the theory would become non-Hermitian.

Also, in these systems, due to the dissipations, the correlation functions show decaying oscillatory behaviors. However, we expect that, in these mixed decaying systems, the JLMS relation \cite{Jafferis:2015del}
\begin{gather}
\langle \chi_i  | H_{\text{mod}}^{CFT} | \chi_j \rangle = \langle \chi_i  | H_{\text{mod}}^{bulk} | \chi_j \rangle ,  \ \ \ \ \forall  | \chi_i \rangle \in \mathcal{H^\psi_{\text{code}}},
\end{gather}
would still hold. In addition, the connections between the modular flowed operators in each region, $\rho_R ^{-is/2\pi} O(x_B) \rho_R^{is/2\pi}$ or $\rho_R ^{-is/2\pi} O(x_B) \rho_R^{is/2\pi}$, and the correlations between the bulk fields on the RT surfaces and operators on each boundary region part, similar to the result of \cite{Faulkner:2017vdd}, could be extended to the mixed symmetric setup of figure \ref{fig:strips}, and one could get the following relation for this setup as
\begin{flalign}
\int^\infty_{-\infty} ds \rho_R ^{-is/2\pi} O(x_A) \rho_R^{is/2\pi} &= 4\pi \Big \lbrack \int_{\partial r_A} dY_{RT_A} \langle \Phi(Y_{RT_A}) O(x_A) \rangle \Phi (Y_{RT_A})\nonumber\\&
+\int_{\partial r_B} dY_{RT_B} \langle \Phi(Y_{RT_B}) O(x_A) \rangle \Phi (Y_{RT_B}) 
-\int_{\partial r_C} dY_{RT_C} \langle \Phi(Y_{RT_C}) O(x_A) \rangle \Phi (Y_{RT_C})\nonumber\\&
-\int_{\partial r_D} dY_{RT_D} \langle \Phi(Y_{RT_D}) O(x_A) \rangle \Phi (Y_{RT_D})\Big \rbrack,
\end{flalign}
where we have used the relation of  mutual information between $A$ and $B$, i.e, $I=2S(l)-S(D)-S(2l+D)$,  and also the equation 1.6 of \cite{Faulkner:2017vdd}. Note that $RT_A$ is the RT surface for only region $A$, $RT_B$ is the RT surface for only region $B$, $RT_C$ is the RT surface for the part $D$ which passes from the point $m$, and $RT_D$ is the RT surface for the part $2l+D$ which passes through the point $m^\prime$.

 The structure of zero modes in the mixed states could further be studied by writing the equation 4.33 of \cite{Faulkner:2017vdd} for the mixed setups. For a free theory and for Gaussian states, their result could be extended to our specific mixed setup as  
 \begin{flalign}
 \Phi_0(Z_X)=&2 \int_{RT_A} \sqrt{h_{I_{RT_A}} } dY \Big(f_\Pi (Y) \Pi (Y)+f_\Phi(Y) \Phi(Y) \Big)
  -\int_{I_{RT_C}} \sqrt{h_{I_{RT_C}} } dY \Big(f_\Pi (Y) \Pi (Y)+f_\Phi(Y) \Phi(Y) \Big)\nonumber\\&
  - \int_{I_{RT_{D}}} \sqrt{h_{I_{RT_D}} } dY \Big(f_\Pi (Y) \Pi (Y)+f_\Phi(Y) \Phi(Y) \Big),
 \end{flalign}
 where $h_I$ is the induced metric on the HRT surface $I_I$ and $\Pi=n^\mu \partial_\mu \Phi$.  Using lattice models, the properties of these relations and the behavior of zero modes versus length $l$, distance $D$ or dimension $d$ could then be studied numerically. Note that for the points close to the RT surfaces, where $Y\to Y_{RT}$, the modular flow of the operator would behave as $\lim_{Y\to Y_{RT}} f_\Pi (Y)= -2\pi \langle \Phi(Y_{RT}) \Phi (X) \rangle$.

\subsection{Quantum recovery channels versus modular flows in mixed states}

Since both modular hamiltonians and quantum recovery channels could model the entanglement wedge reconstruction, we would expect physical connections between their formalisms. One should note that both of these approaches would establish a map between the algebras that are localized in different subregions of the system.

The modular Hamiltonian has specifically been used in formulating many holographic quantum measures such as relative entropy and also for bulk reconstruction. On the other hand, the main point of using a universal quantum recovery channel and quantum error correction formalism would be related to reconstructing information from the damaged information, as some partial parts of the information of the system would be damaged due to the noise. As one would expect that modular evolution has memory, one could imagine that the damaged information are still encoded in the modular flow which could then be extracted, and therefore this signals the connection.

First, note that the ``classical'' channel is a conditional probability distribution $\{p_{Y | X} (y  | x) \} _{x \in \mathcal{X}, y\in \mathcal{Y}} $, where $X$ is the input system and $Y$ is the output system which acts over the alphabets $\mathcal{X}$ and $\mathcal{Y}$. Then, the reversal channel could be written as \cite{Gilyen:2020gmg}
 \begin{gather}
p_{X|Y} (x | y) = \frac{p_X(x) p_{Y | X}  (y|x)    } { \sum_x p_X(x)   p_{Y | X} (y |x)    }.
 \end{gather}

A ``quantum'' channel $\mathcal{N}$ is also a completely positive, trace-preserving quantum ``map''. It is reversible if there would be another quantum channel $\mathcal{R}$, known as the recovery channel, which makes the composition $\mathcal{R} \circ \mathcal{N}$ to act as an identity in the form of $\mathcal{R} \circ \mathcal{N}[\rho]=\rho$. The Petz map is specifically an example of a recovery channel and is a function of the quantum channel $\mathcal{N}$, where $\mathcal{N}$ is the generalization of the classical map $p_{Y | X} (y | x)$. Also, the input state to the channel, $\sigma$, is the generalization of $p_X(x)$.  The Petz map could also be written as
\begin{gather}
\mathcal{P}_{B \to A} ^{\sigma, \mathcal{N}} (\omega_B) := \sigma_A^{1/2} \mathcal{N}^\dagger \big ( \mathcal{N} (\sigma_A)^{-1/2} \omega_B \mathcal {N} (\sigma_A)^{-1/2} \big ) \sigma_A ^{1/2},
\end{gather} 
which is a function of the quantum state $\sigma_A$. It is also a function of the quantum channel $\mathcal{N}_{A \to B}$ which takes system $A$ to $B$. Also, $\omega_B$ is the input density operator.  The Petz map is the composition of three completely positive (CP) maps which would lead to an indirect procedure for the bulk reconstruction.

In \cite{Cotler:2017erl}, the \textit{twirled Petz map} has also been written in the form of
\begin{gather}
\mathcal{R}_{\sigma, \mathcal{N}} := \int_{\mathbb{R}} dt \beta_0(t) \sigma^{-\frac{it}{2}} \mathcal{P}_{\sigma, \mathcal{N}} \big [ \mathcal{N} [\sigma]^{\frac{it}{2}} (.) \mathcal{N} [\sigma]^{-\frac{it}{2}} \big] \sigma^{\frac{it}{2}},
\end{gather}
where $\mathcal{P}_{\sigma, \mathcal{N}}$, is the normal \textit{Petz map}, i.e, $\mathcal{P}_{\sigma, \mathcal{N}} = \sigma^{1/2} \mathcal{N}^* \big [ \mathcal{N} [\sigma]^{-1/2} (.) \mathcal{N}[\sigma]^{-1/2} \big] \sigma^{1/2}$,  and $\mathcal{N}^*$ is the adjoint of channel $\mathcal{N}$. Also $\beta_0$ is the probability density which is $\beta_0(t):= \frac{\pi}{2} (\cosh(\pi t)+1)^{-1}$. For the bulk reconstruction, any of these two structures could be used, so the connections between modular Hamiltonian and quantum recovery channels could be traced using any of these formulations.

In \cite{petz1986}, it has been shown how to construct a recovery channel $\mathcal{R}: S(\mathcal{H_B}) \to S(\mathcal{H_A})$ from any reversible channel $\mathcal{N}: S(\mathcal{H_A}) \to S(\mathcal{H_B})$, where $\mathcal{H_A}$ and $\mathcal{H_B}$ are two Hilbert spaces. Any recovery channel would act approximately as $\mathcal{R} \circ \mathcal{N}  [ \rho ] \approx \rho$, $\forall \rho \in S(\mathcal{H})$, so it should be able to reconstruct the original density matrix with sufficient precision. Since the quality of this approximation would depend on the behavior of relative entropy under the action of the channel $\mathcal{N}$ \cite{Cotler:2017erl}, then, from the connections between relative entropy and modular Hamiltonian,  the connection between the quantum recovery channel and modular flows could be noticed.

When the two subsystems in our setup become closer to each other, the noise would get greatly increased as more information would be damaged. Then, from the modular Berry flow point of view, one could check that the rate of change of elements of modular Hamiltonian would increase and the flows would become more chaotic. Another important point is the monotonicity of the relative entropy which indicates that acting any quantum channel $\mathcal{N}$ on any two states would decrease the relative entropy between two states, so we have $D (\rho  | \sigma) \ge D(\mathcal{N} [\rho] | \mathcal{N} [\sigma])$, where again $D (\rho  | \sigma)  := \text{Tr} \rho \log {\rho} - \text{Tr} \rho \log {\sigma}$ is the relative entropy between $\rho$ and $\sigma$.

Then the approximate version developed in \cite{cite-key3} would be
\begin{gather}
D( \rho | \sigma)- D( \mathcal{N} [\rho] | \mathcal{N} [\sigma] ) \ge -2 \log F (\rho, \mathcal{R}_{\sigma, \mathcal{N}} \circ \mathcal {N} [\rho] ), 
\end{gather}
where $F(\rho, \sigma) := | \sqrt{\rho} \sqrt{\sigma} |_1 $ here is the fidelity.  This inequality then can put a constraint on the holographic bulk curvatures and modular chaos modes. It worths to mention here that the parameters such as dissipation, quantified by the mass of graviton $m$, or the same-sign charge of the system $q$, would make the term $F (\rho, \mathcal{R}_{\sigma, \mathcal{N}} \circ \mathcal {N} [\rho] )$ smaller, as the recovery channel works with less precision.

Another connection between the modular Hamiltonian and quantum recovery channel in the bulk reconstruction could be derived by combining the results of \cite{Cotler:2017erl} and \cite{deBoer:2019uem}. In \cite{Cotler:2017erl}, it has been shown that a boundary operator could be computed as the response of the modular Hamiltonian of the specific subregion to a perturbation of the average code state, which is in the direction of the bulk operator. Also, a bulk operator could be examined by the response of the boundary region's modular Hamiltonian to a perturbation of the bulk state in the direction of the bulk operator. This is the noncommutative version of Bayes' rule and has a representation in terms of modular flows.  Our aim is to check what the recovery channel would tell us about the properties of modular flows along the minimal wedge cross section.

For all the bulk operators $\phi_a$, with support in the entanglement wedge $a$, one could write, \cite{Cotler:2017erl},
\begin{gather}
\mathcal{O}: = \mathcal{R}^*[\phi_a]=\frac{1}{d_{\text{code}}} \int_{\mathbb{R}} dt \beta_0(t) e^{\frac{1}{2} (1-it) H_A} \text{Tr}_{\bar{A}} \big [ J (\phi_a \otimes \mathbbm{1}_{\bar{a}} ) J^\dagger \big ] e^{\frac{1}{2} (1+it) H_A},
\end{gather}
where $H_A=- \log (J \tau J^\dagger)_A$ is the boundary modular Hamiltonian of the subregion $A$ which is associated with the maximally mixed state $\tau$ on the code subspace.  In another way, it could be written as the logarithmic directional derivative as
\begin{gather}
\mathcal{O}_A= \mathcal{R^*} [\phi_a]= -\frac{1}{d_{\text{code}}} \frac{d}{dt} \Big |_{t=0} H_A [\tau_{\text{code}} +t \phi_a \otimes \mathbbm{1}_{\bar{a}} ]. 
\end{gather}

Using the results of \cite{deBoer:2019uem} where a bound on the infinitesimal perturbation of the modular Hamiltonian have been introduced, one could find a constraint on the perturbative properties of the quantum channels.  Since we have $dH_{\text{mod}} \propto ( d \tau_{\text{code}} +ds (\phi_a \otimes  \mathbbm{1}_{\bar{a}})  ) $ and we can get the following bound 
\begin{gather}\label{eq:modbound1}
\Bigg | \frac{d}{dt} \Big{ |}_{t=0} \log \bra{\chi_j} ( J (\tau_{\text{code}} + t \phi_a \otimes \mathbbm{1}_{\bar{a}} ) J^\dagger ) \ket{\chi_j} \Bigg | \le  \frac{2\pi}{d_{\text{code}}}.
\end{gather}
This means that the boundary operator corresponding to $\phi_A$ is related to the response of acting the boundary modular Hamiltonian $H_A$ on the perturbation of the maximally mixed code state in the direction of the operator $\phi_a$. We also expect that having the maximum modular flow through the minimal entanglement wedge cross section would correspond to the most efficient quantum error correction codes. The relation \ref{eq:modbound1}, could also specify a bound on the maximum density of bit threads along the minimal entanglement wedge cross section in our setup.

 In \cite{Gilyen:2020gmg}, the connections between Petz recovery channels and pretty good measurements have been discussed. Additionally, using pretty good measurements which is a special case of Petz recovery channel, various aspects of the connection could be studied. This channel could allow for near-optimal state discrimination. 
The error probability of pretty good measurement (PGM) would satisfy the following inequality $P_e^{PGM} \le \sum_{i\ne j} \sqrt{p_i p_j}  F(\sigma_i, \sigma_j)$, where $\{ \sigma_i \}$ is a set of density matrices and $p_i$ is the probability that a quantum state $\rho$ is in state $\sigma_i$ and $F$ is the fidelity function. As the fidelity and complexity are proportional which has been shown in \cite{Alishahiha:2015rta}, using volume complexity, this inequality then could put a bound on the probability PGM. Therefore, quantities such as capacity of quantum recovery channels would also be bounded by the computational complexity of a state. Also, using the gradient of modular flow, the compressibility of quantum messages and also the capacity of quantum channels could be estimated.

Another observation is that the bound on quantum error correction for every channel $\Lambda$ as
\begin{gather}\label{eq:bounderror}
\underset{ \ket{\psi} \in \mathcal{C}^{\otimes 2} }{\text{min}}   \underset{\mathcal{D}}{ \text{max}} \bra{\psi} ( \mathcal{D}  \circ    \Lambda \otimes I) ( \ket{\psi} \bra{\psi}) \ket{\psi} \ge 1-\epsilon,
\end{gather}
would be related to the bound on modular Hamiltonian
\begin{gather}\label{eq:modbound}
||  e^{ - i H_{\text{mod}}  s} e^{i (H_{\text{mod}} +\epsilon \delta H_{\text{mod}}  ) s} || \le 1,
\end{gather}
for the strip $-\frac{1}{2}  \le \text{Im} \lbrack s \rbrack \le 0 $. This would indicate the connections between the upper bound on the changes of the modular scrambling modes and the maximum precision of quantum error correction codes. Quantities which supress the modular scrambling modes would decrease the precision of the code.

Next, the relations between quantum error corrections and chaos could be considered. In \cite{Brandao:2017irx}, the connections between chaos, eigenstate thermalization hypothesis (ETH) and quantum error correction have been discussed. The eigenstate thermalization hypothesis could be written in the form
\begin{gather}
| \bra{E_l}  O \ket{E_l} - \bra{E_{l+1}} O \ket{E_{l+1}} |  \le \exp(-c_1 N), \ \ \ \ \text{and}  \ \ \  \ | \bra{E_k}  O \ket{E_l} | \le \exp (c_2 N),
\end{gather}
where $c_1, c_2 >0$ are two constants and $E_l$ are the energy eigenstates. The constraint on the modular scrambling modes growth rate would then also be related to the approximate version of the Knill-Laflamme condition in the form of $\bra {\psi_i} E \ket{\psi_j} = C_E \delta_{ij}+ \epsilon_{ij}$, where the index $i$ specifies the codewords that span the code space as $\mathcal{C}= \text{span} (\{ \ket{\psi_1}, ...,\ket{\psi_{2^k} } \} ) $. The error in \ref{eq:bounderror} would also have the bound of $\epsilon \le 2^{2(k+d)}$ where $d$ here is the number of qubits in the system.

By increasing $d$ (the system size), the average error would decrease as the error here would be proportional to $\frac{1}{\sqrt{d}}$. For the modular Hamiltonian, this has the implication that by increasing the system size, the number of modular zero modes would increase, which leads to a smoother bulk geometry, and also the bound \ref{eq:modbound} would get closer to one. Therefore, the connections between quantum error corrections, chaos, and modular chaos could be seen from ETH as well. Considering ``all" of the modular flows that pass through entanglement wedge cross section, corresponds to a ``perfect"  quantum error correction code. Note that for the mixed quantum systems with a \textbf{connected} bulk geometry, which also satisfy ETH, one would have a richer family of eigenvalues which then would need a bigger approximate quantum error correction code (AQECC). The AQECC for the connected versus two disconnected bulk systems, should perform differently, and so it could detect the phase transitions.

The difference between the codes in the disconnected versus connected case could be further examined by considering the theorem 1 of \cite{Brandao:2017irx}. If we imagine that in each strip there are $N$ sites, in the initial state, there would be a set of energy eigenvalues close to $\mathbf{E_1}$ as $S_{\mathbf{E_1}} := \{ E_k : E_k \in \mathbf{E_1}- \sqrt{N} , \mathbf{E_1}+ \sqrt{N} \} $. Then, when the two strips get closer enough, as the mutual information and EoP become non-zero, the sets of energy eigenvalues get mixed, and a bigger set which its components form around  $\mathbf{E_2}$, as $S_{\mathbf{E_2}} := \{ E_j : E_j \in \mathbf{E_2}- \sqrt{2N} ,  \mathbf{E_2}+ \sqrt{2N} \} $, would appear. Since there would be more eigenstates and the distances between them get smaller, the error of AEQCC would become smaller as well. In this case the distance of the code would change by a factor of $\Delta \sim \log(2)$.

When the two subregions are far from each other and the mutual information between them is zero, in each region, the eigenstates which have close energies, would form an approximate quantum error correcting code (AQECC) which all together reconstruct their own corresponding bulk dual space. When the two regions get closer enough to each other, more eigenstates would be available and a bigger set for AQECCs could emerge, which also is the case for the linking part of the two subregions, and therefore the dual ``connected'' entanglement wedge in the bulk could be constructed.

There are two reasons one expects that the discussion of \cite{Brandao:2017irx} would work for the mixed setup and connected entanglement wedge reconstruction as well. One is the invariance of states under \textit{modular flow} for any subregion, which could be written as $\delta H_{\text{mod}} \ket{\psi}=0$ or $G_\pm \ket{\psi} \approx 0$. This would lead to the approximate local isometries of $\mathcal{L_{\zeta_{(\pm)}   } } g_{\mu \nu} (x^\alpha=0, y^i)= O(e^{-2\pi \Lambda})$ along \textbf{all} the RT surfaces and even those connecting the two regions \cite{deBoer:2019uem}. In fact the existence of this symmetry could ensure the ``uniformity'' of spreading of information along the the minimal wedge cross section $\Gamma$, and so this symmetry reassures us that even after mixing, the code would not be corrupted by noise completely and so is still capable of bulk reconstruction.  The other point is the finiteness of the correlation lengths, which even in the mixed setup still ensures that the spreading of information would not diverge and just will be uniform enough to forge a nice bulk geometry from the mixed correlations.

 Another interesting point about the effects of charges can be found  by studying the connections between quantum error correction and symmetries in holography. In \cite{Faist:2019ent}, the authors studied the approximate error correction in the presence of continuous symmetries and Haar-random charged systems and they found a bound on the recovery error as $\epsilon \gtrsim \frac{Q}{n}$  where $Q$ is the total charge of the state and $n$ is the number of physical subsystems where the system is made of. Again one could see that by increasing the charge $Q$, the minimum error for constructing states would increase so in the bulk reconstruction, less quantum recovery channels and fewer gates would be available for calculation and therefore once again one could see that charge would decrease the complexity of purification as it also decrease the correlations among mixed systems.

We could get further connections from other symmetry generators. The observation in \cite{deBoer:2019uem}, was that the chaotic properties of the dual boundary CFT theories would lead to the symmetry generators in the bulk.  Near the RT surfaces in the bulk, as found in \cite{deBoer:2019uem}, the modular Hamiltonian would act as a geometric boost, i.e, $\lbrack \delta H_{\text{mod}}, \phi(x^\alpha, y^i) \rbrack \propto 2\pi \left (\zeta^\mu_{(+)} -\zeta^\mu_{(-)}\right )$. So the zero modes close to the RT surfaces would have translation invariance. If the ETH would be appliable in the system too, then using the result of \cite{Brandao:2017irx}, for the $2d$ case one could show that an AQECC would exist. This point then would indicate further connections between modular chaos and quantum error correction codes.  The formation of quantum error correcting codes (QECC) in chaotic systems which eigenstates exhibit the eigenstate thermalization hypothesis could be related to the saturation of modular scrambling modes, so one expects that in the bulk reconstruction formalism by quantum recovery channels, the inequality found in \cite{deBoer:2019uem} for modular chaos somehow appears.

As ETH have many significant implications for QECCs, its implications for EoP and CoP would be notable too. Then the dynamical properties could also point out to other connections. For instance, in \cite{PhysRevX.9.031029}, the \textit{fluctuation theorem} has been applied to quantum recovery channels. Their complex-valued entropy production could detect the relation between the forward and backward processes through the quantum channel. So, one could propose that the imaginary part of the complexity of purification is also related to the symmetry breaking while passing through the quantum channel.
 
By changing the parameters of the system, further connections could be revealed. For example, the parameter $m$ increases the dissipations in the channel and therefore decreases the correlations among the two subregions. In this case, then less modes could pass through the quantum recovery channels and so increasing $m$ would decrease the imaginary parts of entanglement and complexity of purification as we observed in \cite{Ghodrati:2019hnn}. The same argument could be applied for the case where the two subregions have a same sign charge $q$. As these two quantities increase the errors in AQECCs, and suppress the flow of modular modes, and so they decrease EoP and CoP linearly. This can also be seen from a simple model of Jaynes-Cummings models in a cavity QED \cite{doi:10.1080/09500349314551321},
\begin{gather}
H_{i-f}=\frac{1}{2m_a} [ p-q A(x) ]^2 + U(x)+  \hbar \omega \left( a^\dagger a +\frac{1}{2} \right )+H_{el}.
\end{gather}
 where $p$, $q$, $x$ and $m_a$ are momentum, charge, position and mass of the atom and $H_{el}$ is the Hamiltonian of the electronic state of the atom. As charge decreases the element of this Hamiltonian linearly, we expect its effect would be linearly decreasing of EoP, CoP, entropy production \cite{PhysRevX.9.031029}, coherence in quantum channel, and the transition rate between diagonal and off-diagonal of the system density matrix. It worths to mention here that an interesting question would be to specify the share of the dual of real versus imaginary entropy production, from the holographic bulk perspective.

Now we could investigate the relations for the modular zero modes and the connections with EoP and CoP in more details. For each subsystem, the zero modes are those operators $Q_i^A$ which satisfy the following relation 
 \begin{gather}\label{eq:zeromode}
 \lbrack Q_i^A , H_{\text{mod}, A } \rbrack=0,
 \end{gather}
 where $i$ indexes the zero-mode sub-algebra. Due to the equivalence between bulk and boundary modular flows \cite{Jafferis:2015del}, for holographic CFTs which satisfy \ref{eq:zeromode}, the dual Bulk operators would be on the RT surfaces which are anchored on $\lambda$.

For the mixed case, one should specify how the correlation functions ``outside'' of a CFT subregion $A$ would change under the unitary evolution which are generated by the modular zero modes of the system $A $, i.e, $Q_i^A$.
If all the correlation functions stay \textit{within} the subregion $A$, then the evolution is invariant and we get \ref{eq:zeromode}. However, since here we are interested in understanding how the correlation functions and their relative bundles connect the subregion $A$ to subregion $B$, in the mixed state setup, this relation then should be modified for the case of entanglement of purification. One then expects that for the full setup of EoP, one would need to consider  the whole normal modes and not just the zero modes.

So, for the mixed setup, we expect the commutative relations would be more complicated. For the case where the distance between the two subsystems is bigger than the critical distance $D \ge D_c$, the mutual information is zero, $I=0$ and therefore $\text{EoP}=0$. In this case, for each of the two systems of $A$ and $B$, still $\lbrack Q_i^A, H_{\text{mod},A} \rbrack =0$ and $\lbrack Q_i^B, H_{\text{mod},B} \rbrack =0$. However, for $D \le D_c$ when the two systems become correlated, the commutations become non-zero, i.e., $
 \lbrack Q_i^A, H_{\text{mod},A} \rbrack  \ne 0,\  
 \lbrack Q_i^B, H_{\text{mod},B} \rbrack  \ne 0,\ 
 \lbrack Q_i^A, H_{\text{mod},B} \rbrack  \ne 0,\ 
 \lbrack Q_i^B, H_{\text{mod},A} \rbrack  \ne 0,
$ as they would depend on the amount of mixing and correlations between the subsystems, or the mutual information shared among them, and therefore on the parameters such as $l$, $D$, $d$, etc, which totally could be quantified by a mixing parameter $\mu$. We later study this parameter further.  Note that, here, $Q_i^A$ or $Q_i ^B$ are the generators of the unitary evolutions. For the modular Hamiltonian of the total system and its zero modes, we expect though that again the commutation relation would vanish.

 Looking for more connections, we turn again to \cite{Czech:2019vih}, where it was suggested that the modular Hamiltonian could be written as $H_{\text{mod} } = U^\dagger \Delta U$,  where $\Delta$ has the information of the spectrum and the unitary operators $U$ have the information of the basis of the eigenvectors. Its derivative with respect to the modular parameter $\lambda$ has been written there as
 \begin{gather}\label{eq:modederivative1}
 \dot{H}_{\text{mod}} = \lbrack \dot{U}^\dagger U, H_{\text{mod}} \rbrack+ U^\dagger \dot{\Delta} U,
 \end{gather}
 and we expect that this relation would still work for the case of mixed setup as well.

 Now again the main point we want to look for here is to determine how by changing the charge $q$, or the dissipation rate, i.e, changing $m$, each term of this relation would be changed as we would like to understand better the connections between the procedure of quantum recovery channels in entanglement wedge reconstructions and modular flow by studying the effects of charge and dissipation rate on each method. For studying this problem, similar to \cite{Khemani:2017nda}, we could consider an out-of-time-order correlator (OTOC), but now using modular Hamiltonian, and in the setup of mixed states. So instead of the zero modes $Q_0$, we need the operators $Q_x$, which has support near the position $x$. Then, the OTOC would be \cite{Khemani:2017nda}
 \begin{gather}
\mathcal {C} (x,s)= \frac{1}{2} \text{Tr} \  \rho^{eq} \lbrack H_{\text{mod}} (s) , Q_x \rbrack ^\dagger \lbrack H_{\text{mod}} (s), Q_x \rbrack, 
 \end{gather}
where $\rho ^{\text{eq}}$ is a Gibbs state. Using this relation, the Lieb-Robinson bound in the setup of modular chaos could be considered and it could be connected to modular chaos bound. 

To make the study simpler we can use the model of the spin-$1/2$ chain of length $L$ where the spreading operator could be written in the basis of $4^L$ Pauli string operators $\mathcal{S}$ which are some products of Pauli matrices on distinct sites. Similarly it could be done for modular Hamiltonian which controls the evolution of modular scrambling modes and so it could be written as $H_{\text{mod}}(s) = \sum_{\mathcal{S}} a_{\mathcal{S}} (s) \mathcal{S}$. So the modular Hamiltonian would be the combination of some string operators which by evolution of the modular time $s$ grow in the spatial extent. The OTOC then would be zero at first, when the two subregion are far away from each other, but it becomes non-zero as the two subregion get closer. The effects of mass and charge on modular Hamiltonian and modular chaos modes could then be seen by considering their effects on these string operators.  The modular Hamiltonian would satisfy a type of conservation law;  the operator norm $\text{Tr} \lbrack H_{\text{mod}}^\dagger H_{\text{mod}} \rbrack $ should be conserved which leads to the fact that the total weight of these Pauli strings $\sum_{\mathcal{S}} \big | a_{\mathcal{S}} \big |^2 $ would be conserved as well. So for the modular scrambling modes, similarly one could imagine a hydrodynamical spreading picture. Then, the effects of charge or dissipation could be observed by using this model and just by considering their effects on the string operators.  

When the left and right systems become close enough to each other that we get $I \ne 0$ and $\text{EoP} \ne 0$, then the zero modes would mix with each other, and the strings or the Wilson lines between them would become screened. The dissipation and (same-sign) charge would shrink the string operators. The charge and dissipation also make the scrambling time shorter and also the term $U^\dagger \dot{\Delta} U$ in relation \ref{eq:modederivative1} would reach to its maximum value faster.

 In \cite{Czech:2019vih}, a projector operator into the zero modes, $P_0^\lambda$, has also been proposed which could separate the contribution of spectrum changing in relation \ref{eq:modederivative1}. It could be written as
 \begin{gather}
 P_0^\lambda \lbrack V \rbrack \equiv \lim_{\Lambda \to \infty} \frac{1}{2 \Lambda} \int_{-\Lambda}^{\Lambda} ds \ e^{i H_{\text{mod}} (\lambda) s }   V  e^{-i H_{\text{mod} } (\lambda) s  },
 \end{gather}
 or
 \begin{gather}\label{eq:zeromodeop2}
 P_0^\lambda \lbrack V \rbrack \equiv \sum_{E, q_a, q^\prime_a} \ket{E,q_a} \bra{E,q_a} V \ket{E, q^\prime_a} \bra{E, q^\prime_a},
 \end{gather}
where in \ref{eq:zeromodeop2}, $\ket{E, q_a}$ would be the simultaneous eigenstates of modular Hamiltonian $H_\text{mod}$  and a set of commuting zero mode operators $Q_a$. Also, $E$ is the eigenvalue of $H_\text{mod}$,  and $q_a$ would be the eigenvalue of $Q_a$.We would like to raise here, the similarities between the procedure of applying the projector operator $P_0^\lambda$, quantum recovery channels and also the Wilson line formulations. As mentioned, the quantum recovery channel
 \begin{gather}
 \mathcal{P}^{\sigma, \mathcal{N}} _{B \to A} (\omega_B) := \sigma_A^{\frac{1}{2}} \  \mathcal{N}^\dagger \big( \mathcal{N} (\sigma_A)^{-\frac{1}{2}}  \tilde{\omega}_B \mathcal{N} (\sigma_A)^{-\frac{1}{2}} \big ) \sigma_A^{\frac{1}{2}},
 \end{gather}
 is a combination of three maps \cite{Gilyen:2020gmg},
 \begin{gather}
 i)\ \ (.)  \to \lbrack  \mathcal{N} (\sigma_A) \rbrack^{-1/2} (.) \lbrack \mathcal{N} (\sigma_A) \rbrack^{-1/2},\  \ \ 
  ii)(.) \to \mathcal{N} (.), \  \ \ \ 
  iii) \ \ (.) \to \sigma_A^{1/2} (.) \sigma_A^{1/2}.
 \end{gather}
 The combination of the first and third one is similar to the applying of the projection operator into zero modes. So we could think of the projector operator applied into the zero-mode sector of $H_{\text{mod}} (\lambda) $ as a quantum recovery channel.

Also, the flux of zero modes along the minimal wedge cross section could be considered by the integral below
\begin{gather}
\int_{m}^{m'} dz \frac{d \left(U^\dagger \dot{\Delta} U \right) }{dz} \Big |_\Gamma= \int_{m}^{m'} dz \frac{d \left(P_0^\lambda \lbrack   \dot{H}_{\text{mod}} (\lambda)   \rbrack \right)}{dz} \Big |_\Gamma.
\end{gather}
 So the change of spectrum of zero modes could also be written in terms of projection operator and rate of change of modular Hamiltonian. Then, the spectrum complexity part of complexity of purification \cite{Agon:2018zso,Ghodrati:2019hnn} and its growth rate could be written in terms of the modular Hamiltonian. We explain further this point in section \ref{sec:complexitymod}. Note also that the mass parameter $m$ and charge, would suppress the rate of growth spectrum complexity by suppressing the operators $P_0$ and $\dot{H}_{\text{mod}}$ through the suppression of the eigenvalues $E$ and $q_a$.

One last point that we would like to mention in this section is the connection between remaining in the code subspace along the minimal wedge cross section, which follows the equation $H_{mod}=P_{code} H_{mod}^{exact} P_{code}$, and staying within the distance smaller than $D_c$ in our mixed setup, where the mutual information is still non-zero, i.e, $I=S(l)+S(D)-S(2l+D) \ne 0$. One could argue that if the non-local effects outside the code subspace could be considered, the singularities of the first-order phase transitions shown in figure \ref{fig-D-COP} could be removed. In particular, considering the effects of quantum tunnelings through the Berry potential could improve the relation for the mutual information and remove the sudden drop in the phase diagrams of entanglement and complexity of purification and make those figures get vanish in late times smoothly. The quantum mutual information would satisfy the inequality  $I(A,B) \ge \frac{\mathcal{C} (M_A, M_B)^2}{ 2 ||M_A||^2 ||M_B||^2}$, where $\mathcal{C}(M_A,M_B):=  \langle { M_A \otimes M_B} \rangle   -   \langle M_A \rangle  \langle  {M_B} \rangle  $  is the correlation function of $M_A$ and $M_B$, \cite{Wolf:2007tdq}. So this way by adding the quantum effects to the relations of mutual information, EoP and CoP the sudden drops in their phase diagrams will be removed.

Finally, we would like to mention that we expect $H_{\text{mod}} (\text{mixed})$ would have more zero modes than the initial Hamiltonian and also when the system become mixed, the corresponding gauge groups would get larger. Also, we expect that for the mixed setup we get $\lbrack Q_i, P_{\text{code}} H_{\text{mode}(\text{mixed})} P_{\text{code}} \rbrack \ne 0$. Also, charge and dissipation (the term $m$) would change the size and the behavior of projection operators and the \textit{code subspace} which could be studied further.

\subsection{Complexity, Berry phase and modular Hamiltonian}\label{sec:complexitymod}

The question of what information modular Berry phase yield could be studied in the setup of $\text{AdS}_3/\text{CFT}_2$ where it has been shown in \cite{Czech:2017zfq} that it has deep connections with the entanglement and bulk reconstruction. In the bulk and in the mixed systems, the modular phase is related to the complexity of purification introduced in our previous work \cite{Ghodrati:2019hnn}.

For the case of $\text{AdS}_3$, as we have seen, CoP would be constant and its absolute value would be $\pi$. For higher dimensions it would change similar to the volume as in figure \ref{fig:CoPvolume}. It becomes much bigger as the dimension of space-times increases which is similar to the behavior of Berry phase. Also, the behavior of EoP shown in figure \ref{fig:Eopp1122}, could show such connections between the correlations and modular Berry connection in various dimensions. However, as we discussed, complexity and CoP would be better probes of correlations in mixed setup, therefore, one should also  find the connections between CoP and Berry phase. In \cite{Akal:2019hxa}, the connections between complexity measures in the path integral optimization proposal and Berry phase have been depicted. Now here we would find more connections between the complexity of purification, basis complexity and spectrum complexity (defined in \cite{Agon:2018zso}) and the Berry phase by varying the modular Hamiltonian through changing $m$ and $q$.

Simply as the modular Hamiltonian has been employed in the calculation of relative entropy, entropy bounds, or determining the statistical properties of vacuum CFTs, one would expect it would be useful in studying holographic computational complexity as well.  In references \cite{Resta2000REVIEWAM,RevModPhys.82.1959}, it has been shown  that the Berry curvature and Berry phase would also play an important role in studying many electronic properties in molecules and solids. The Berry's phases of the many-electron wavefunction have been related to several observable phenomena and measurable effects such as the polarization in the material, various manifestations of Hall effects, orbital magnetism and also quantum charge pumping. One could add to those studies by connecting the Berry phase of many-body systems to the complexity and complexity of purification of the system. Specially, connections between quantum charge pumping, the direction of bit threads, the behavior of correlations between two mixed systems and purification could be related.
  
  As mentioned,  the modular Hamiltonian could be considered as the Hermitian operator on the CFT and can be decomposed as $ H_{\text{mod}}= U^{\dagger} \nabla U$
 where in this relation $\nabla$ is a diagonal matrix which determines the spectrum and it would  be connected to the spectrum complexity. The unitary $U$ specifies the basis of eigenvectors and it would be connected to the basis complexity. Then from that, one could get $
 \dot{H}_{\text{mod}} = [\dot{U}^\dagger U, H_{\text{mod}}]+ U^\dagger \dot{\nabla} U$,
 where the dot is the derivative with respect to the quantity $\lambda$, $\dot{} \equiv \partial_\lambda$,  which reparametrizes CFT. Then, in \cite{Czech:2019vih}, the modular Berry connection is defined as
\begin{gather}
\Gamma(\lambda^i, \delta \lambda^i) = P_0^\lambda [\partial_{\lambda^i} U^\dagger U] \  \delta \lambda^i,
\end{gather}
where $P_0^\lambda$ is the projector which acts only on the zero-mode sector of $H_{\text{mod} } (\lambda^i)$.  One should note that unlike the case in \cite{Czech:2019vih}, in the mixed setups, to connect $A$ and $B$, the transformation $U$ would be generated not only by the zero modes but by the whole modes as $U_Q'=e^{-i \sum_i Q'_i s_i}$ . Therefore, the form of $H_{\text{mod}}$ for mixed states would not completely be preserved, but still the change in its form could be calculated and it would be related to complexity of purification.

Indeed, it would be interesting to rewrite various quantum information measures in terms of modular Hamiltonian. For instance, here one could then connect modular Hamiltonian to the definition of complexity of purification defined in \cite{Agon:2018zso} which is a summation of two parts, basis complexity and spectrum complexity. As we have mentioned, the complexity of purification could be written as
\begin{gather}\label{eq:changeCoP1}
CoP= \mathcal{C}_B+ \mathcal{C}_s,
\end{gather}
and in the boundary CFT, the change of modular Hamiltonian could also be decomposed into the change of basis and change of spectrum as \cite{Czech:2019vih}
\begin{gather}\label{eq:changeCoP2}
\underbrace{\dot{H}_{\text{mod}}}_{ \propto \ \text{CoP} } = \underbrace{ \lbrack \dot{U}^\dagger U, H_{\text{mod}} \rbrack }_{\propto \ \mathcal{C}_B} + \underbrace{ P_0^\lambda \lbrack \dot{H}_{\text{mod}} \rbrack} _{\propto \ \mathcal{C}_s},
\end{gather}
leading to the desired relations. The part which corresponds to the change of the spectrum of modular Hamiltonian and therefore corresponds to spectrum complexity part of CoP, could also be written in the form of modular flow as \cite{Czech:2019vih}
\begin{gather}
P_0^\lambda [V] \equiv \underset{\Lambda \to \infty }{\text{lim}} \  \frac{1}{2 \Lambda} \int^\Lambda_{-\Lambda} ds \ e^{i H_{\text{mod} } (\lambda) s } V  e^{-i H_{\text{mod} } (\lambda) s }.
\end{gather} 

Then, in the bulk, this relation could be written in the following form
\begin{gather}\label{eq:changeCoP3}
\underbrace { \delta_\lambda \zeta_{\text{mod}}^M (x; \lambda)}_{\propto \ \delta \text{CoP} }  = \underbrace{ \lbrack \xi (x; \lambda, \delta \lambda) , \zeta_{\text{mod}} (x; \lambda) \rbrack ^M} _{\propto \ \text{rotation of the basis} } + \underbrace{P_0^\lambda \lbrack \delta_\lambda \zeta_{\text{mod}}^M (x;\lambda) \rbrack}_{\propto \ \text{change of spectrum}}.
\end{gather}
These relations could show how changing the modular Hamiltonian would change basis, spectrum and purification complexities.  Also, the operator $\dot{U}^\dagger (\lambda) U(\lambda)$ which corresponds to the basis component of complexity of purification could depict the interrelationship of information and the infinitesimal shape variation.

Note that entanglement entropy and relative entropy previously have been written in terms of modular Hamiltonian as well. For instance, for a ball-shaped region $A$ in the CFT, the first law of thermodynamics would be
  \begin{gather}
  \frac{d}{d\epsilon} ( \langle  H_A \rangle -S_A) = \frac{d}{d\epsilon} S (\rho_A \big|\big| \rho_A^{(0)}),
  \end{gather}
  where $\rho_A^{(0)}$ is the density matrix of region $A$ without perturbation and  $S (\rho_A \big|\big| \rho_A^{(0)})$ is the relative entropy between the perturbed and unperturbed states. This could then be extended to define a first law of EoP in terms of modular Hamiltonian as
\begin{gather}
\partial_\lambda \Delta \langle H_0 \rangle \big |_{\lambda=\lambda_0} = \partial_\lambda \Delta \text{EoP} (A,\lambda) \big |_{\lambda=\lambda_0}.
\end{gather}
The gradient of EoP also could be written in term of the modular Hamiltonian as
\begin{gather}
\Delta \langle H_0 \rangle (A, \lambda)= \partial_\lambda \Delta S (A, \lambda) \big |_{\lambda=\lambda_0} \tilde{\lambda} +\mathcal {O} (\tilde{\lambda^2}).
\end{gather}

 Another example is the coherent states, where their modular Hamiltonian is equal to the canonical energy. These would not change the bulk von Neumann entropy of subregion, due to the relation $\Delta S_{\text{bulk}} =0 $ \cite{Jafferis:2015del}, which then would lead to
\begin{gather}
S_{\text{bdy}} (\rho  {||} \sigma) = S_{\text{bulk}} ( \rho ||  \sigma ) = \Delta K_{\text{bulk}}- \Delta S_{\text{bulk}}= \Delta K_{\text{bulk}}= E_{\text{canonical}}.
\end{gather}
So in the case of excitations of coherent states, under the action of their specific quantum channel, modular Hamiltonian would not increase. Additionally, the first law of entanglement entropy in terms of the modular Hamiltonian could be written as
  \begin{gather}
  \Delta S= S(\rho^1)-S(\rho^0)= \langle H \rangle_1 -\langle H \rangle_0= \Delta \langle H \rangle.
  \end{gather}

Note that when the Hamiltonian of the system evolves adiabatically, the system would remain in the n-th eigenstate of the Hamiltonian, but it would gain a phase factor. So, If we replace the Hamiltonian of the system with the modular Hamiltonian,  similar to the studies of \cite{deBoer:2019uem, Czech:2019vih, Czech:2018kvg, Czech:2017zfq}, we could depict the connection between the Berry curvature and the modular scrambling modes but for the \textbf{mixed setup}. For doing that we could take into account the picture we got in \cite{Ghodrati:2019hnn} for the minimal wedge cross section through the bit thread formalism. When the modular evolution is cyclical, the modular Berry phase would be invariant and could be an observable of the system and the whole change could be characterized by this phase term.  Using the adiabatic approximation, the coefficient of the nth eigenstate under such adiabatic process would be
\begin{gather}
C_n(s)= C_n(0) \text{exp} \Big[ - \int_0^t \langle \psi_n (s') \big | \dot{\psi_n} (s') \rangle \Big ] =C_n(0) e^{i \gamma_m (s)}= C_n(0) e^{i \gamma_m (s)}, 
\end{gather} 
where $\gamma_m (s)$ is the modular Berry phase with respect to the modular parameter $s$. One could change the variable $s$ into the generalized parameter and then write the modular Berry phase as
\begin{gather}
\gamma [C] = i \oint_c   d\lambda(s)  \  \langle \lambda,s \big | ( \nabla_\lambda \big | \lambda, s) \rangle, 
\end{gather}
where $R$ here parameterizes the cyclic adiabatic process. The term $V_n =i \langle \lambda,s \big | ( \nabla_\lambda \big | \lambda, s) \rangle$ is the modular Berry potential we expect that considering the \textbf{quantum tunneling} through it would smooth out the phase diagrams of MI, i.e, fig  \ref{fig-D-COP} and EoP.

So when the modular time $s$ varies in a sufficient slow manner, if the system was initially in the eigenstate $ \ket{n(\lambda(0) ) } $, it would remain in the instantaneous eigenstate $\ket{n(\lambda(s)) } $ of the modular Hamiltonian $H(\lambda(s))$ up to a phase. However, the complexity of state would change, therefore in this case, the only parameter of the state which the complexity could be proportional to would be the Berry phase. This result would also be related to  the Chern theorem, as the Berry phase could be written in terms of the integral of Berry curvature $\omega_n(\lambda)=\nabla_\lambda \times V_n (\lambda)$, in the form of $\gamma_n= \int_S dS . \  \omega_n (\lambda)$, where this integral would be quantized in units of $2\pi$ (Chern number), pointing a connection between $ e^{i \gamma_n}$ and complexity of purification (specifically the basis complexity).

The Chern theorem states that the integral of the Berry curvature over a closed manifold is quantized in units of $2\pi$.  We found a similar result for the complexity of purification (CoP) in \cite{Ghodrati:2019hnn}. Even the multipartite complexity of purification would be an integer multiplet of $2\pi$. The reason that complexity of purification, Berry curvature and Chern number are connected could be explained by the mechanisms that the zero modes in the setup of \cite{Czech:2019vih} would create the curvature of the bulk. In other words, if a state changes from $\ket{\psi_n(a)}$ on a path $\gamma(s)$ where $\gamma(0)=a$, , then one could write $\ket{\psi(s) } = e^{ i \theta_n (s)} \ket{\psi_n(\gamma(s))}$. The phase could be divided into two pieces, $\theta_{n,dynamic}$ and $\theta_{n, geometric}$. The conjecture is that the dynamical part of the change of phase corresponds to the spectrum complexity part of CoP,  and the geometric part, which is related to the rotation of the basis and so to Berry connection and Berry phase, would be related to the basis complexity part of CoP, which could also be noted from the relations \ref{eq:changeCoP1}, \ref{eq:changeCoP2} and \ref{eq:changeCoP3}. The gauge transformation relation $\ket{\tilde{n} (\lambda(s)) }=e^{-i \beta (\lambda) } \ket{n (\lambda(s)) }$, which for an open-path gives the Berry phase $\tilde{\gamma_n} (s)=\gamma_n(s)+\beta(s)-\beta(0)$ and for a closed path would give $\beta(T)-\beta(0)= 2\pi m$ ($m$ is an integer), then leads to the result that the Berry phase $\gamma_n$ by modulo $2\pi$ would be invariant. This fact points to another connection between the properties of complexity of purification (see also the results of \cite{Ghodrati:2019hnn}) and the Berry phase.

Another connection between complexity and Berry phase could come from the results of \cite{Camargo:2019isp} where it has been suggested that the deformation of Euclidean path integral which prepares a state and is related to the Berry phase could provide a new formulation for complexity with a standard gate counting notion. So this way, the Liouville action would be related to the Euclidean analogue of the Berry phase. Since the connections between Liouville action and complexity have already been established in \cite{Caputa:2017yrh, Ghodrati:2019bzz}, therefore this  would point out to our desired connections between Berry phase and complexity. This is because from the change in the measure of the path integral, one could get the exponent of the Liouville action. The procedure would be to act with the operator $\rho_\beta$ in the form of
$\rho_\beta= \text{exp} (-\beta H)$, on the the vacuum state.  Note that $H$ is the physical Hamiltonian operator of a $2d$ CFT living on a line. One could also write $\rho_\beta$ as a circuit in the form of
 \begin{gather}
 V=\mathcal{P} \text{exp} \Big \{ - \int_{t_i}^{t_f} dt \int dy [a(t,y) h(y)+ i b(t,y) p(y)] \Big \}. 
 \end{gather}
 So acting by this operator on the vacuum state would produce states which could be parametrized by the circuit parameter $t$. These states would be different from the vacuum state themselves but will end on the vacuum state at $t=t_f$ and the change in the measure of path integral would be the exponent of the Liouville action. This would lead to the connection between Liouville action and Berry phase \cite{Berry:2019vih}.

As we found the connections between modular Hamiltonian and complexity, the bound that has been found for modular scrambling modes could be used for complexity of purification,  as the bound of $2\pi$ has been observed for both system, so we get
\begin{gather}
\Big | \frac{d}{ds} \log F_{ij}(s) \Big | \propto \text{CoP}, \ \ \ \ \ \ 
\text{where again:} \ \ F_{ij}(s)= \Big | \langle \chi_i \big | e^{i H_{\text{mod}} s} \delta H e^{- i H_{\text{mod}} s}  \big | \chi_j \rangle \Big |. 
\end{gather}

In \cite{Ghodrati:2019hnn}, as mentioned, we also found that in $2d$ CFTs the complexity of purification between two mixed states would be smaller than $2\pi$ where quantities such as mass $m$ or charge $q$ would decrease CoP. Here we make the conjecture that this bound is related to the bound for modular Hamiltonian for two regions as $
||  \Delta_\psi^{is} (R_2) \Delta_\psi^{-is} (R_1) || \le 1 \ \  \text{for}  \ \  -\frac{1}{2} \le \text{Im} \lbrack s \rbrack \le 0$. We also expect that dissipation (mass) and charge would decrease this bound by changinf the complex modular time $s$. This means that by the effects of the mass of graviton and charge, the growth rate of modular Hamiltonian would become suppressed and the internal modular time would ``click'' more slowly. This then affects entanglement and complexity of purification between the two mixed states.

Furthermore, recently in \cite{Akal:2019hxa}, more direct connections between circuit complexity and Berry phase have been discussed, as the computational cost function has been related to Berry connection, and the Berry phase could be written in terms of Virasoro circuit complexity. For a general path, this relation has been written as
\begin{gather}
\mathcal{B}_{h,c} [g] (\tau) = - \mathcal{C}_{h,c} [g] (\tau)-i \log \langle h \big | \mathcal{\hat{U} } [ ( g^{-1} (0), 0) \ . \ (g(\tau), 0 ) ] \big | h \rangle .
\end{gather}
In fact, the computational cost function has been related to the Berry connection in the unitary representation of the Virasoro group. This can be extended to complexity of purification and a corresponding Berry connection for mixed states could be proposed. The relations between Berry phase and the group representation of other field theories, such as Kac-Moody algebra for the case of warped CFTs as in \cite{Ghodrati:2019bzz,Ghodrati:2017roz} could also be studied.

The links between the geometric phase and complexity could further be understood intuitively. For instance,  when an electron is spiraling along a wormhole passing through one side of a thermofield double state to the other side, the geometric phase that it would pick would be related to the size of the wormhole and therefore to the volume complexity.  On the other hand Berry curvature is also the only gauge-invariant quantity related to the geometric properties of the wavefunctions in the parameter space, and so the links could be evident. In addition, using the Berry phase, and similar to the Bohr-Sommerfeld quantization condition, complexity could also be quantized.  We could have the relation
\begin{gather}
\hbar \oint d {s} \ .  {k} -e \oint d s \ . \ V + \hbar \gamma= 2 \pi \hbar (n+1/2),
\end{gather}
where $\gamma$ is the geometric phase which the electron picks up along the closed loop of the cyclotron orbit and $V$ is the Berry potential. For free electrons, we get $\gamma=0$, while for the electrons in graphene, it would be $\gamma =\pi$. In terms of the energy level, these values are  related to $\alpha=1/2$ for free electrons in the vacuum with the relation $E=(n+\alpha) \hbar \omega_c$,  or $\alpha=0$ for electrons in graphene with the relation for the energy levels as $E=\nu \sqrt{2 (n+\alpha) e B \hbar}$. 

Also, as we mentioned before, the modular Berry curvature could be written as the sum over all the eigenstates as
\begin{gather}\label{eq:indexzeromoded}
\omega_{n, \mu \nu}(\lambda(s)) = i \sum_{n' \ne n} \frac{\langle n | (\partial H/ \partial \lambda_\mu) | n' \rangle \langle n'   | (\partial H / \partial \lambda_\nu )  | n \rangle - (\nu \leftrightarrow \mu)    }{(\epsilon_n - \epsilon_{n'} )}. 
\end{gather}
For a general background which is \textit{mixed}, considering ``all'' \ the eigenvalues with the right degeneracies would be essential for the reconstruction of bulk geometry, similar to the case of reconstructing any function using Fourier expansions. So one would expect that instead of the definition of \textit{index} just based on the number of zero-modes, which would be invariant under all continuous transformations, for the case of mixed states and their purifications, one should add all the modes to define the index which subsequently could be used for bulk reconstruction and defining EoP and Cop. For the case of the symmetric setup of figure \ref{fig:Bitthread2}, however, one could just ``add'' the two terms in eq. \ref{eq:indexzeromoded}, then divide the result by two, in order to find a new quantity which contains all the normal-modes and would be more suitable in reconstructing the wedge cross sections of mixed states. Additionally, in QCD and in heterotic string theory, the index is the number of generations minus the number of antigenerations of leptons and quarks. So, for mixed setups, to consider the effects of all EPR pairs, one needs to make the sum over all the eigenstates in \ref{eq:indexzeromoded}, and then the result divided by two would be proportional to complexity of purification in \cite{Ghodrati:2019hnn}. Note also that using the ``Atiyah-Singer index theorem'', one could specify the relations between fermionic zero modes, the topology of spacetime with various genera, and the anomalies. Moreover, note that the zero-modes could tell how the object moves in space, or superspace in the case of fermions. In the case of instantons, its zero modes would determine how the size or shape would change and this is related to the holographic complexity.

From other parameters of the CFT one could get further information. For instance, the behavior of EoP and CoP for various values of the mass of graviton $m$, has been shown in figures, \ref{fig-D-EOP}, \ref{fig-D-COP}, \ref{fig-D-EOP2}, and  \ref{fig-D-COP2}. So the mass, similar to charge would decrease both EoP and CoP and this could imply that in a gravity background when the graviton is massive, the modular Berry connection, Berry curvature and Berry phase would be smaller than the massless case. This is because in such systems, when varying the Hamiltonian $H(\lambda(t))$, the charge or mass could decrease the rate of the process. Comparing the diagrams of EoP versus CoP, one could also deduce that the modular Berry connection is more interconnected to CoP than to EoP, as CoP probes deeper in the bulk. That is the reason that at $d=2$, $q=0$ and $m=0$, CoP is very close to $\pi$ but it would not be the case for EoP. By increasing the charge $q$, for the case of three-dimensional bulk metric, CoP decreases from $\pi$ to a lower value. This could indicate that modular Berry connection creates a potential wall which its height decreases by increasing the charge of the system. This observation has been established by studying complexity growth rate in charged black holes, \cite{Brown:2018kvn,Ghodrati:2018hss}, so charge would decrease the modular Berry curvature. Additionally, using the idea of \cite{Czech:2019vih}, which considers modular Berry connection as a sewing kit for the entanglement wedge, and also the idea of quantum error connections \cite{Dong:2016eik,Pastawski:2015qua,Almheiri:2014lwa} and the conditions for having a well defined spacetime \cite{Das:2019qaj}, one could see that the patches of spacetimes which contain the same sign charge would become less correlated. This is also true for the backgrounds where the graviton is massive, as the dissipations also make the patches less correlated and lower the modular Berry connections compared to the scenarios with the massless gravitons.

In the setup of mixed states of two strips, there would be a lot of degeneracies. The Berry transformations, acting by automorphisms of energy eigenspace, can rotate these degenerate eigenstates into one another \cite{Wilczek:1984dh, Czech:2017zfq}. The charge or mass could then add additional terms to the initial gauge due to the zero modes and therefore could increase or decrease the holonomy and so the Berry curvature. Any conserved charge would produce new sets of modular zero-modes \cite{Czech:2019vih} and then these modes holographically would be mapped to the edge modes of the dual bulk gauge fields and then, the changes in the modular Berry curvature would change the local field strengths of the gauge fields along the HRRT surface \cite{Ghodrati:2019hnn, Ghodrati:2015rta}. So as we study the changes of EoP and CoP when a gauge field turn on, we also study how these conserved charges would change the modular zero-modes along HRRT surfaces or even along the minimal wedge cross section $\Gamma$ and their effects on the correlations, modular curvature, complexity and CoP. The behavior of gravitational edge modes and soft modes along the minimal wedge cross section would be dual to the behavior of the correlations of \textit{mixed states} in the boundary. Also, if one considers the case of instantons in gauge theories, additional terms would be added and new operators would be built which have new sets of zero-modes and normal-modes. In the region where the instantons are located, the zero modes will be nonzero and nontrivial. So the instantons will induce multi-product interactions which in the bulk can change the holonomy of space time in a non-trivial way and back in boundary side would be translated as the  change in entanglement and complexity of purification as we observed in \cite{Ghodrati:2019hnn}.

As a side note, in \cite{Nakata:2020fjg}, a quantity using the ratio of $\frac{N}{M}$ where $N$ comes $(\ket{\psi}_{AB} \bra{\psi})^{\otimes N}$ and $M$ from $(\ket{\psi}_{A} \bra{\psi})^{\otimes M}$ (for one of the regions $A$ or $B$ in the symmetric case) has been used to define a new quantity for mixed entanglement/complexity.  In \cite{Nakata:2020fjg}, the minimal area in Euclidean time dependent case has been dubbed pseudo entanglement and pseudo complexity.  In addition to complexity of purification (CoP), other quantities such as these new measures or quantities such complexity of randomness which is also called unitary t-design, could be related to modular chaos and modular scrambling modes and bulk reconstruction.

In \cite{deBoer:2019uem}, the modular Berry transport which is being generated by two operators $G_{\pm}$ and the modular Berry scrambling modes which could be written as $ e^{-i H_{\text{mod}}} \delta H_{\text{mod}} e^{i H_{\text{mod}}s} \sim e^{\lambda s} G_+$, has been discussed. If similar to \cite{deBoer:2019uem}, the Hermitian operator which encodes the ``stripped'' \ matrix elements of $\delta H_{mod} (s) $ at large $s$ is being considered, then one could see that the dissipation would suppress these ``stripped'' \ matrix elements $G_+$, as it would also suppress CoP.  The relation $[H_{\text{mod}}, G_+] \approx e^{\lambda s} G_+$, however, would still work for the mixed systems.
  
In our setup, the holonomy for the space of RT surfaces for this specific case which has a gradient, see figure \ref{fig:Bitthread2}, could then be calculated. This holonomy would depend on the geometric component of the modular Berry curvature \cite{Czech:2019vih}.  With an infinitesimal Virasoro excitation, the modular Hamiltonian would be perturbed with the element of $Y+\bar{Y}$, which $Y$ has the form $Y=\int dx^+ f(x_+) T_{++}(x^+)$, and similarly for $\bar{Y}$. When considering the massive and charged case, where $m$ is the mass of graviton and $q$ is the charge on each boundary, we expect to get
\begin{gather}
Y_\lambda=\int dx^+ (1-\cos x^+)^{\frac{1+f(m)+g(q)}{2}}  (1+\cos x^+)^{\frac{1-f(m)-g(q)}{2} } T_{++} (x^+).
\end{gather}

In the next sections, we discuss the specific form of modular Hamiltonian in several cases in more details.

\subsection{Modular Hamiltonian, connected vs. disconnected regions}
One might wonder what would be the modular Hamiltonian for connected versus the disconnected regions.  The modular Hamiltonians for Euclidean path integral states have been studied recently in \cite{Balakrishnan:2020lbp}, and the exact formulation derived, could point out to some analytical expressions for these two cases.

\begin{figure}[ht!]
 \centering
  \includegraphics[width=6cm] {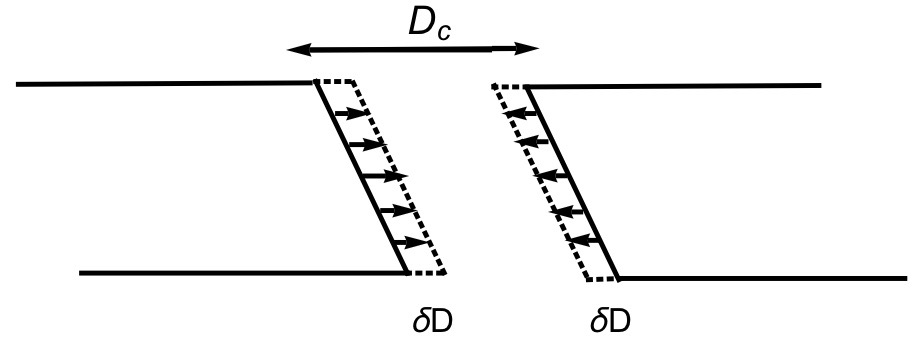}
  \caption{Perturbing the modular Hamiltonian and modular flow by bringing the two strips closer to each other infinitesimally and slowly.}
 \label{fig:twostripsmixperturb}
\end{figure}

To get intuition of the correlations behavior and modular flow in our setup, first we could imagine that both regions just be the half space and in the beginning they are far away. The subregion in the left could be considered as the sources  $\lambda_i$ for the local operators $O_i$ which are distributed along the half-space, left strip, and they excite the modular Hamiltonian and change the modular flow of the subregion in the right half, as shown in figure \ref{fig:twostripsmixperturb}. As found in \cite{Balakrishnan:2020lbp}, the operators then could only be the function of Lorentzian time. Then, if we assume that the two regions move very slowly toward each other as in figure \ref{fig:twostripsmixperturb}, we could use the results from the shape deformation section of \cite{Balakrishnan:2020lbp} to get an intuition of how modular Hamiltonian and modular flow would change.

First, note that the modular Hamiltonian and modular flow for the two disconnected regions would have a stable structure. When one of two regions move toward the other, if they are close enough but still the distance be bigger than $D_c$, the modular flow could ``oscillate'' but the scrambling modes of the components of modular Hamiltonian would still have an upper bound of $2\pi$. When some parts of the two regions get closer than $D_c$ to each other, the modular flow of one region would ``flow'' toward the other one to form a narrow bridge connecting the two regions. At this stage of phase transition, from the disconnected entanglement wedge to the partially connected case, the speed of formation and the rate of change of the components of modular Hamiltonian would be greater than $2\pi$. When the entanglement wedge becomes connected and the states becomes mixed, the non-local part of modular Hamiltonian would get mixed and produces a bigger holonomy in the bulk.

\vspace*{6px}

In the perturbation regime, the expression found for the modular Hamiltonian is  \cite{Balakrishnan:2020lbp}
\begin{gather}
K_\lambda=c_\lambda+K+\sum_{n=1}^\infty \frac{1}{n!} \delta^n K,\nonumber\\
\delta^n K= n! \frac{  (-i)^{n-1} }{ (2\pi)^{n-1}}  \int d\mu_n \int_{-\infty}^{\infty} ds_1 . . . \int_{-\infty}^{\infty} ds_n f_{(n)}(s_1+i \tau_1 . . ., ,s_n+i \tau_n) O(s_1, Y_1) . . . O(s_n, Y_n),
\end{gather}
and $K$ is the vacuum modular Hamiltonian, $Y$ is the spatial coordinate on the half-space subregion, $c$ is constant and $\tau$ is the angular coordinate around the entanglement cut which parameterizes the vacuum modular flow. Also, $d\mu_n = \prod_{i=1}^{n}d\tau_i d^{d-1} Y_i \lambda(\tau_i, Y_i)$ contains $n$ powers of the source $\lambda$. The function in the modular flowed operator relation also would be 
\begin{gather}
f_{(n)}(s_1, . . . s_n)=\frac{1}{2^{n+1}}   \frac{1}{ \sinh (\frac{s_1}{2})      \sinh (\frac{s_2-s_1}{2}). . .   \sinh (\frac{s_n -s_{n-1} }{2})    \sinh (\frac{s_n}{2}) }.
\end{gather}
Using the above relation, the difference between the connected versus disconnected case could be studied. One could see that the main factor in changing the modular Hamiltonian during the phase transition is the singularities of the function $f_{(n)}$ which make $\delta^ n K$ discontinuous.

As another observation, note that the dissipation parameter $m$ and charge $q$ would increase the boost or Rindler times $s_i$ and also the Euclidean angular or modular times $\tau_i$ and therefore they would decrease the function $f_{(n)}$ and therefore the smearing function $F_n^{(\lambda)}$ and so $m$ and $q$ would suppress the excitation terms in the modular Hamiltonian as one expected. For this configuration, the operators could also be replaced as
\begin{gather}
O(s_i) \to a \big ( -e^{s_i} T_{++}+ e^{-s_i} T_{+-} \big ) e^{s_i+ i \tau_i} (+\delta D)+  a \big ( -e^{-s_i} T_{--}- e^{-s_i} T_{+-} \big ) e^{-s_i- i \tau_i} (-\delta D).
\end{gather}
We could assume that as this deformation is null, the modular Hamiltonian could be written as
\begin{gather}
K=2\pi  \int d^{d-2} \vec{x} \int_{\delta \vec{D}} ^\infty dx^+ \big(x^+ -\delta \vec{D} \big) T_{++}(x^+, 0, \vec{x})+c_V.
\end{gather}
The first order correction of modular Hamiltonian would be
\begin{gather}
\delta^1 K= - 2\pi \int d^{d-2} \vec{x} \int_0^\infty dx^+ V^+(\vec{x}) T_{++}(x^+, \vec{x})-2\pi \int d^{d-2} \vec{x} \int_0^{-\infty} dx^- V^-(\vec{x}) T_{--}(x^-, \vec{x}).
\end{gather}
The derivative of this relation, would be related to the Berry phase and therefore to the complexity of purification.

Also note that when the eigenstate of the initial Hamiltonian changes from $\ket{E}= \ket{E(\lambda(0))}$ to
\begin{gather}\label{eq:berryeq2}
\exp (-i \int_0^T E(\lambda(t')) dt') \times \exp (i \oint \Gamma_\lambda d\lambda) \ket{E},\text{where} \ \ \ \ \ \ \Gamma_\lambda=i \langle E(\lambda) \big | d/d\lambda \big | E(\lambda) \rangle,
\end{gather}
the mass and charge would decrease both the Berry connection $\Gamma_\lambda$, and the second phase of the above relation of the closed integral, which is the Berry phase. This could simply be explained by the fact that when varying the system adiabatically, if one has a charged or dissipative system, the gradient $d/d\lambda \ket{E(\lambda)} $ would be smaller. Note that as mentioned in \cite{Czech:2017zfq}, the second factor of \ref{eq:berryeq2}, would indeed arise from the ``\textit{precession}'' of the instantaneous Hamiltonian eigenbasis. So when the graviton is massive or for same sign-charged background, this precession would be smaller which leads to a smaller Berry phase.

\subsection{The effects of dissipation and charge on CC flow and kink transform}

Now we turn to other mathematical tools for bulk reconstructions, such as Connes cocycle flows, OPE blocks and Uhlmann holonomy.

At the cut between two CFTs, a stress tensor shock by the Connes cocycle (CC) flow could appear. The boundary CC flow or one-sided modular flow is dual to the bulk kink transform as shown in \cite{Bousso:2020yxi}. In the dual bulk then there would be a Weyl tensor shock. This unitary cocycle could be written as
\begin{gather}
U_{\psi_2 \psi_1} (s) \Delta^{is}_{\psi_1} O \Delta_{\psi_1}^{-is} U^\dagger_{\psi_2 \psi_1} (s) = \Delta_{\psi_2}^{is} O \Delta_{\psi_2}^{-is},  \ \ \ \ \ \ \forall O \in \mathcal{A},
\end{gather}
The operator $\Delta_\psi= S_\psi^\dagger S_\psi$ is a positive operator called \textit{modular operator} and $S_\Psi$ is an  anti-linear operator acting as $S_\Psi O \ket{\Psi}= O^\dagger \ket{\Psi},  \ \forall O \in \mathcal{A}(R)$. Note that the unitary cocycle flow would also act as a quantum recovery channel and these two would be related through the Hadamard three-line theorem and complex interpolation. The applicability of using Petz map for recovering quantum information which would be dual to remaining in the code subspace would also be related to structure of quantum Markov chain, meaning for the cases close enough to the code subspace,  then we could model the local correlations similar to the quantum Markov chains. The connections between cocycle flow and quantum recovery channel could also be observed from the result of \cite{Levine:2020upy}, where they found that the commutator
\begin{gather}
C_{\text{see}}= \bra{\psi}  \lbrack \tilde{O}, \Delta^{-is}_{\psi; \mathcal{A}} \phi \Delta_{\psi ;\mathcal{A}}^{is} \rbrack \ket{\psi} 
= \bra{\psi}   \Delta^{is}_{\psi; \mathcal{A}}  \Delta^{-is}_{\Omega; \mathcal{A}} O \Delta^{is}_{\Omega; \mathcal{A}}  \Delta^{-is}_{\psi; \mathcal{A}} \ket{\psi},
\end{gather}
could detect the information beyond the causal wedge. All of these cocycle flows would be added together to build up the final modular flow. The quantum recovery channels could also be modeled using them. Therefore, similar to the work of \cite{Chen:2019iro}, the quantum recovery channels in the bulk reconstruction procedure, could also bring operators out from entanglement islands.

From the symmetries of the modular scrambling modes, $\delta H_{\text{mod}} \ket{\psi} =0$, which would lead to $G_\pm \psi  \approx 0$, and their duals which are proposed to be the local Poincare symmetry groups of the bulk, one could see that the change in modular Hamiltonian and the application of cocycle flow, is dual to the change in the curvature of the bulk.  The bigger the change of modular flow,  the deeper the quantum circuit would be with higher complexity, and with more quantum recovery channels which then would lead to bigger curvatures in the bulk. Therefore, black holes are the systems with the ability to change the modular flow fastest, which then means highest cocycle flows. By passing more ``modular time'', $s$, extracting the geometric quantities would become more precise and one could get closer to the RT surface and could zoom-in further.

Using the  boundary connes cocyle flow of $\ket{\psi}$ as the toy model a family of states $\ket{\psi_s}$ could be generated as  $\ket{\psi_s} =( \Delta^\prime_\Omega)^{i s} ( \Delta^\prime_{\Omega | \psi})^{-i s} \ket{\psi}$, where the relative modular operator is $\Delta_{\psi | \Omega} \equiv S^\dagger_{\psi | \Omega}  S_{\psi | \Omega}$. Note that the modular Hamiltonian here could be written as $\widehat{K}_{V_0}= - \log \Delta_{\Omega; \mathcal{A}_{V_0}}$. The parameters such as charge or dissipation could change this flow through changing the modular parameter $s$ and the matrix elements of $S$. For the arbitrary cuts of Rindler horizon one would have the simpler relation
\begin{gather}
\Delta \langle K^\prime_{V_0} \rangle= -2\pi \int dy \int^{V_0}_{-\infty} dv [ v-V_0 (y)] \langle T_{vv} \rangle_\psi,
\end{gather}
where the effects of dissipation and charge on modular Hamiltonian, only through the effects on ``energy momentum tensor'' components could be observed, and then their effects on CC flow could be found out. One could see that both of these parameters would \textit{suppress} the CC flow.

We could write the following relation for the CC flow of stress tensor as
\begin{gather}
\bra{\psi_s} T_{vv} \ket{\psi_s} |_{v < V_0} = e^{-4 \pi s} \bra{\psi} T_{vv} \big ( V_0+ e^{-2\pi s} (v-V_0) \big) \ket{\psi} |_{v< V_0}.
\end{gather}
So, one could see here too that the dissipation and same sign charge would increase $s$ and matrix elements of $T$ and therefore the CC flow. Also, the shape derivatives of modular Hamiltonian which could be written as
\begin{gather}
\frac{\delta \langle K_V^\prime \rangle_\psi }{\delta V} \Big |_{V_0}=2\pi \int^{V_0}_{-\infty} dv \  \langle T_{vv} \rangle_\psi,
\end{gather} 
would be suppressed, since the components of energy momentum tensor would become smaller due to the effects of dissipation and same-sign charges.

Then, note that the dual of CC flow which is the bulk ``kink transform'' \cite{Bousso:2020yxi} could be written in the form of 
\begin{gather}\label{eq:kink1}
(K_\Sigma)_{ab} \to (K_{\Sigma_s})_{ab}= (K_\Sigma)_{ab} - \sinh (2\pi s) x_a x_b \delta (\mathcal{R}).
\end{gather}
These transformations then, by combinations of modular operators could generate some sequences of global states $\psi_s$. These then could point out to its relation to complexity of building states using the intuitions from \cite{Chen:2020nlj}.
The rapidity, $2\pi s$, which is being formed by the relative boost which glues entanglement wedges $A$ and its complement $A'$ would be connected to the value of $2\pi$ in $2d$ that we found for the complexity of purification. The effects of the dissipation parameters $m$ and the same-sign charge could be also be seen using the relation \ref{eq:kink1}. Again one could see, through their effects on ``$s$'', these two parameters would suppress the kink transform.

  \subsection{Entanglement wedge cross section from OPE blocks}

  In this section, we investigate  the ``geodesic operator/OPE block dictionary'' to understand the complexity of purification, mixed state correlations of two strips, the bit thread structure and wedge reconstruction further.

  On the boundary CFT, the zero modes of $H_{mod} (\lambda)$ would make the OPE blocks, which come from two spacelike separated local operators in the form of
  \begin{gather}
  \mathcal{O}_L (x_L) \mathcal{O}_R (x_R)= \sum_\Delta \big | x_R-x_L \big |^{-\Delta_L -\Delta_R} c_{LRi} \big | x_R-x_L \big |^\Delta (\mathcal{O}_\Delta+ \text{descendants}),
  \end{gather}
where the OPE blocks are the second part, $\big | x_R-x_L \big |^\Delta (\mathcal{O}_\Delta+ \text{descendants} )$. For a finite transformation $s_0$, the OPE blocks transform as $B_\Delta^\kappa (\lambda) \to e^{s_0 \kappa} B_\Delta^\kappa (\lambda)$ which is a change of normalization.
  
The holographic dual of a scalar OPE block is a bulk operator in the form \cite{Czech:2017zfq, Czech:2016xec,daCunha:2016crm}
 \begin{gather}
 B_\Delta^\kappa (\lambda)= N\int_{[\lambda]} ds \phi_\Delta (s) e^{-\kappa \varsigma},
 \end{gather}
 where here $\varsigma$ is the proper length parameter along the bulk geodesic $[\lambda]$ (in the units of $L_{\text{AdS}}$) and $\phi_\Delta$ is the bulk operator dual to $\mathcal{O}_\Delta$.

Any product scalar of two operators $O_1(x_1)$ and $O_2 (x_2)$ could be expanded in terms of the OPE blocks. These blocks are some primary operators that are being smeared in a causal diamond $\diamond_{12}$ and so could be written as \cite{Czech:2016xec},
 \begin{gather}
 \mathcal{B}_k (x_1,x_2)= \frac{\Gamma(2h_k) \Gamma (2 \bar{h}_k) }{\Gamma(h_k)^2 \Gamma (\bar{h}_k)^2}\int_{\diamond_{12}} dw d\bar{w} \left( \frac{ (w-z_1)(z_2-w)    }{z_2-z_1}  \right)^{h_k-1}    \left( \frac{ (\bar{w}- \bar{z}_1)(\bar{z}_2-\bar{w})    }{\bar{z}_2- \bar{z}_1}  \right)^{\bar{h}_k-1},
 \end{gather}
 where  the left and right moving conformal weights are $h_k= \frac{1}{2} (\Delta_k + \ell_k) $ and $\bar{h}_k= \frac{1}{2} (\Delta_k - \ell_k) $ and $\Delta_k$ and $\ell_k$ are the scaling dimension and spin of the quasiprimary operators $\mathcal{O}_k$. The parameter which is being changed by the mass of graviton $m$ or charge $q$ would be $h_k$ which changes the nature of OPE blocks in the bulk reconstructions formalism.  
 
 The complexity of purification for the mixed states would then be related to the number of quasiprimary operators, $ \mathcal{O}_k$,  which are needed to produce $\diamond_{12}$ with enough, desired precision. The sewing of patches of various causal diamonds $\diamond$ would create the bulk entanglement wedge. So the similarities between the summation of causal diamonds and other bulk reconstruction methods such as recovery channels and modular flows could be seen here.

Note that in the bulk, the dual of the boundary causal diamond would be a geodesic operator. Then, there is the equality of OPE blocks and the X-Ray/ Radon transforms, which even in higher dimensions, and for time-like separated pairs, the OPE blocks could be considered as a surface Witten diagram which could be written as
\begin{gather}
g_k(u,v) =\frac{1}  {(c^\varsigma_\Delta)^2 } \int_{\sigma_{12} } d^{d-2} z \int_{\sigma_{34}} d^{d-2} z' G_{bb} (z, z'; m_k).
\end{gather}
Here, $x_1$, $x_2$ and $x_3$, $x_4$ are the endpoints of the two intervals. This non-local quantity, could be considered similar to our notion of volume interval (VI) \cite{Ghodrati:2019hnn} as a measure of complexity of purification of mixed states. Note that the appearance of OPE blocks in various measures such as EoP \cite{Tamaoka:2018ned} or odd entanglement entropy \cite{Tamaoka:2018ned} has been discussed. Therefore, the connections between these new quantities and modular flow, quantum recovery channels and CoP could be considered. We expect that the integral of bulk fields along the minimal wedge ``cross section'', and therefore the OPE blocks would be proportional to the entanglement of purification as we have
\begin{gather}
c_{\Delta} \mathcal{B}_k(x_1,x_2) = \tilde{\phi}_k(\gamma_{12}) = \int_{\Gamma} ds \ \phi(z),
\end{gather}
 where $c_\Delta= \Gamma(\frac{\Delta}{2})^2/ 2\Gamma(\Delta)$ here is a constant. Then, the relationship between the space of causal diamond which is a coset space and is being defined as
 \begin{gather}
 \mathcal{K}=\frac{SO(2,2) }{SO(1,1) \times \overline{SO(1,1)}}=\frac{ {SO(2,1)}  }{{SO(1,1)} } \times \frac{ \overline{SO(2,1)}  }{\overline{SO(1,1)} },
 \end{gather} 
  and the minimal entanglement wedge cross section and the flow of modular zero modes $B_i (\lambda) $ and Berry flow could then be considered. We conjecture that the tips of all causal diamonds have a decreasing flow from the point $m$ to $m'$ and also as the OPE block transform as $\mathcal{B}_\Delta^\kappa (\lambda) \to e^{\varsigma_0 \kappa} \mathcal{B}_\Delta^\kappa (\lambda)$, the parameter $\varsigma_0$ becomes smaller when moving from the point $m$ to the point $m'$.

Another point worths to mention is that a $2d$ causal diamond would be stabilized by an $SO(1,1) \times \overline{SO(1,1)}$ group. The anti-symmetric combination of these two $SO(1,1)$ labeled $P_D(\lambda)$ in \cite{Czech:2017zfq}, satisfy the relation $\lbrack P_D, \mathcal{O}_L \mathcal{O}_R \rbrack = i \kappa \mathcal {O}_L \mathcal{O}_R$, where $\kappa= \Delta_R - \Delta_L$. Note that the dissipation parameter and  the same sign charge would suppress $\kappa$ as they could suppress the dilatation and transformation generated by $P_D$.

A related point is that when considering the light-cone cuts as in \cite{Engelhardt:2016crc,Engelhardt:2016wgb}, in the bulk of  two points of $p$ and $q$, i.e, $C^-(p)$ and $C^-(q)$ would be intersected at a single point $X$ and they would be regular at that point, then these two points $p$ and $q$ are null-separated. From the data of the boundary mixed CFTs, then one could determine the cut in the bulk, as whether it would be a connected one if its radial coordinate is above the point $m'$, i.e, $z_{m^\prime}<z<z_m$, or disconnected, if $z<z_m$, or non-existent if $z>z_{m^\prime}$.

   \begin{figure}[ht!]
 \centering
  \includegraphics[width=4.5cm] {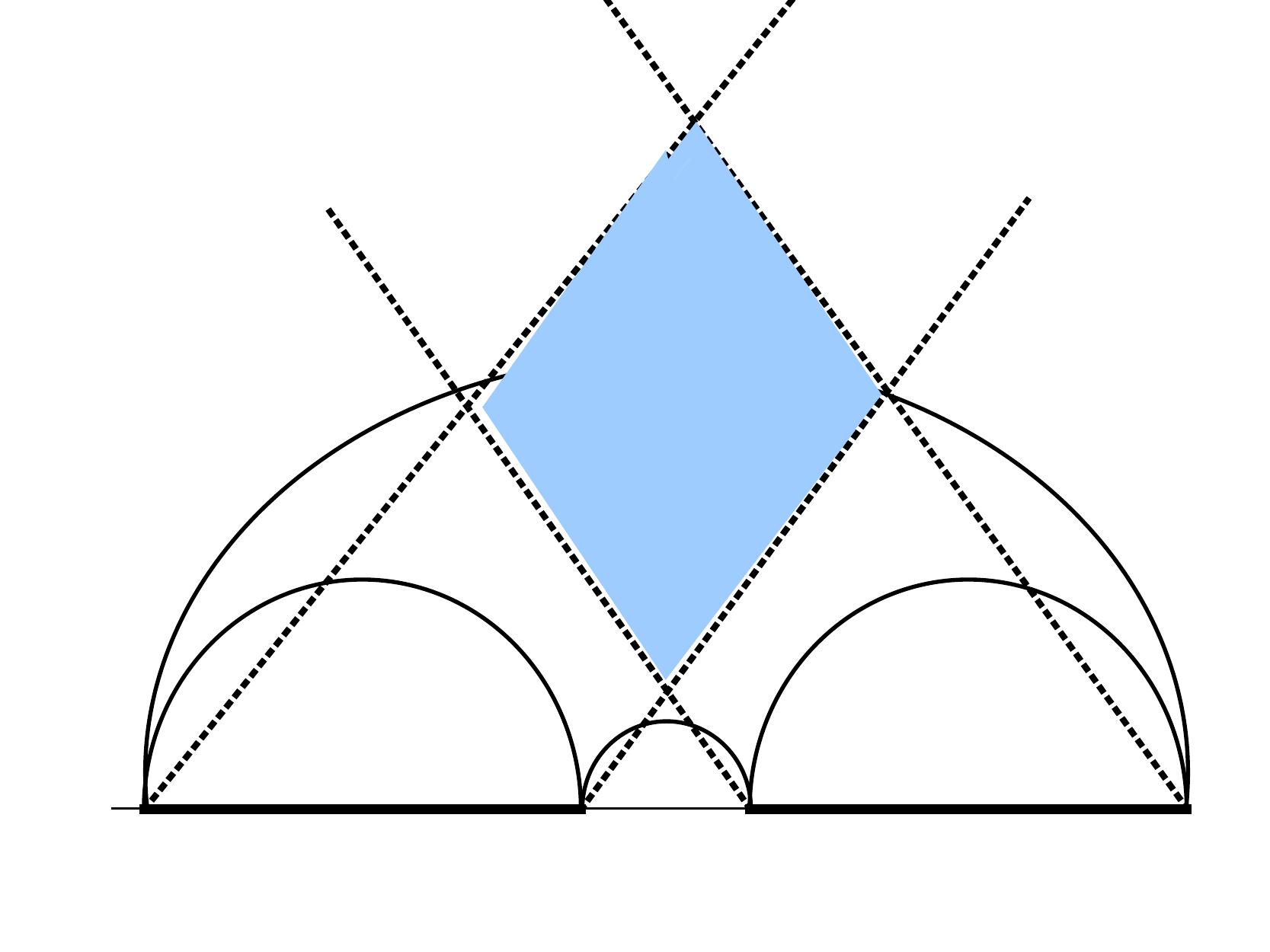}
    \includegraphics[width=4.5cm] {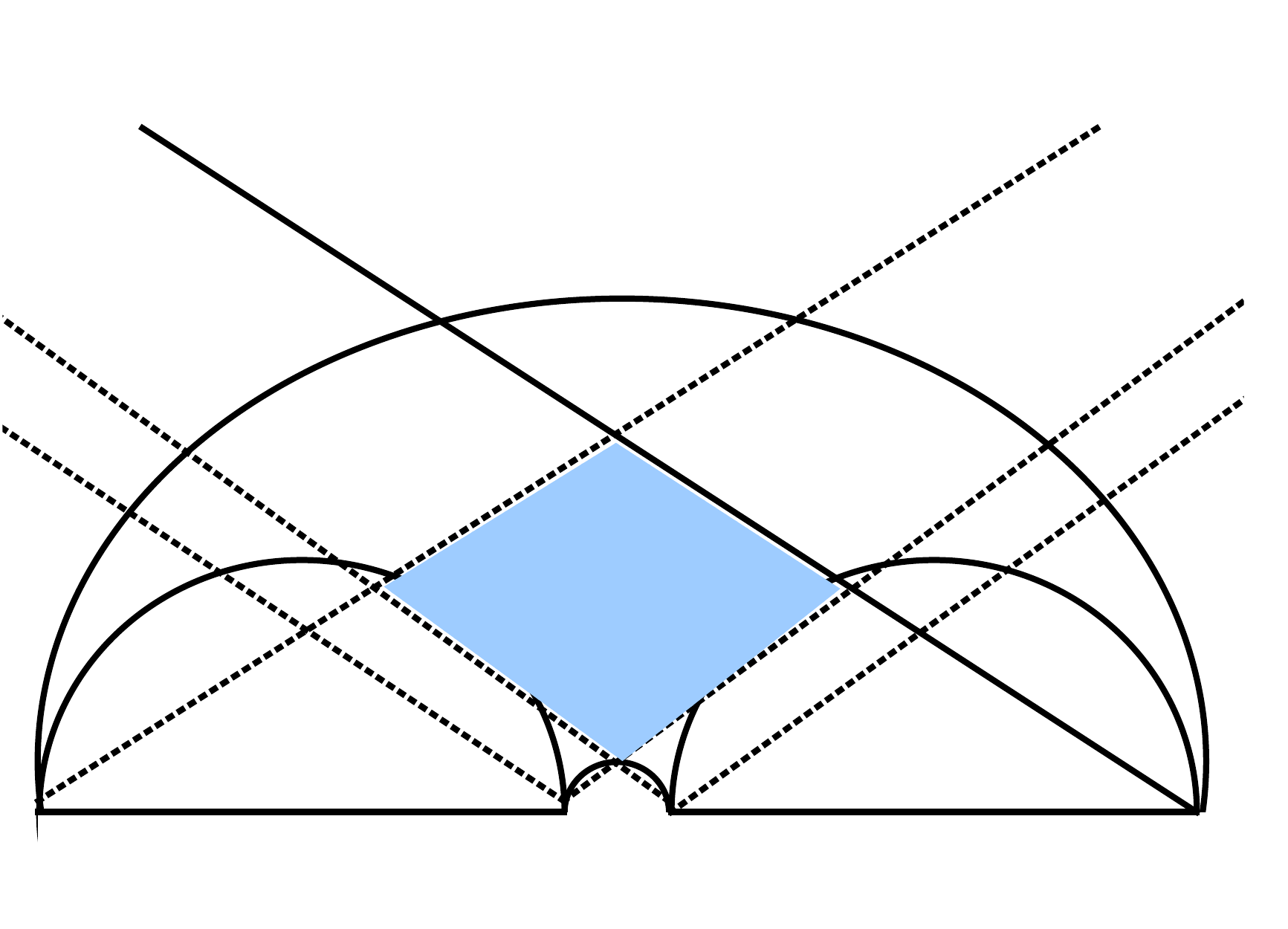}
  \caption{Various conditions for the shape of mutual diamond between two subregions. It could either be completely inside the entanglement wedge or partially inside it, depending the size of strips and the distance between them. }
 \label{fig:wedgecone}
\end{figure}

The connections between the causal wedge and mutual information could also point out the way to reconstruct the dual of mixed states. In a recent paper \cite{May:2019odp}, the connections between entanglement wedge reconstruction and mutual information through the holographic scattering has been discussed. Using the null vectors, the light cone cuts could be constructed for the boundary strips which would lead to various forms of the \textit{mixed diamonds} between the two states as shown in figure \ref{fig:wedgecone}. Using the OPE block structure the position of mutual diamond inside the entanglement wedge could be studied. We leave the detailed numerical calculations of these studies to future works.

 \subsection{Uhlmann holonomy for mixed states}

In this section, we turn to another tool in studying the wedge reconstruction. In \cite{Kirklin:2019ror}, it has been proposed that the dual of the symplectic form of the bulk fields in any entanglement wedge would be the curvature of the Uhlmann phase. In another word, for the mixed states, the symplectic form of the bulk fields would be dual to the Uhlmann holonomy of the parallel transport of purifications of density matrices of the boundary which is the maximization of the transition probabilities.

The symplectic form of the fields in the bulk is in the form of $\Omega= \int_\Sigma \omega$, which is the integral of 
\begin{gather}
\omega  \lbrack \phi, \delta_1 \phi, \delta_2 \phi \rbrack=\delta_1 (\theta \lbrack \phi, \delta_2\phi \rbrack ) -\delta_2(\theta \lbrack \phi, 
\delta_1 \phi \rbrack ) -\theta \lbrack \phi, \lbrack \delta_1 \phi, \delta_2, \phi \rbrack \rbrack,
\end{gather}
and the parameters could be defined by the variation of the Lagrangian as
\begin{gather}
\delta L= L \lbrack \phi+\delta \phi \rbrack - L \lbrack \phi \rbrack= \delta \phi \  . \ E + d \theta.
\end{gather}
On the other hand, in the dual boundary theory, for two states with parallel purifications, the fidelity would be $ | \braket{\psi_2| \psi_1}| = \text{tr} \left ( \sqrt{ \sqrt{\rho_1} \rho_2 \sqrt{\rho_1}  } \right )$, and the replicated fidelity could be defined as $F_k=\text{tr} \left ( ( \sqrt{\rho_1} \rho_2 \sqrt{\rho_1})^k \right )$. Considering a closed curve $C: S^1 \to \mathcal{H} $ and the sequence of $n$ states in the form of $\ket{\psi_1}, \ket{\psi_2}, . . . ., \ket{\psi_n} $, for the limit of $n \to \infty$ where the states $\ket{\psi_i}$ covers the curve $C$, one would get $\langle \psi_1 \big | \psi_n \rangle       \langle \psi_n \big | \psi_{n-1} \rangle .\ . \ . \ \langle \psi_3 \big | \psi_2 \rangle  \langle \psi_2 \big | \psi_1 \rangle    \longrightarrow \text{exp} (i \gamma)$, where the Berry phase (or the Uhlmann phase) is $\gamma= \oint_C a$ and $a$ is defined as $a=i \langle \psi \big | \mathrm{d} \big | \psi \rangle,$ which is a real 1-form.  The Berry connection is defined by the transformation $a \to a-\mathrm{d} f$ which is a $U(1)$ connection, and the Berry curvature which is gauge-invariant is defined as $da=i \mathrm{d} \langle \psi \big | \land \mathrm{d} \big | \psi \rangle$. One question is that how much this phase (in the bulk side) would change with changing the position of the curve $C$ with respect to the two correlated intervals that we consider. 

Using holography, the boundary replicated fidelity for the case of $k=1/2$, could be written in terms of an on-shell gravitational action where the sources are on the boundary. Therefore, the bulk symplectic form and the boundary Uhlmann phase would be connected to each other holographically. The curvature of the integral of the abelian connection which is the Berry phase along the Uhlmann parallel path would then be the symplectic form of the entanglement wedge and consists of flux tubes between the mixed states. The relative (total)-mode frames would then be related to the modular Uhlmann connection. In each side, the effects of dissipation or $m$ and charge, $q$, on the bulk symplectic form and on boundary Uhlmann phase could then be considered. Since these two would suppress $\delta \phi$, they would then suppress $\delta L$ and then the fidelity between two states and as the result the Uhlmann phase would be suppressed which again the duality is being checked this way.

In \cite{Viyuela:2016ror} also a method for observing the topological Uhlmann phase with superconducting qubits for topological insulators has been reported. They used an ancillary system and some particular interferometric techniques.  The single qubit density matrix could be written in terms of $\theta(t) {\big |}^1_{t=0}$ as $\rho_\theta=(1-r) \ket{0_\theta}\bra{0_\theta}+r  \ket{1_\theta}\bra{1_\theta}$, where $r$ quantifies the mixedness between the two states of $\ket{0_\theta}$ and $\ket{1_\theta}$. These two states are kind of a transmon qubit introduced in \cite{PhysRevA.76.042319}. The evolution of the purification of $\rho_\theta$ would then be in the following form
\begin{gather}
\ket{\Psi_{\theta(t)}} = \sqrt{1-r} U_S(t) \ket{0}_S \otimes U_A(t) \ket{0}_A
+\sqrt{r} U_S(t) \ket{1}_S \otimes U_A(t) \ket{1}_A,
\end{gather}
where $\ket{0} = \begin{pmatrix} 1\\0 \end{pmatrix}$ and $\ket{1} = \begin{pmatrix} 0\\1 \end{pmatrix}$ are the basis, $S$ stands for the system and $A$ for the ancilla. Then, the Uhlmann parallel transport condition is satisfied when the distance between the two infinitesimally close purifications, $\big{\Vert} \ket{ \Psi_{\theta(t+dt)}} -\ket{\Psi_{\theta(t)}} \big {\Vert}^2 $, becomes minimum, which physically means that the accumulated phase which is the Uhlmann phase $\Phi_U$ would be completely geometrical and not dynamical. So due to the holographic duality, this interferometric technique of observing Uhlmann phase could show a way to measure the properties of the symplectic form and quantum gravity characteristics in the bulk. Therefore, this method could paint the CFT entanglement structures and the bulk gravity curvature properties in more details, as we also did some few steps in \cite{Ghodrati:2020mtx}.

 These studies could then point out to some specific intrinsic properties of space-time and gravity and specially topological gravity, which is independent of the dynamics of the system as Berry phase and Uhlmann phase are so too. For example the connections between some differential geometric properties of bulk could further be examined by studies of Uhlmann phase. Case in point, the extremum of the K$\ddot{a}$hler potential, $\mathcal{K}=\log \langle \alpha \big | \alpha \rangle$, corresponds to the minimum of the entanglement wedge cross section for two mixed states. The connections between K$\ddot{a}$hler potential and Berry curvature and complexity then could be specified.  The K$\ddot{a}$hler potential is a real-valued function, being denoted by $f$ and is defined on a K$\ddot{a}$hler manifold for which the K$\ddot{a}$hler form $\omega$ could be written as $\omega=i \partial \bar{\partial} f$, where $\partial= \sum \frac{\partial}{\partial z_k} dz_k$ and $\bar{\partial}= \sum \frac{\partial}{\partial z_k} d \bar{z}_k$. The Berry curvature in terms of the K$\ddot{a}$hler potential could then be written as
\begin{gather}
\mathcal{A}= \frac {i}{2} \partial_{\alpha_i} \mathcal{K} d \alpha_i - \frac{i}{2} \partial_{\alpha_{{ i^\ast}}} \mathcal{K} d \alpha_{i^\ast}=\frac{i}{2} (\partial- \partial^\ast) \mathcal{K}.
\end{gather}
For the minimum entanglement wedge cross section, this Berry curvature becomes zero.

Also, note that the relative modular frame along the RT surface is encoded in the connection of the relevant bundle. For the mixed states, the mode frames along the minimal wedge cross section $\Gamma$ would create the modular Uhlmann connection. However, the strength of this connection depends on the mutual information and therefore on the distance between the two subregions.  So the gradient of the modes along $\Gamma$ depends on $\frac{dI }{d D_x}   \big | _\Gamma$ or $\frac{dE_W }{d D_x}   \big | _\Gamma$. This pattern is the same pattern of the Wilson lines which is also being dictated by the pattern of entanglement of purification and mutual information among the subsystems. The form of the dressing of gauges, which is dual to the pattern of EoP, would dictate the structure of curvature in the bulk.

The pattern of the mutual information and the entanglement of purification between the two subsystems are being determined by the pattern of the physical Wilson lines ``dressed in gauge theories''. Wilson lines could be written in the form of
\begin{gather}
U \lbrack x_i,x_f; C  \rbrack=\mathcal{P} \exp \Big ( i \int_{\tau_i}^{\tau_f} d\tau \frac{dx^\mu}{d\tau} A_\mu(x(\tau)) \Big )=\mathcal{P} \exp  \left ( i \int_{x_i}^{x_f} A \right ). 
\end{gather}

Moreover, as mentioned in \cite{Czech:2017zfq}, the modular Berry transformations could recognize the bulk operators which are localized on the RT surfaces. So for two subregions of $A$ and $B$ which correspond to two mixed states, the RT surfaces which probe the entanglement wedge cross section between them as shown in the figure \ref{fig:Bitthread2}, where  the bit threads and the bulk operators are located, would be related to the CFT modular Berry transformations. So this way the modular Berry transformation would be related to the bit thread structures as well.

 \section{Correlation Measures for Mixed States and Quantum Discord}

 In this section, we aim to study the nature of correlations among mixed systems further. Using some new correlation measures, we also study the effects of dissipation and charge in our setup as well. As a first step, note that the relative entropy which is a measure of distinguishability between two states which could be a reference vacuum state $\sigma$ and another state $\rho$ is defined as $S(\rho | \sigma)= Tr [ \rho \log \rho-\rho \log \sigma]$. This quantity would have some connections with strength of correlation and therefore the purification (EoP/CoP). Note that this quantity would be related to the free energy difference between $\rho$ and vacuum at temperature $\beta=1$. For stronger correlations, the free energy that one would be able to extract, would be lower. Therefore, EoP and relative entropy would have an inverse behavior relative to each other.

 One then can define various measures for quantum correlations. For instance, in \cite{Liu:2019ent}, the Uhlmann fidelity was proposed for Gaussian states. The form of their quantity $N_F^{ \mathcal  { G  }, A } (\rho_{AB}) $ is
\begin{gather}
N_F^{ \mathcal  { G  }, A } (\rho_{AB})= \text{sup}_{U \in \mathcal{U}_{\rho_{AB}} }  C^2 (\rho_{AB} , (U \otimes I) \rho_{AB} ( U^\dagger \otimes I )) \nonumber\\
= \text{sup}_{U \in \mathcal{U}_{\rho_{AB}} } \{ 1- F(\rho_{AB}, (U \otimes I) \rho_{AB} (U^\dagger \otimes I)) \},
\end{gather}
where as mentioned before, the Uhlmann fidelity, $F(\rho, \sigma) = ( \text{Tr} \sqrt{ \sqrt{\rho} \sigma \sqrt{\rho}  })^2$, is a measure of closeness between the two states, the sine metric is $C(\rho, \sigma) = \sqrt{1-F(\rho, \sigma)}$, and also the supremum should be taken over all Gaussian unitary operators, $U \in \mathcal{U_{\rho_{AB}} }$.

After applying any Gaussian channel, $\Phi$, $N$ would decrease as  $N_F^{ \mathcal  { G  } } ( I  \otimes \  \Phi   (\rho_{AB}))  \le    N_F^{ \mathcal  { G  } } (\rho_{AB}) $, which means that this measure is non-increasing under any Gaussian quantum channel. For our setup of two strips of $A$ and $B$, the parts which are further away from each other and therefore less correlated would have smaller  $N_F^{ \mathcal  { G  } } ( I  \otimes \  \Phi   (\rho_{AB}))$ which corresponds to the regions where more quantum channels have been applied to the subsystems and the density of modular flow would be lower there.

\begin{figure}[ht!]
 \centering
  \includegraphics[width=4.5 cm] {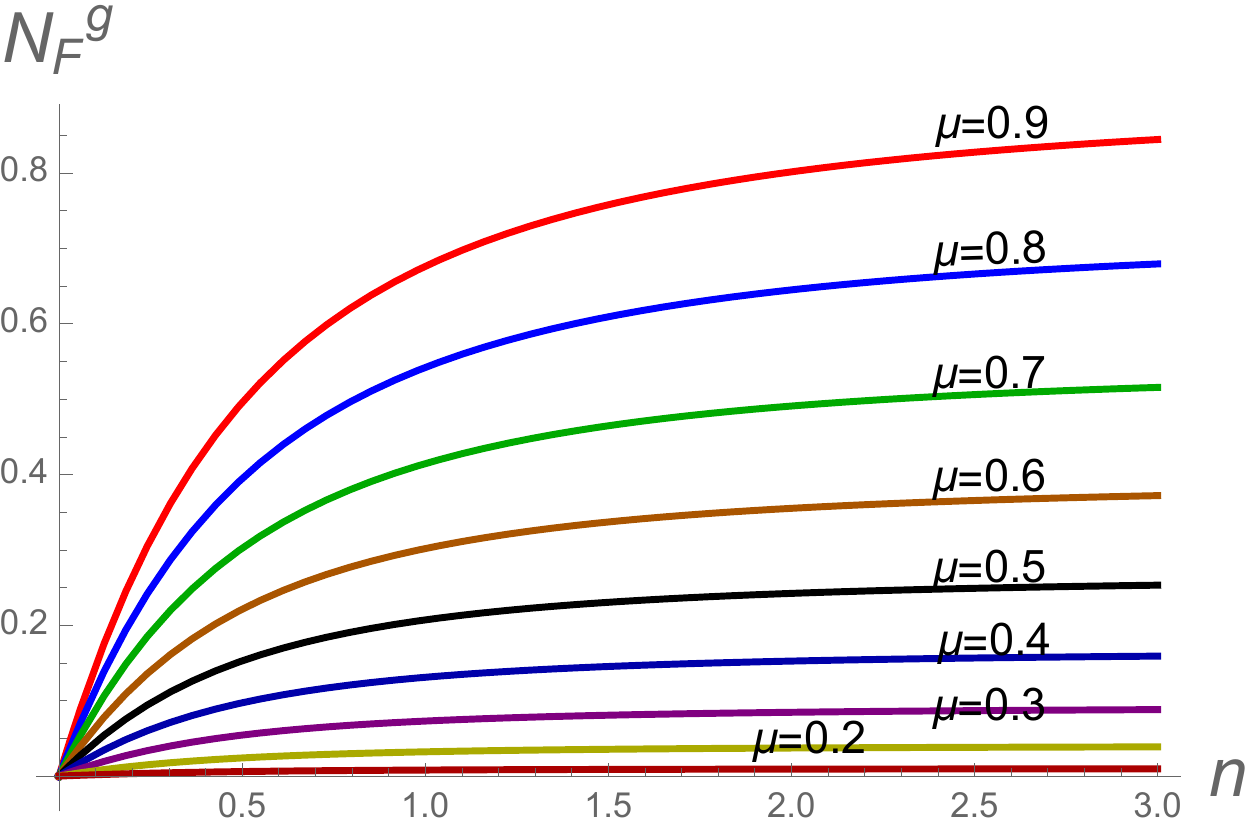} 
  \includegraphics[width=4.5cm] {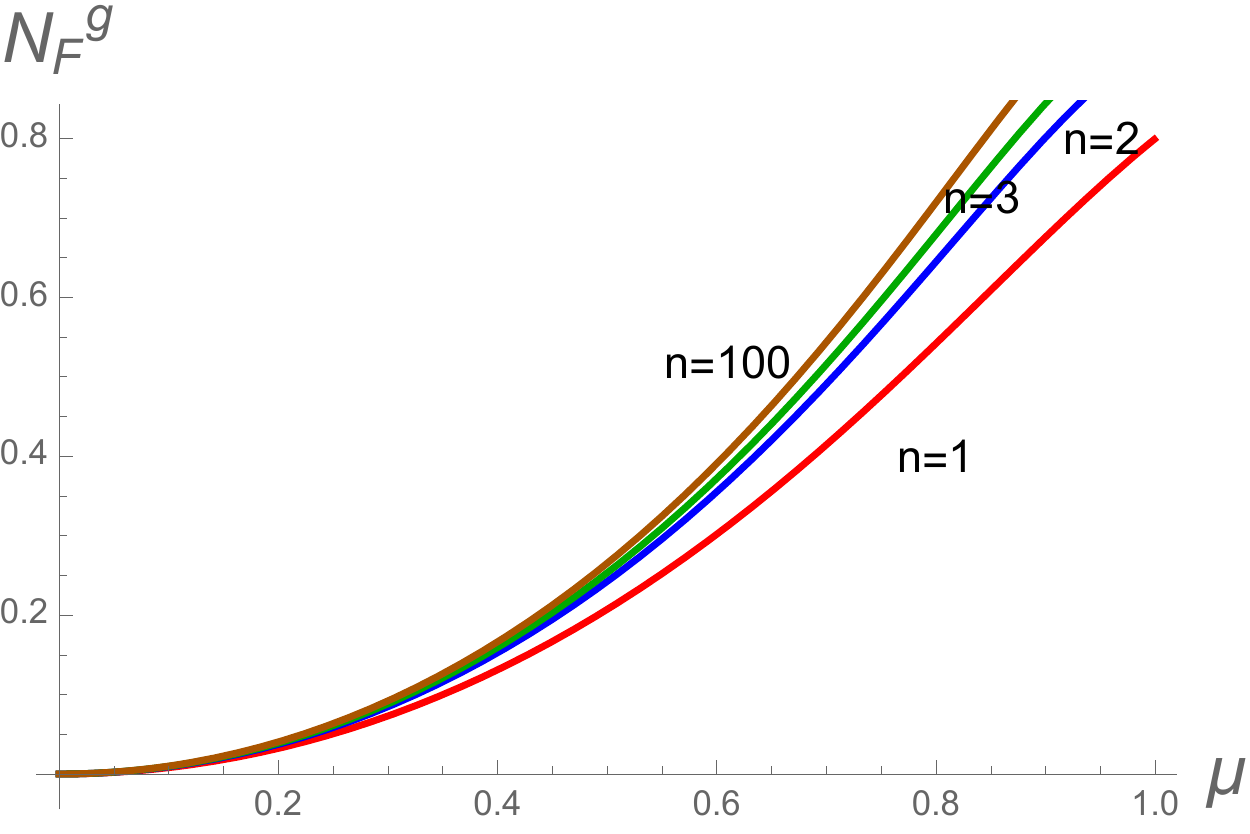}
  \includegraphics[width=5cm] {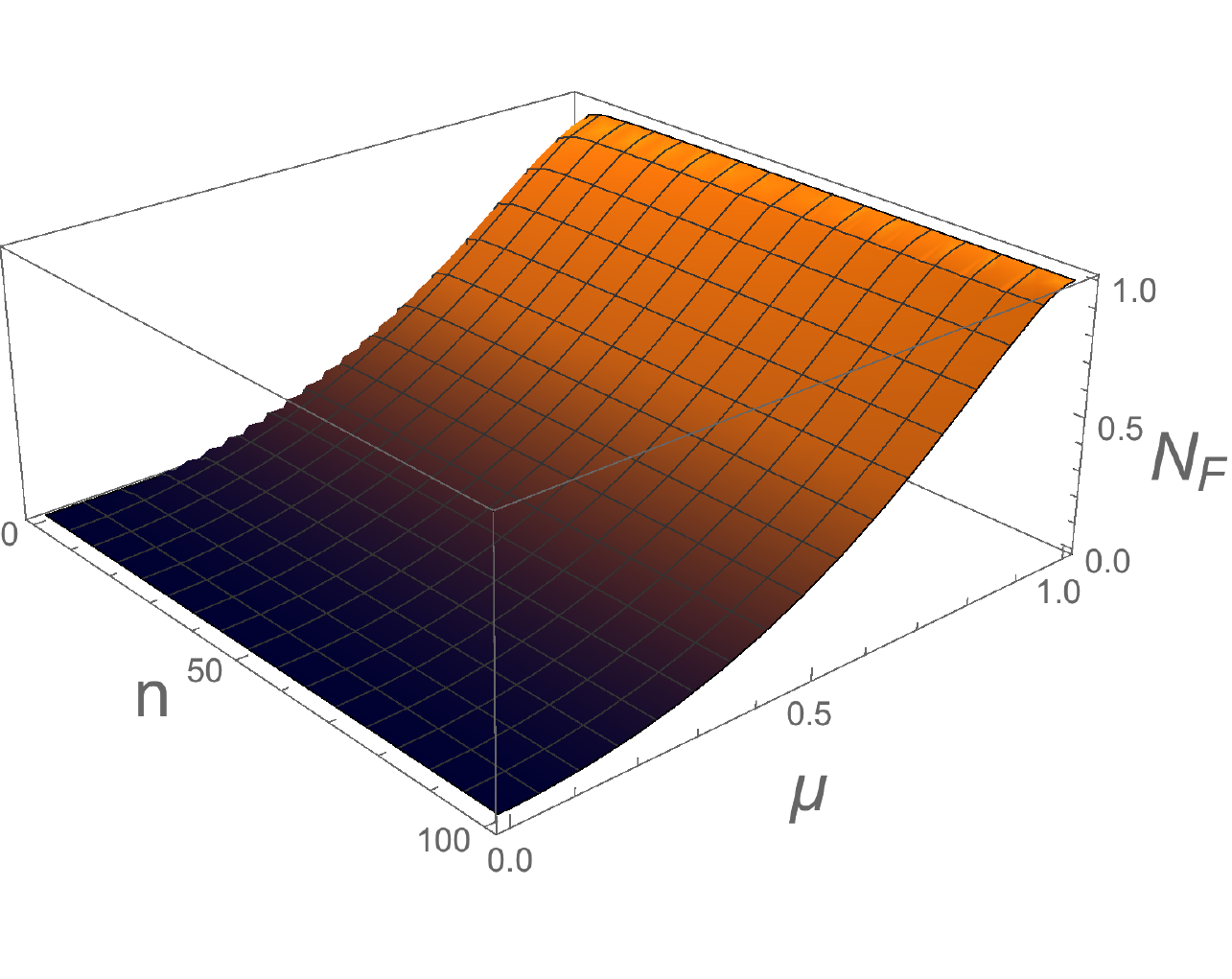}
  \caption{The relation between the correlation measure $N_F^{\mathcal{ G  },A}$ versus $\mu$ and $n$ .}
 \label{fig:correlationmeasureUhlmann}
\end{figure}

Then, for the case of $(1+1)$-mode ``symmetric squeezed thermal state (SSTS)'' $\rho_{AB} (n,\mu)$ of \cite{Liu:2019ent}, in their equation 4,   the relations between the correlation measure $N_F^{\mathcal{ G  },A}$ and the two parameters denoted by $n$ and $\mu$ have been derived.  Actually,  $\mu$ is the mixing parameter where $0 \le \mu \le 1 $ and $n$ would be the mean photon number for each part. In figure \ref{fig:correlationmeasureUhlmann}, we present several plots to show the relations between the correlation measure $N_F^{\mathcal{ G  },A}$ versus the parameters $\mu$ and $n$.

\vspace*{6px}

 It could be seen that the mixing parameter $\mu$ would increase this correlation measure. We expect that parameters such as mass of graviton modeling dissipation and the same sign charge, which can decrease the mixing parameter, would also decrease this specific correlation measure. As both EoP/CoP are two other measures of correlation between mixed systems, we see that since $m$ and $q$ would decrease $\mu$, therefore, these quantities also decrease EoP and CoP as well \cite{Ghodrati:2019hnn}.

Another point comes from the definition of the channel capacity which is in the form of $C_E(\mathcal{N})= \underset{ \text{all \ } p_i  , q_i }{\text{max}}  I(A:B)$, where the maximization is over all input ensembles of the mutual information between the two systems.  As the parameters $m$ and $q$ of the boundary would decrease $I(A:B)$, and also the mixing parameters, we could see that they also decrease the capacity of the quantum channels.  Therefore, EoP and CoP behave the same way as the quantum channel capacity under changing charge and dissipation, as these would decrease the mixing parameter.  Decreasing the quantum channel capacity then decreases the modular flows between the two subregions. 

This could also be seen from the definition of modular flow being written as
\begin{gather}
U_\sigma (s) \mathscr{A}_L U_\sigma^\dagger (s)=\mathscr{A}_L, \ \  \ U_\sigma (s) \mathscr{A}_R U_\sigma^\dagger (s)=\mathscr{A}_R, \ \ U_\sigma (s)\equiv \Delta_\sigma^{-is},
\end{gather}
where
\begin{gather}
K_\sigma^R = - \log \rho_\sigma^R,  \ \ \ K_\sigma^L= - \log \rho_\sigma^L,\ \ \ K_\sigma^L= J_\sigma K_\sigma^R J_\Omega, \ \ \ \Delta_\sigma=\rho_\sigma^L\otimes (\rho_\sigma^R)^{-1}.
\end{gather}
Generally dissipation makes the eigenvalues of the density matrix smaller. Also, the same sign charge suppresses the operator $\Delta$  as it decreases the entanglement between the physical modes among $L$ and $R$, and therefore $q$ would suppress the modular flow. This is also the case for the relative entropy between two states $\ket{\Psi}$ and $\ket{\Omega}$, $D(\rho | \sigma)= - \bra{\rho} \log \Delta_{\sigma \rho} \ket{\rho}$.

These results about the effects of $m$ and $q$ are also true for the difference between two modular operators, which could be seen from relations 3.34 - 3.43 of \cite{Lashkari:2018oke}, as charge and dissipation increase the ``distance'' between the clicks of modular time and therefore decrease the terms $\frac{g_\epsilon (t_{i+1}-t_i) }{\cosh(\pi t_i) \cosh (\pi t_{i+1} )}$.

\begin{figure}[ht!]
 \centering
  \includegraphics[width=4.5 cm] {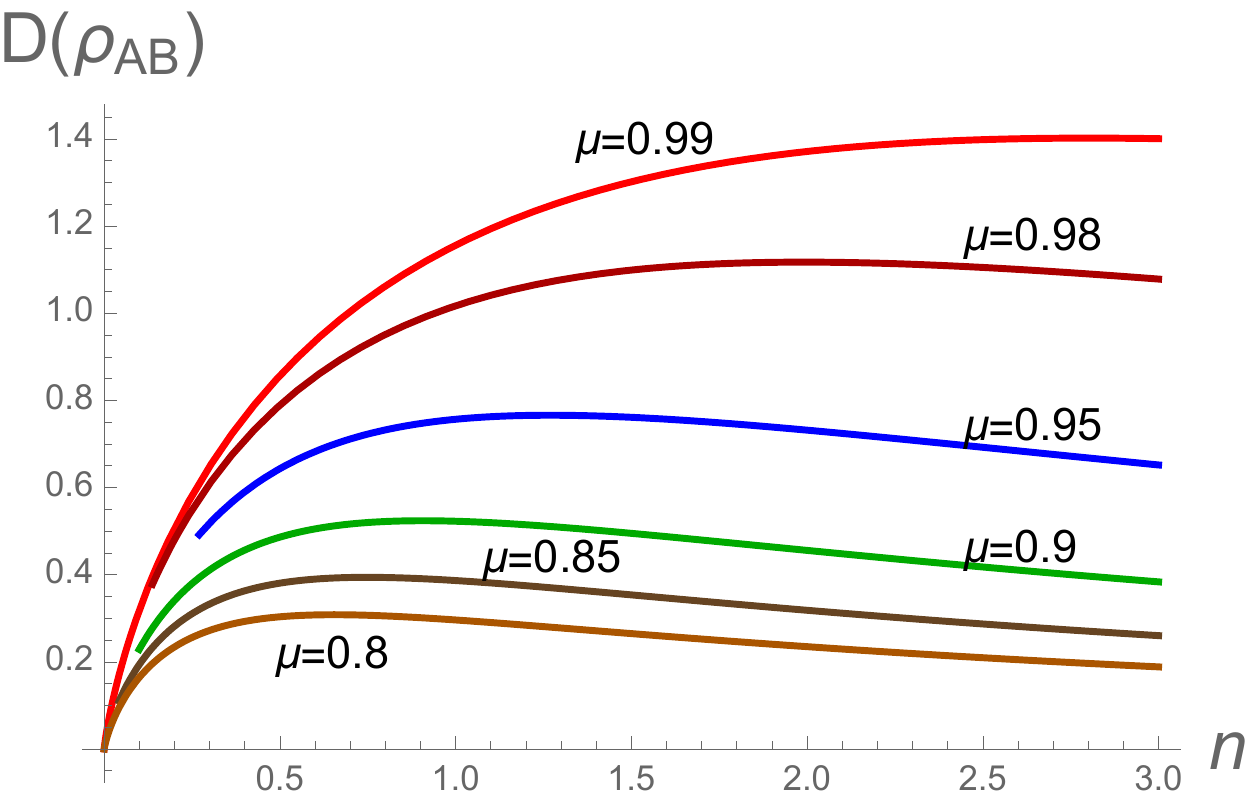} 
  \includegraphics[width=4.5cm] {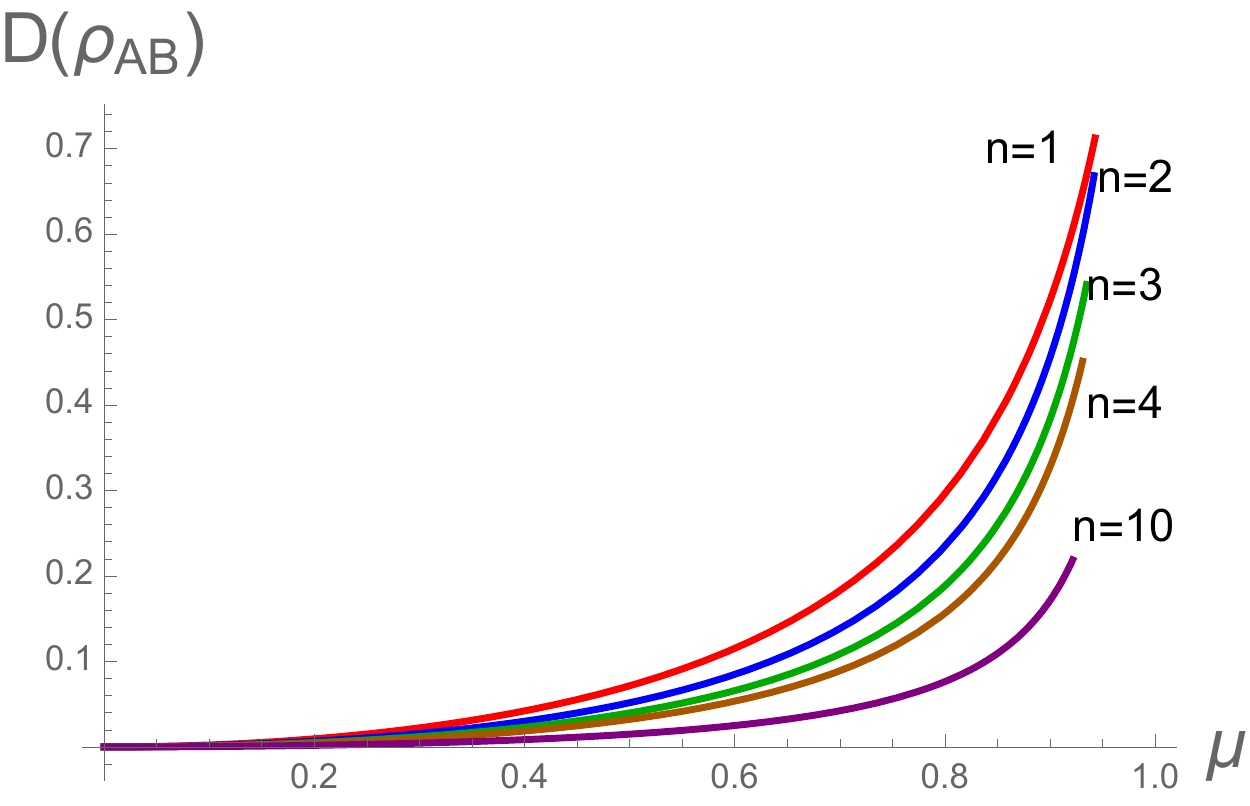}
  \includegraphics[width=5cm] {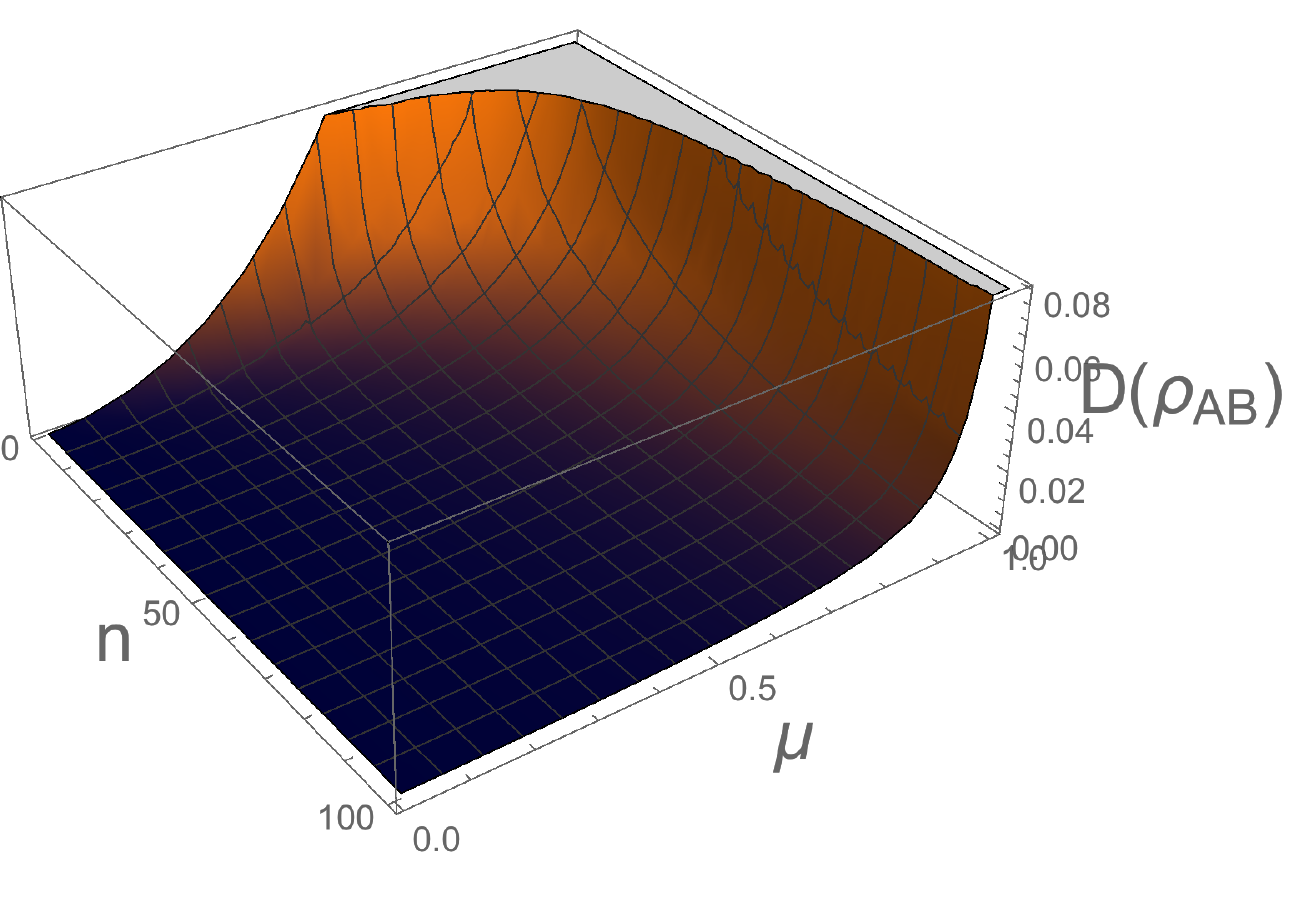}
  \caption{The behavior of the geometric quantum discord versus $\mu$ and $n$ .}
 \label{fig:discordPlot}
\end{figure}

Also, as $N_F^{\mathcal{ G  },A}$ qualitatively behaves very similar to various measures of quantum discord, we expect that these discussions also apply for them too. This could be seen from figure \ref{fig:discordPlot}, which as one could check, the behavior is qualitatively similar to $N_F^{\mathcal{ G  },A}$.

Another quantity, the Gaussian quantum discord could also be written as \cite{Liu:2019ent}
\begin{gather}
D(\rho_{AB})= S(\rho_A)-S(\rho_{AB})+ \underset{\Pi}{\text{inf}} \int p(z) S(\rho_B (z)) dz,
\end{gather}
where $\Pi= \{ \Pi(z) \}$ is a collection of positive operators in the form $\Pi(z) = D(z) \tau D^\dagger (z)$. Here $D(z)$ are the Weyl operators and $\tau$ is a n-mode Gaussian state. Additionally, $p(z)= \text{Tr} ( \rho_{AB} \Pi(z) \otimes I)$. 
\vspace*{6px}
From these plots again one could see that increasing mixing parameter would increase the discord. So as  dissipation and same sign charge would decrease the mixing parameter $\mu$, therefore they decrease the quantum discord between the two systems which then would decrease both the EoP and CoP. However, the behavior versus the mean photon number is opposite of the case of $N_F^{\mathcal{ G  },A}$.

\section{Modular Hamiltonian in QCD}
When the system is mixed, it could be simulated using some aspects of QCD models. Some interesting properties of these models could be seen in the connections between CFT characteristics and bulk reconstruction models. For instance, in these models, for various quantities such as Roberge-Weiss (RW) periodicity in lattice QCD or periodicity for $\mu_I/T=\theta$, or for pressure, entropy density, etc \cite{Ghoroku:2020fkv}, always a $2\pi$ periodicity is being observed, similar to CoP. These mixed states could be modeled by dual quark condensations as \cite{Ghoroku:2020fkv} $\sigma^{(n)}= \int_0^{2\pi} \frac{d\phi}{2\pi} \sigma(\phi) e^{i n \theta} d\phi$, where $\sigma (\phi)$ is the chiral condensate and the phase of the boundary condition is in the range of $0 \le \phi \le 2\pi$. Note that this phase is related to the dimensionless imaginary chemical potential, $\phi=\theta+\pi$. Some notion similar to the imaginary chemical potentials could also be defined for modular flows and modular Hamiltonians.

 The mechanism of getting information from the confining phase, using modular flow, would be similar to what has been employed in \cite{Chen:2019iro}. In the exact moment of transitions between confining and deconfining phases, the quark-gluon plasma phase could be considered as the island, and the de-confined surrounding gas as the heath bath. So the mechanisms of extracting information from the island to the bath using modular flow could give information about the interactions of information and modular flow in QCD as well. Note the quasi-localness of modular Hamiltonian would have a significant role in the behavior of QCD. The non-local piece of modular Hamiltonian for free fermions has been found as
\begin{gather}
H_0=-2\pi i   \sum_{l=1}^n \frac{1}{x-y} \left ( \frac{dz}{dy} \right )^{-1} \delta(y-x_l (z(x)), \ \ \ \ x_l(z(x)) \ne x,
\end{gather}
which agrees with our expectations from the behavior of entanglement structures for mixed states and intuitions from the bit threads construction.

\begin{figure}[ht!]
 \centering
  \includegraphics[width=7.4cm] {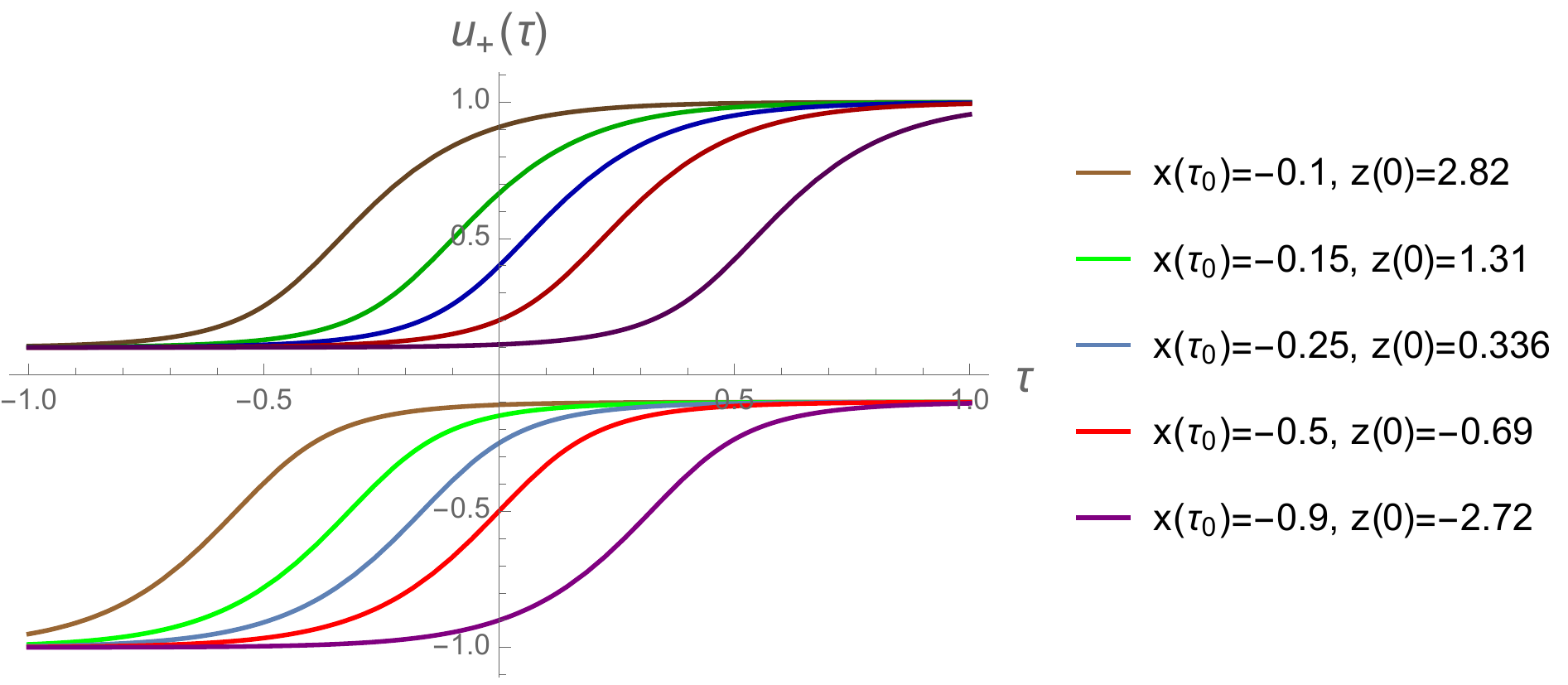}
    \includegraphics[width=7.4cm] {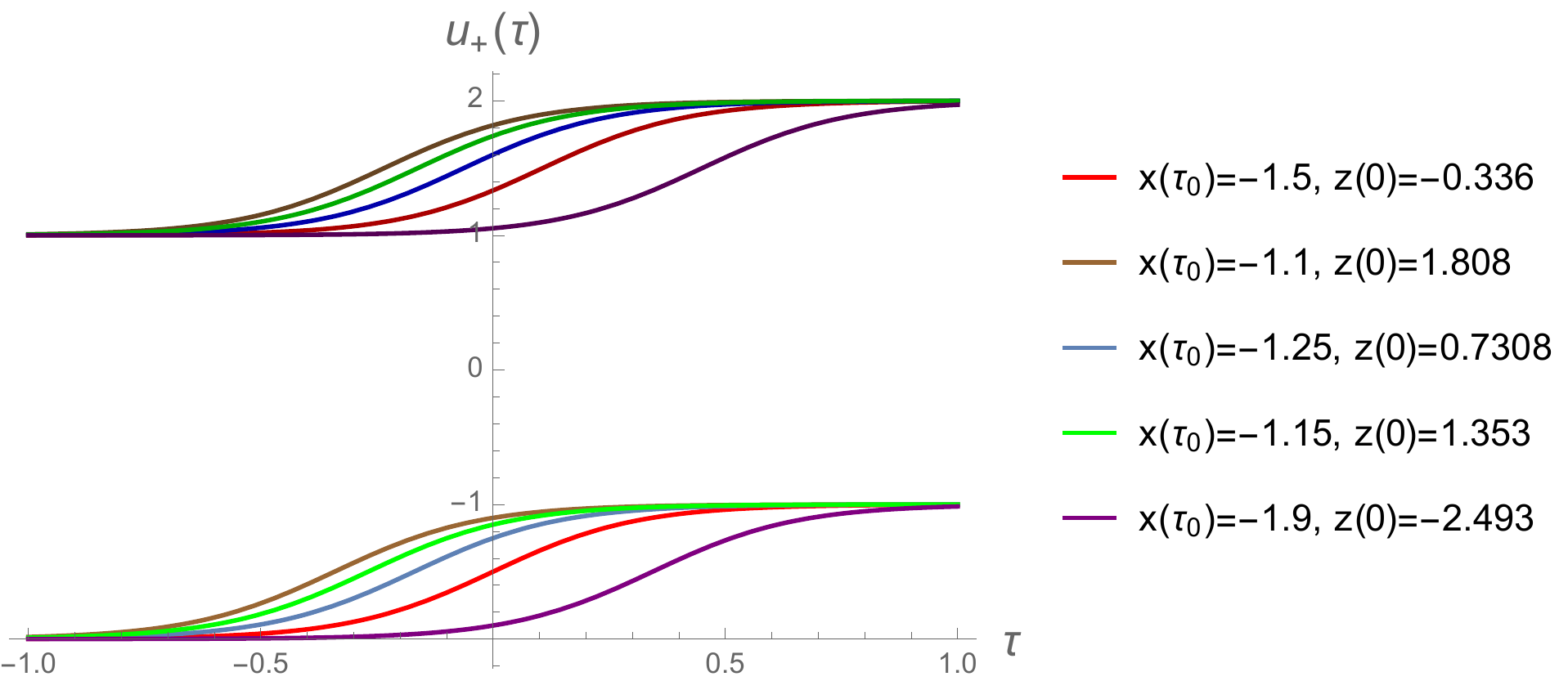}
        \includegraphics[width=7.4cm] {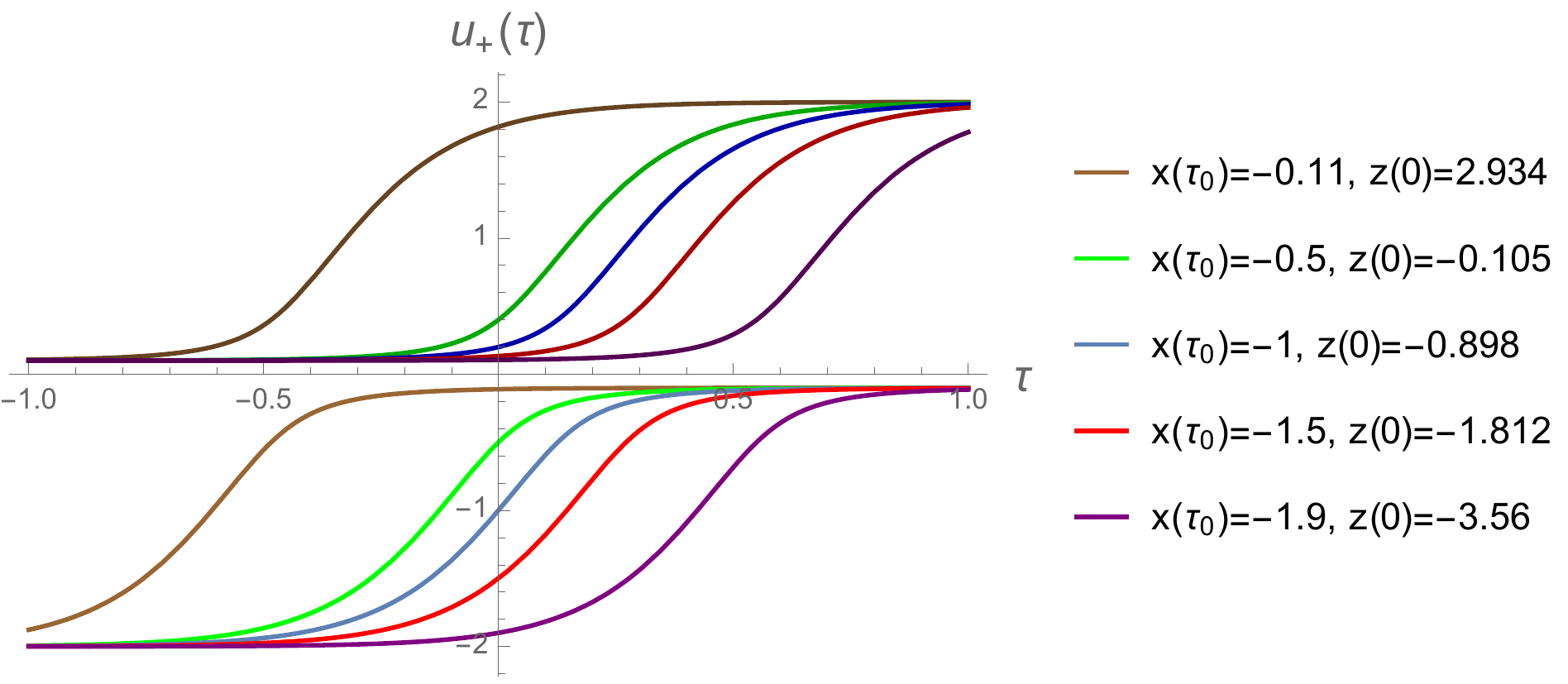}
            \includegraphics[width=7.4cm] {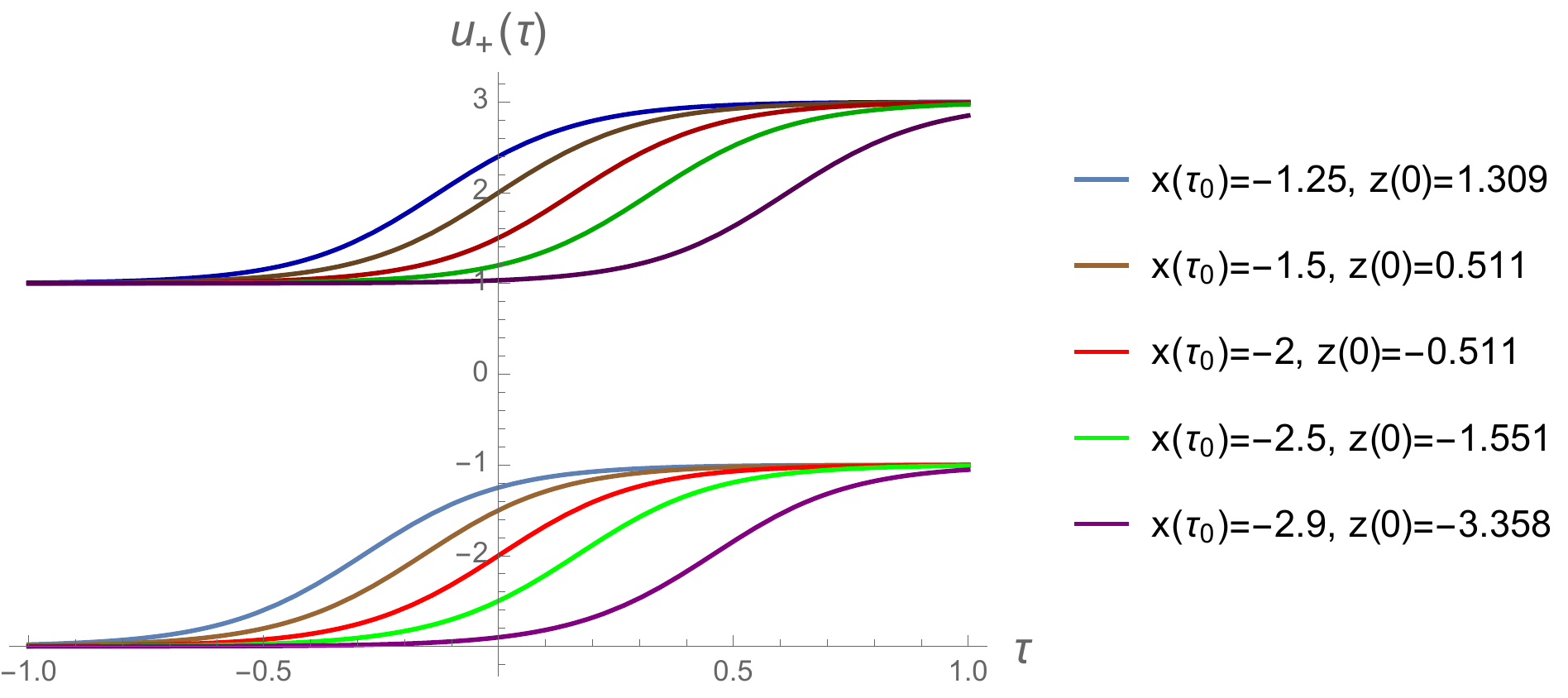}
  \caption{Various trajectories of left-moving operators during the flow, with different initial values for $x(\tau=0)$ and $z(0)$ for the setup of   two intervals of $\lbrack -1,-0.1 \rbrack \cup [0.1,1]$ in top left part, $\lbrack -2, -1 \rbrack \cup [1,2]$ in the top right part, $\lbrack -2, -0.1 \rbrack \cup [0.1,2]$ in the down left part and $\lbrack -3, -1 \rbrack \cup [1,3]$ in the down right part. }
  \label{fig:flowinitials}
\end{figure}

Note that many aspects of QCD could also be modeled by many-fermion systems. For instance in \cite{Cheamsawat:2020awh}, it has been shown that the Casimir energy of free fermions and holographic CFTs (which correspond to strongly coupled systems) behave very similarly. As the Casimir energy and modular Hamiltonian are directly related \cite{Wong:2018svs}, one could propose that modular Hamiltonian of QCD system, which could be derived by using holography, could be modeled by modular flow and modular Hamiltonian of the free fermion systems on two infinite strips which are close enough to each other to form a non-zero mutual information. The modular flow of such system could approximately model modular flow of QCDs.  Using the results of \cite{Casini:2009vk,Chen:2019iro} for free fermions in $2d$, the plots of various trajectories of left-moving operators during the flow have been shown in figure \ref{fig:flowinitials}. One interesting observation is that the modular flowed operators and the correlations among them could frame the structures of entanglement and complexity of purification. This could be seen from the relations 4.2, 4.5 and 4.9 of \cite{Balakrishnan:2020lbp} and  19, 21, and 57, 58 of \cite{Takayanagi:2017knl}. For our case, we should consider various operators at the same modular times but distributed along the minimal wedge cross section $\Gamma$, where the distribution function of the source have the form
\begin{gather}
g(s, s+i \tau')=-\frac{1}{2} \frac{\sinh (\frac{s} {2} ) }{\sinh (\frac{s+i \tau'}{2} ) \sinh (\frac{i \tau' }{2})}.
\end{gather}

If we consider two operators in the two regions, they are time-like separated and we could have still the following relations for the modular time as
\begin{gather}
s^\prime_{*, \pm}=s+ \log (\alpha \pm \sqrt{\alpha^2-1}),  \ \ \ \ \ \ \alpha= \frac{x_1^2+z_1^2}{x_1 z_1}.
\end{gather}
Note that the entanglement cut could be approached by setting $x^1 \to 0$ and $\alpha \to \infty$.  On the minimal wedge cross section, the modular time would have the same structure as those approaching the entanglement cut, $x^1 \to 0$ , so it would be like a boost operator but multiplied by a factor of two.  Also, modular flow would act as a local boost on the minimal wedge cross section line, $\Gamma$. For the two regions in pure $\text{AdS}_3 $, this then, would lead to the following relation for the entanglement of purification as
\begin{gather}
E_w(\rho_{AB})= \frac{c}{6} \log (1+2z +2 \sqrt{z(z+1)}),  \ \ \ \ \text{where} \ \ z=\frac{(a_2-a_1)(b_2-b_1)}{(b_1-a_2)(b_2-a_1)},
\end{gather}
where the two regions are $A= [a_1,a_2]$ and $B=[b_1,b_2]$ where $a_1 < a_2 < b_1 < b_2$. The fact that, along the imaginary axis, the modular cuts would be repeated with period of $2\pi$ is related to the fact that in $2d$, the CoP would be $2\pi$. This would also be related to the periodicity of QCD potential which is proportional to $2 \pi$, as we will explain further in the next part.

For more examinations, our specific setup would be two intervals of $\lbrack -1,-0.1 \rbrack \cup [0.1,1]$. This simple example could act as a toy model to further understand the behaviors of modular flow and the mechanisms of extraction of information from the islands. When the intervals are closer to each other, or when their widths are bigger, the modular flows are stronger. The system with a stronger flow is shown in the down-left part of figure \ref{fig:flowinitials} and the weaker one is shown in the top-right.  As for the QCD, these observations could denote that the parameters which make the confined phase more compressed, such as pressure, then make pulling information out of islands more difficult.

\vspace*{2px}

The relation for the angle $\theta (\tau) $ which determines how the fermion operator would flow under modular Hamiltonian as in \cite{Chen:2019iro} is
\begin{gather}
\theta(\tau)= \arctan \frac{(b_1+b_2-a_1-a_2) x_1(\tau)+( a_1 a_2-b_1 b_2) } { \sqrt{(b_1-a_1)(a_2-b_1)(b_1-a_1)(b_2-a_2)}}-
\arctan \frac{(b_1+b_2-a_1-a_2) x_1(0)+( a_1 a_2-b_1 b_2) } { \sqrt{(b_1-a_1)(a_2-b_1)(b_1-a_1)(b_2-a_2)}},
\end{gather}
where $x_1(\tau) \in \lbrack a_1, b_1 \rbrack$ and $z(\tau)= z(0) + 2 \pi \tau$. 

\vspace*{2px}
 \begin{figure}[ht!]
 \centering
  \includegraphics[width=10cm] {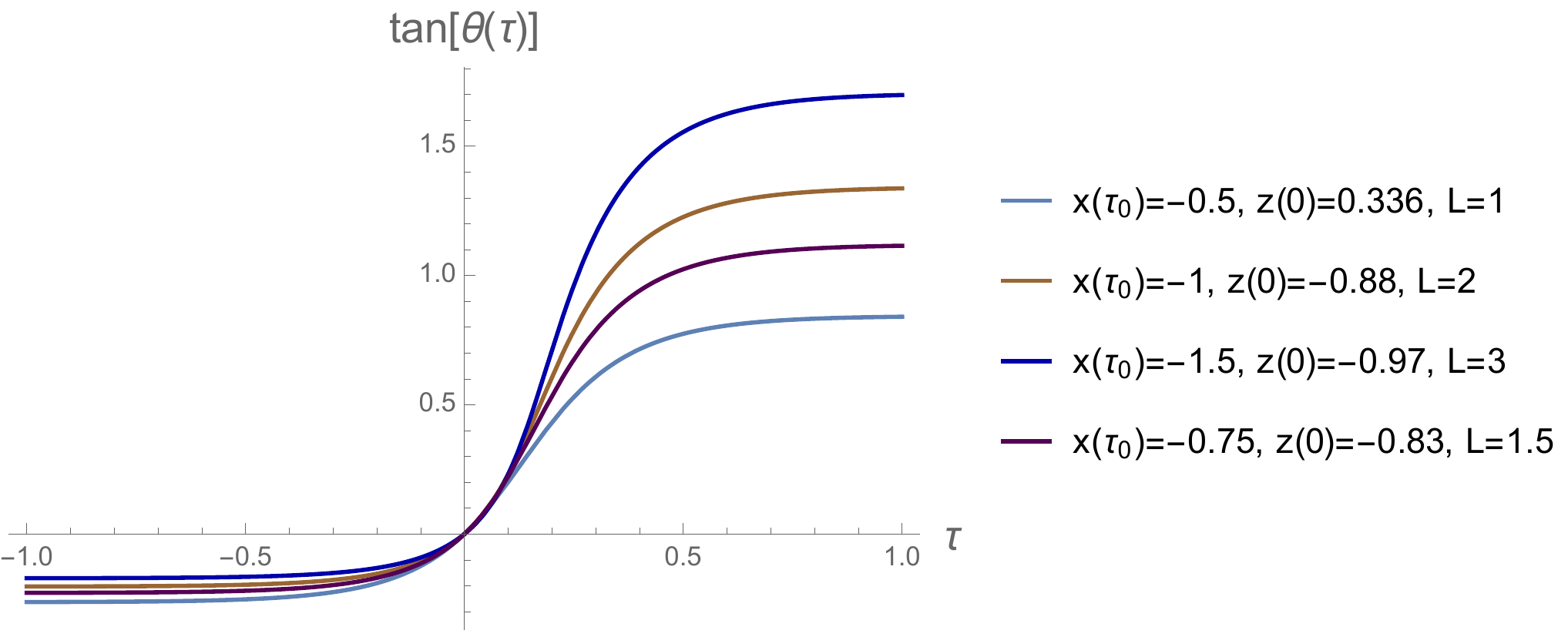}
  \caption{The plot of $\tan[\theta(\tau)]$ of \cite{Chen:2019iro} for various lengths of strips where the flow left mode and is passing from the midpoint of the strip. }
  \label{fig:flowplot}
\end{figure}
From figures \ref{fig:flowinitials} and \ref{fig:flowplot}, it could be seen that when the intervals are closer to each other or when the widths of them are bigger, the maximum point of $\tan \lbrack \theta(\tau)\rbrack$ would become higher. This then means that more information would be pulled out from one interval into the other, which then leads to the bigger values for entanglement and complexity of purification, as the modular flow through the minimal entanglement wedge cross section would be higher. So EoP and CoP would be proportional to $\theta$. From figure \ref{fig:flowplot}, it could be also seen that increasing the length of the intervals would increase the maximum of $\tan[\theta(\tau)]$, but it does not change the minimum much. However, one could note that still bigger intervals would have slightly bigger minimums.

Next, note that the operator reconstruction in entanglement wedge using modular evolved boundary operators would be written as
\begin{gather}
\tilde \Psi_1 (z(0))= \frac{1}{\sin \theta (\tau)} e^{- i \mathbf{K} \tau}  \tilde{\Psi}_2 (z(\tau)) e^{i \mathbf{K} \tau}- \frac{1}{\tan \theta (\tau)} \tilde{\Psi}_2 (z(0)) + \mathcal{O} (1/c).
\end{gather}
The structure of this relation is again two maps which used the modular flow and modular hamiltonian and also a projection. This combination is what we have seen for the bulk reconstruction from quantum recovery channels such as Petz map and connes cocycle flows. These operations indeed bring the information out to reconstruct the bulk. The main point is that, the bigger the angle $\theta$, the easier these projections would be performed and therefore the easier the modular flow and the bulk reconstruction would be. For the Petz map, these projections would be done by $\sigma_A$. So if the angle between $\omega_B$ and $\sigma_A$ be bigger, it would be easier to ``see'' the information inside the peninsulas and therefore the Petz map and recovery channel could be implemented easier. This makes EoP and CoP bigger as the result.

 \section{Dynamics of Correlation Exchanges} 
 
For studying the ``dynamics'' of correlations among mixed systems, various models such as shockwaves, void formation, numerical models of quenches, etc, could be used which here we employ some of them to get various results in our setup. One motivation to study the dynamics would be to get further information on how gauge connections and curvature in the bulk would be related. For instance, in various setups such as \textit{quenches} with different speeds, one could study the emergence of the dynamics of gauge fields and compare with the dynamics in the bulk. Using the connections between modular chaos and Riemann curvature, one could also understand better how information between mixed states would propagate through the quantum channels and the recovery ones.

 For studying the dynamical setup of \cite{Czech:2019vih} and how the excitations of CFT states would affect the modular Hamiltonians in the future causal cone, the Berry connection and the associated Berry phase, one could study the change of complexity of purification during quench similar to \cite{Yang:2018gfq}.  Then, using the quench setup, one could check how Berry connection could  be promoted to a dynamical object. One could also do a similar study for the QCD case and check how confinement could change the holonomy and modular Berry curvature. Again, we expect that studying the EoP and CoP for this case, gives us further information on the relations between gauge connection and bulk curvature.

  \subsection{Information speed in mixed setup}
 
 First, we consider various measures for the speed of information and correlation exchanges among the mixed system.  For example, our setup of moving these two strips closer to each other could be modeled by passing a shockwave. Then, using the results of \cite{Maldacena:2015waa}, we could notice that the holonomy of the edge modes would measure the soft graviton component of the two shockwaves commutators. The spreading of information inside quantum channels is in ``ballistic'' form, with the speed of butterfly velocity `$`v_B$''. The speed of spreading of the modular chaos would then be lower than $v_B$, since first these modes should create the bulk spacetimes and then the entanglement could spread inside it. For tracking of the spreading of quantum information in mixed and pure states, we could also use properties of modular scrambling modes and its relation with the speed of information $v_I$, entanglement speed $v_E$, and the butterfly speed $v_B$. So one could try to depict a connection between these various speeds and the properties of the modular flow and modular Berry curvature.

Our setup of figure \ref{fig:strips} could be modeled by considering the subsystem $A$ as the input one where via the quantum channels the information contained in it would expand ballistically toward the subsystem $B$ and bounce back via the recovery channels. These spreads of modular scrambling modes would be with the speed of $v_B=\sqrt{\frac{d}{2(d-1)} }$, and entanglement and mutual information would spread with the speed of $v_E= \frac{\sqrt{d} (d-2)^{\frac{1}{2}-\frac{1}{d} }   }  { \lbrack 2 (d-1) \rbrack ^{1-\frac{1}{d}} } $. It is expected that the information and modular chaos modes scramble with velocity $v_B$, but the minimum entanglement wedge and the cross section $\Gamma$ would be created with velocity $v_E$, as it is related to the network depth and ``vertical'' direction of the channel. Note that in most cases we have $v_E \le v_B$. Also, the signature of the information speed of the dynamics of bulk curvature could be detected on the boundary CFT.
 
Similar to the studies of \cite{Yang:2018gfq}, the change of complexity and complexity of purification during a thermal quench could be considered, which give intuitions on the speed of correlation exchanges between mixed systems. In \cite{Yang:2018gfq,Zhou:2019jlh}, entanglement of purification in the background of Vaidya-AdS spacetime has been worked out, where the metric is
 \begin{gather}
 ds^2=\frac{1}{z^2} \Bigg[ -f(v,z) dv^2-2 dz dv+dx^2+\sum_{i=1}^{d-2} dy_i^2 \Bigg], \ \ \
 f(v,z)=1-m(v) z^d.
 \end{gather}
The AdS radius is set to one and the mass function is taken as $m(v)=\frac{M}{2} \left (1+ \tanh \frac{v}{v_0} \right)$ where $v_0$ determines the quench speed.

In \cite{Yang:2018gfq}, for this background, the authors found that during the thermal quench, the critical separation $D_c$ would initially increase with time and then decrease to a smaller value in the late times. They have also found that the holographic mutual information at first grows by time but then decreases to a smaller value than the initial one. These process could be explained by void formations which we will explain in the next subsection. Initially, voids would be created when the system could exchange information but then they would get absorbed.  These process could also be explained using the zero modes and modular flows. So, first, the flow of modular zero modes increases and it would reach to a maximum and then when most of the mutual information have been exchanged, it would decrease. As has been found in \cite{Yang:2018gfq}, the equilibrium time would be approximately $l +D/2$, which is the time needed for the HRT surface of $2l+D$ and so $S_{2l+D}(t)$ to reach the equilibrium.  At this time the voids would have the maximum sizes. During the thermal quench, one could also track the changes in the \textit{holonomy} and the change of the modular flow and as the result the change in the Berry curvature of the bulk, which in this case would be similar to the passing of a shockwave in the geometry of the bulk, making the curvature to increase at first and then it would decrease.

By considering the process using the behavior of quasiparticles, one could get into more details to check how information would become scrambled between the two intervals. In \cite{Alba:2019ybw}, the behavior of quantum information scrambling after a quantum quench has been studied. It was shown there, that the behavior of the decay would be different for the integrable versus non-integrable systems, as for the later case, the decay would be much faster. This then would lead to the point that void formations and the stability of voids would be higher in integrable systems compared to the non-integrable ones. As the quasiparticles are non-stable or have short lifetime, voids would get absorbed faster as well.

 The general point here is to understand how information from the initial state would spread throughout the system. For doing that one could study different quench scenarios in various setups. One important aspect of the dispersion of quantum information is that entanglement and correlations would disperse \textit{globally} and in a non-local way, the fact that we have used in our new definition of \textit{volume interval} in \cite{Ghodrati:2019hnn} as a definition of complexity of purification for mixed states.
 
 In \cite{Alba:2019ybw}, it was proposed that in integrable systems, the ``quasiparticles'' would move ballistically and this way they would spread the initial correlations. This is also what we have observed by using the bit thread picture. However, note that after some time, the initial correlation would get dressed by the ``many-body effects'' and the thermodynamics of the system.

\begin{figure}[ht!]
 \centering
  \includegraphics[width=6.5cm] {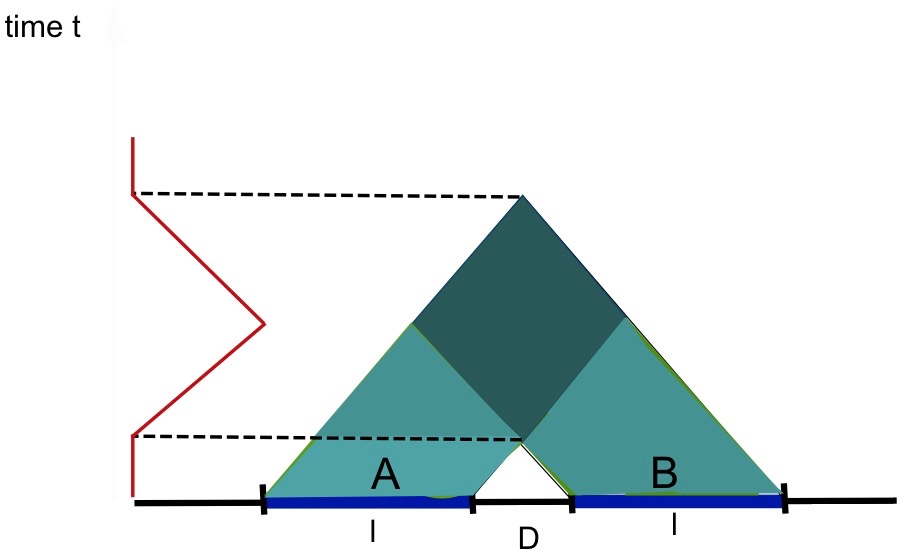} \hspace{1cm}
  \caption{In an integrable model, the mutual information behavior for two separated strips, $I_{A:B}$ has been shown by the red line \cite{Alba:2019ybw}. At each time $t$, $I_{A:B}$ would be proportional to the width of the darker region.}
 \label{fig:wedgeshadow}
\end{figure}

In integrable systems, after a quantum quench, the late time behavior of the quasi-particles could be described by the ``emergent thermodynamic macrostates or \textit{Generalized Gibbs Ensemble (GGE)''} while for the non-integrable systems, it would be a thermal ensemble and the entanglement entropy of EPR pairs become the ``thermodynamic'' entropy of the stationary ensemble. Generally, the non-integrable systems are better scramblers as there are no stable quasi-particles. The connections between quasiparticles and mutual information could then be studied using this. In fact, there are infinite species of quasi-particles in integrable models where in \cite{Alba:2019ybw} they were labelled by an integer $n$ and the quasiparticles of the same species were identified by the parameter ``rapidity'' $\lambda$, which for non-interacting particles would just be the momentum. In various condensed matter systems, where there are many degeneracies, bound Majorana fermions as the quasiparticle excitations could appear which instead of acting as a single particle, would  behave collectively with a ``monoidal'' or non-abelian statistics \cite{Stern2010:2017zfq}. These non-abelian anyons have practical applications in building topological quantum computers using the fractional quantum Hall effect.

In figure \ref{fig:wedgeshadow}, in our setup, we considered the case where each qubit at position $x_i$ has a frequency $\omega_i$ and therefore has a localized lump of energy $E_i= \hbar \omega_i$. 
 The solutions with $\omega=0$, i.e, the zero modes, would lie on the corresponding HRRT surface which is along the thread connecting it to its pair. Going from one bit thread to another needs at least additional energy of $\Delta E= \hbar \Delta \omega$ and therefore one needs to consider all the normal modes. These additional energy then create the curvature gradient in the bulk along the surface $\Gamma$.  Note that the gradient of the flow of the modular zero modes would also be related to MERA and the tensor network structure.  If the qubits have additional charge or mass, then $\Delta E$ would be different and therefore the bulk curvature and its gradient along $\Gamma$ would change and as the result, EoP and CoP would also change.

For a single quasiparticle with fixed velocity $v (\lambda)=v$, the mutual information is
\begin{gather}
I_{A_1: A_2} = \text{max} \Big(\frac{d}{2}, vt\Big) +\text{max} \Big( \frac{d+2\ell}{2}, vt\Big) -2 \text{max}\Big (\frac{d+\ell}{2},vt\Big),
\end{gather}
 which is non-zero only for $ d/(2v) <t< (d+2l)/(2v) $. Its maximum is at $t=(d+\ell)/(2v)$ and its height is proportional to $\ell$ which is the same result coming from holography. For the general case of quasiparticles with non-trivial dispersion, the contribution of all the species of quasiparticles could be derived as
 \begin{gather}
 I_{A_1: A_2} = \sum_{n} I^{(n)} = \sum_n  \int d\lambda s_n(\lambda) \Big[\text{max} \Big(\frac{d}{2}, v_n(\lambda)t\Big)+\text{max} \Big( \frac{d+2\ell}{2}, v_n (\lambda)t\Big) -2\text{max} \Big( \frac{d+\ell}{2}, v_n (\lambda) t\Big) \Big].
 \end{gather} 
It has been shown that in integrable models, the scrambling modes would follow the exponential behavior and the non-integrable cases would follow the algebraic behavior. One would expect that systems which have long-lived, but unstable quasiparticles such as confining models, have a cross over behavior between algebraic and exponential decay which could also be noticed from the behavior of modular Hamiltonian and void formations in QCD models.
 
In \cite{Couch:2019zni}, the speed that quantum information would spread in chaotic systems has been discussed. The three speeds were, information speed $v_I$, entanglement speed $v_E$ which is related to the growth rate of entanglement after a quantum quench, and third the butterfly speed $v_B$ which is related to the growth rate of perturbations in space. In \cite{Couch:2019zni}, it has been shown that the relationship between these speeds would follow the relation $v_I=\frac{v_E (\epsilon, f) }{1-f}$. Here, $\epsilon$ is the energy density of the initial state and $f$ is the entanglement fraction. The entanglement speed $v_I$ has a range between $v_E (f=0)$ to $v_B(f=1)$.  The range for the speed of modular chaos could then be found and its relationship with such speeds could be investigated. In \cite{deBoer:2019uem}, a bound for the speed of spread of modular scrambling modes has been found. As mentioned previously, after a perturbation,  in the limit of $s \to \pm \infty$, the matrix elements of the modular Hamiltonian of a QFT subregion could not grow faster than $e^{2 \pi s}$. For holographic CFTs, during the modular time of $1 \ll 2\pi s \ll \log N $, the growth of code subspace matrix elements of $\delta H_{\text{mod} }(s)$ has the bound \cite{deBoer:2019uem}
\begin{gather}
\underset{1 \ll 2\pi  | s | \ll \log N } { \lim_{1 \ll N} }  \big | \frac{d}{ds} \log F_{ij} (s) \big | \le 2\pi, 
 \ \  \ F_{ij} (s) = \Big | \langle \chi_i  | e^{i H_{\text{mod}} s} \delta H_{\text{mod}} e^{-i H_{\text{mod}} s} | \chi_j \rangle \Big | \nonumber.
\end{gather}
There is a connection between this bound of $2\pi$ on the maximum rate of growth of modular scrambling modes and entanglement and information speed which were mentioned before. Using that, the connections between Hayden-Preskill protocol \cite{Hayden:2007cs}  which tracks information as a function of time, and modular scrambling modes in modular time could be studied.
 
 Finally note that at the critical distance between the intervals, and in the phase transition moment,  the modular change, and the formation of the new structure for the modular flow would happen by the ``tsunami velocity'' which is bounded by the speed of light.  Also as found in \cite{Casini:2015zua}, the spread of the entanglement and also the modular flow would be highly sensitive to the initial entanglement pattern.

 \subsection{Void formation in mixed states}
In \cite{Liu:2019svk}, the void formation in CFTs and its links to black hole entropy and entanglement have been studied. These void formations would have interesting implications for the multipartite structure of entanglement entropy and mutual information. In chaotic systems the distribution of voids would be random, but in our mixed state setup, the structure of such voids would be integrable, or even just plain ``linear''. Therefore, further analysis of this void distribution in mixed states and the connections with EoP and CoP of entanglement wedge would be interesting.

During the evolution of the system, the void formation is responsible for generating mutual information and multipartite entanglement among the disjoint intervals. So it is expected that entanglement and complexity of purification would be directly related to the probability of void formation as well. So the bigger the voids, the higher the mutual information between separated regions and also the higher the entanglement and complexity of purification would be. Also, the volume of the voids, and complexity or CoP, would be interconnected as well. On the other side, void formation could also be characterized and quantified by the number of quantum gates needed to purify the systems and therefore by the EoP and CoP.   Also, for the same-sign charged and also dissipative systems, we would expect that the probability of void formations would be lower. So,  again we would arrive to the results we already got, that these parameters would decrease EoP and CoP as we have observed in \cite{Ghodrati:2019hnn} and also from other models studied here.

The probability of void formation for any operator $O$ to form a void in any subsystem $A$ or $B$ is related to the correlations as \cite{Liu:2019svk}
 \begin{gather}
 P_O^{(A) \text{or}(B)}(t) = \frac{\text{Tr} \big [ (O^{(1)} (t) )^\dagger O^{(1)} (t) \big] }{\text{Tr}(O^\dagger(t) O(t))},
 \end{gather}
  where the time-evolved operator $\mathcal{O}(t)$ could be written as $ \mathcal{ O} ( t )  = \mathcal{O}_1 (t)+ \mathcal{O}_2 (t), \ \mathcal{O}_1(t)= \tilde{\mathcal{O}}_{\bar{A}} \otimes \mathbf{1}_A$.
The probability of evolution of an operator $\mathcal{O}_\alpha$ to another operator $\mathcal{O}_\beta$ and also the capacity of quantum error correction channels are actually related. Under the time evolution we have
\begin{gather}
\mathcal{O}_\alpha (t)= U^\dagger(t) \mathcal{O}_\alpha U(t) = \sum_\beta c^\beta_\alpha (t) \mathcal{O}_\beta,
\end{gather}
and $| c_\alpha^\beta (t) |^2$ is the probability of evolution, which we conjecture that is directly related to the channel capacity, implying the connections between void formation and quantum error correction. This number would also be related to the complexity and complexity of purification. More precisely related to $N_A(t) \equiv \sum_{\alpha \in I} \sum_{\beta \in A} | c_\alpha^\beta (t) |^2$, introduced in equation 2.11 of \cite{Liu:2019svk}, which is the expected number of operators in the set $I$ contained in $A$ after passing the time $t$. It is also connected to the complexity of state with density matrix $\rho_A(t)$. Also, the number  $N(A,B;t) \equiv \sum_{\alpha \in I\cap B} \sum_{\beta \in A} | c_\alpha^\beta (t) |^2$, characterizes the number of initial operators in $I$ from the region $B$ which at time $t$ would become contained in $A$. This number then would be related to the complexity of purification between regions $A$ and $B$.

Another point worths to mention here is related to the linear growth of entanglement and complexity due to the ``ballistic'' operator growth, which would be true in both chaotic and integrable systems. This ballistic behavior of operators is also responsible for the decreasing behavior of EoP and CoP after decreasing the same sign-charges $q$ and the dissipative parameter $m$. Similarly, for other correlation measures, the number $\sum_{\alpha \in I \cap B} P_{\mathcal{O}_\alpha ^{(A)} } (t)$ could be used.

  \begin{figure}[ht!]
 \centering
  \includegraphics[width=3cm] {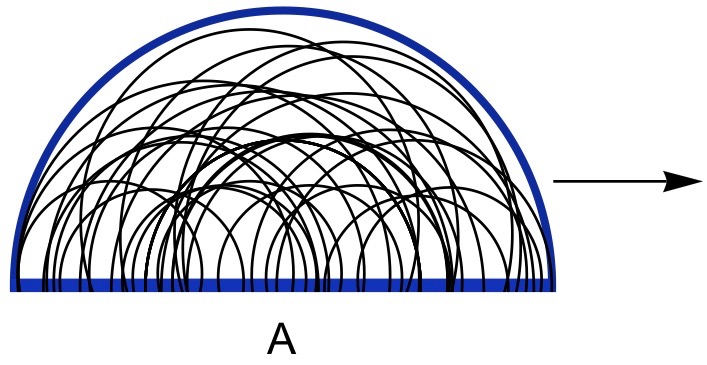} \hspace{0.3cm}
   \includegraphics[width=3cm] {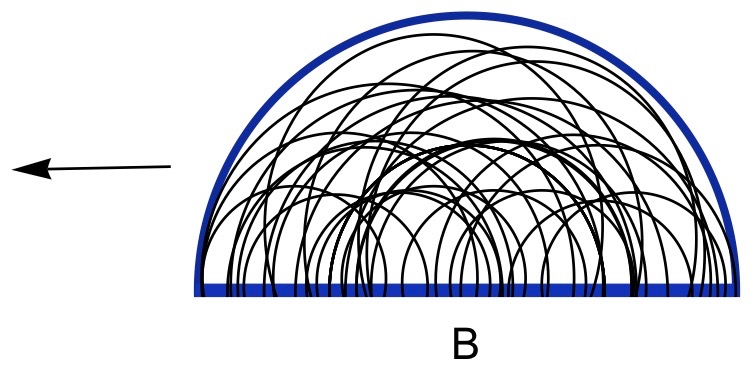} \hspace{0.3cm}
     \includegraphics[width=4cm] {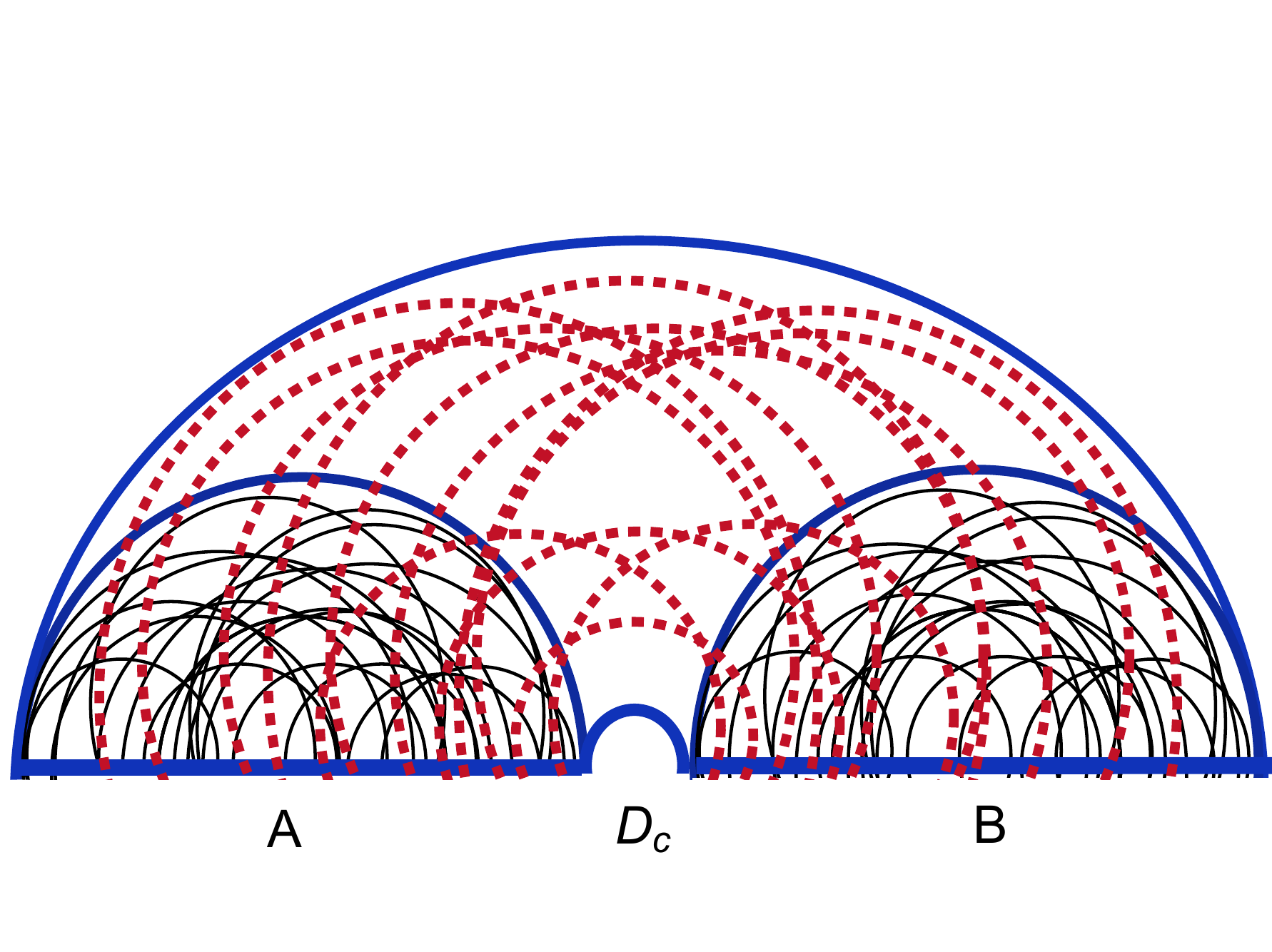} \hspace{0.3cm}
  \caption{The creation of correlations at $D_C$ is shown. This structures could also be explained using the void formation of \cite{Liu:2019svk}. }
 \label{fig:EWvoidformation}
\end{figure}

We can simulate our setup of two strips and the correlation evolution among them using the model of void creation. One could imagine that at first, i.e, $t=0$, when the two subregions are far away from each other, the density operator for each one could be written as
\begin{gather}
\rho_{0A}=\ket{\psi} \bra{\psi}=\frac{1}{d} \mathbf{1}+\hat{\rho}_{0A}, \ \text{Tr}\hat{\rho}_{0A}=0, \ \ \ \ \ \ 
\rho_{0B}=\ket{\psi} \bra{\psi}=\frac{1}{d} \mathbf{1}+\hat{\rho}_{0B}, \ \text{Tr}\hat{\rho}_{0B}=0.
\end{gather}
Then, after they become close enough to each other, similar to \cite{Liu:2019svk}, one could write
\begin{gather}
\hat{\rho}(t)= \mathbf{1}_A \otimes O_B+ O_A \otimes \mathbf{1}_B+\tilde{O}_A \otimes \tilde{O}_B.
\end{gather} 
The first term corresponds to the void formation in $A$, and the second one corresponds to the void formation in the subsystem $B$. Then, similar to the case of void formation between black hole and its radiation in \cite{Liu:2020gnp}, one could write $ \rho_A=\frac{1}{d_A} \mathbf{1}_A+d_B O_A, \ \rho_B=\frac{1}{d_B} \mathbf{1}_B+d_A O_B$. The reduced density matrix in the system $A$ is related to the void formation in its complement part including the system $B$, and vice versa. Then, the EoP, CoP and modular flow evolution could be modeled by the probability plus the higher moments of these void formations. For the EoP case, the probability of void formation between $A$ and $A^\prime$ to find $\rho_{AA^\prime}$ and also between $A$ and $B$ to determine $\rho_{AB}$ can be calculated. The first one would be proportional to $\frac{1}{d_B^2}$ and the second one would be proportional to $\frac{1}{d_{A} d_{B} }$, where $d$ here denotes the dimension of the Hilbert space.

 Similar to the story of the Page curve, the phase transitions between the two structures of RT extremal surface, and the sudden appearance of the $E_W$ case in the lower picture, could be related to the change of dominance between the identity and the void formation parts, as shown in figure \ref{fig:EWvoidformation}. So considering
 \begin{gather}
 e^{-(n-1) S^{(A+B)}_n}=\frac{1}{d^{n-1}_A}+\frac{1}{d^{n-1}_B}+...+d^n_A \text{Tr}_B O^n_B+d^n_B \text{Tr}_A O^n_A,
 \end{gather}
the first two terms would be dominant before the phase transition and the last two terms would dominate after the phase transition when the mutual information and EoP become non-zero as the two intervals get closer to each other to form mixed correlations. So different parts of the reduced density matrix in these series would become dominant at different stages. As one would expect, this behavior of the correlations in mixed setups would be similar to the behavior of replica wormholes in JT gravity studied in \cite{Penington:2019kki}.

The modular operators mixing and the correlation exchange and also the jump in the mutual information between the two strips at the phase transition point, as shown in figure \ref{fig:EWvoidformation}, would be due to the transferring of information from the first strip to the second one and then, one could track information using the Hayden-Preskill like process. The second situation, could be considered as a typical state similar to BH case and the results could be derived by averaging over the Hilbert space of the subsystems. We expect that if the two subregions become close to each other fast enough, some aspects of modular chaos would behave similar to \textit{vortices} where such new mathematical structures could be formulated using vortex dynamics. Also, the butterfly effects in modular chaos could be noted there. For any numerical simulation of the dynamics of exchange of correlations, operators and information between these states, one could model these states using the quantum Markov chains \cite{cite-key} where their patterns of correlations are very orderly. We leave the detailed numerical calculations for the future projects.

 \section{Discussion}
 
 In this work, various models of bulk reconstruction and the connections between them have been studied, specifically for a setup of mixed states of two intervals, entanglement wedge reconstruction through modular Hamiltonian and modular flows and also quantum recovery channels have been investigated. Also, their connections with the behavior of mutual information, entanglement and complexity of purifications have been explored. Specifically, we used the results for EoP and CoP of a charged and massive gravity backgrounds and compared various results with each other. The structures of zero modes and modular flows through minimal wedge cross section, explicitly for cases with dissipations and same sign charges have been probed. The interconnections between quantum recovery channels, in particular the Petz map, and modular flow have also been looked into where ideas such as  eigenstate thermalization hypothesis have been employed.

Furthermore, the links between modular Berry phase and complexity were probed where the already known behaviors of EoP and CoP have been used along the way. Also, we compared the structure of the modular Hamiltonians for the connected versus disconnected entanglement wedges and specially the effects of singularities have been tracked. In addition, the effects of dissipations and charge on CC flows and kink transforms have been studied. Then, models of OPE block and the duality of geodesic operator/OPE block for the bulk reconstruction have been applied. Also, the ``CFT Uhlmann phase/bulk symplectic form'' dictionary has been used to study the mixed entanglement wedge constructions and also the effects of dissipations and charge on each side have been discussed. We also commented on the connections with the quantum capacity and modular Hamiltonian.
 
Next, to get better intuitions of the quantum correlations of mixed states in our setup, we studied a particular correlation measure and studied the effects of parameters such as mixing and mean photon numbers. We also studied some exact forms of modular Hamiltonian in simple models and numerically investigated the effects of parameters such as lengths of strips and the distance between them. We also commented on various information speeds in our mixed setup and the relationships among them and also the connections with the bound on modular scrambling modes. In the study of the dynamical behaviors of correlations, we also considered void formations and the role they play in the phase transitions.

 Our studies was specifically for CFTs. These studies could also be done for warped CFTs as well. So, for instance the holonomy of Berry connection along the path in the $U(1) \times SL(2,R)$ group manifold and the Berry phase on Virasoro-Kac-Moody orbits \cite{Oblak:2017ect} could be studied. In \cite{Oblak:2017ect}, using the Maurer-Cartan form of the Virasoro group, the Berry phase had been computed. Using these calculations, the same could be done for the case of Kac-Moody as well and the results  could be compared with what we have found for complexity in warped CFTs \cite{Ghodrati:2017roz, Ghodrati:2019bzz}.

Also, recent ideas of the relationships between connecting CFTs and the dual domain walls in the AdS \cite{Simidzija:2020ukv,Ooguri:2020sua}, in our setup of entanglement wedge reconstruction, right in the moment when the two boundary regions become close enough could be studied.  One could imagine that the two regions are separated by some type of codimension-one brane and when the two regions become close enough, the two corresponding branes collapse and merge with each other as shown in \cite{Simidzija:2020ukv}. Similar to \cite{Simidzija:2020ukv}, the bulk region dual for each CFT could have matters and at the interface, one could consider a Gibbons-Hawking-York boundary term and another action for the matter on the brane. In \cite{Simidzija:2020ukv} the tension of the interface brane is constant, but based on our analysis in \cite{Ghodrati:2019hnn}, this could not be true and the tension should have a profile with a decreasing gradient.

These modular flows could also have some similarities with the ``\textit{fracton}'' quantum matter states.  The implications of Majorana islands  could also be investigated for the case of modular flows in the mixed setups. Specially for the case of massive gravity, the interactions between the fractons and gravitons would be very interesting, see \cite{PhysRevD.96.024051,PhysRevB.99.155126}.   The connections between these models and the quantum universal recovery channels and modular Berry flows and also the bit-thread models could also be investigated.

Defining a notion of entanglement ``monodromy'', specially for studying the structure of entanglement around singularities, similar to the notion of entanglement ``holonomy'' of  \cite{Czech:2018kvg}, would also be interesting. The connection between emergence of space, modular Berry flow and other novel and interesting ideas such as AdS/Deep Learning  or AdS/CFT as a deep Boltzmann machine \cite{Hashimoto:2019bih} could be studied. There should also be a connection between the applicability of replica trick and the specific properties of modular Hamiltonian which lead to the entanglement wedge reconstruction and the validity of Hayden-Preskill decoding criterion \cite{Penington:2019npb,Chen:2019gbt}. The existence of multiple replicas, the ability of modular Hamiltonian to sews field theories, and the connectivity of geometries could be studied. The Markovian properties of Hawking radiation and the modular Hamiltonian for the vacuum would support our guess.  The symplectic form for the correlation between multipartite systems or between wormholes could also be studied. The connections between all these arguments and the bulk reconstruction using Hartle-Hawking wavefunction \cite{Jafferis:2017tiu} would be compelling. The Wheeler-DeWitt wave function and its formalism could also be another method of bulk reconstruction. In the picture of traversable wormhole of \cite{Gao:2016bin}, the coupling in the form of $ V=\frac{g}{n} \sum_{i=1}^n Z_i^L Z_i^R $ has been added where $g$ is small and $n$ is large which in the bulk has the net effect of pushing the signal down and make the teleporting between the two boundaries possible. The effects of such terms in our structure, for transformation of information between the two subsystems and the net effects on the phase transitions could also be studied. Very recently, in \cite{Longo:2020amm}, a new formula for the massive modular Hamiltonian of a unit space ball has been found. It has been noted that the mass parameter there would also decrease the matrix elements of modular Hamiltonian. It would be interesting to check how in their setups the mass parameter also affects various mixed correlations, CC modular flow, scrambling modes, quantum recovery channels. Also, note that there the Green's function has a form of Yukawa potential $e^{-m r}/ 4\pi r$ which could then be used for removing the singularity of phase diagrams of mutual information. These methods of entanglement wedge reconstruction and emergence of bulk spacetimes could also be connected to the new studies of mechanism of precision microstates counting of black hole entropy using topologically twisted index \cite{Benini:2015noa},  localization techniques, Bethe-ansatz formulation and $\mathcal{I}$-extremization. This point came to mind because of the procedure they employed as they uses imaginary chemical potentials to find the number of black hole microstates which then would lead to the imaginary result for the complexity of purification of mixed states. We hope to address some of these problems in the future works.

\section*{Acknowledgement}
I would like to thank Mahdi Torabian and Xiaomei Kuang for their help and supports and for useful discussions. This work has been supported by Iran's National Elites Foundation (INEF) and Chinese Postdoctoral Science Foundation.

 \medskip

\bibliography{BerryPRD}
\bibliographystyle{JHEP}
\end{document}